\documentclass[nofootinbib,aps,superscriptaddress,preprint,floatfix]{revtex4}

\usepackage{amsmath,amssymb}
\usepackage{graphicx}
\usepackage{cancel}
\usepackage{textcomp}

\def\beq{\begin{equation}}
\def\eeq{\end{equation}}
\def\beqa{\begin{eqnarray}}
\def\eeqa{\end{eqnarray}}
\def\bea{\begin{eqnarray}}
\def\eea{\end{eqnarray}}
\def\beq{\begin{equation}}
\def\eeq{\end{equation}}
\def\tev{\,{\rm TeV}}
\def\gev{\,{\rm GeV}}
\def\to{\rightarrow}
\def\lsim{\mathrel{\raise.3ex\hbox{$<$\kern-.75em\lower1ex\hbox{$\sim$}}}}
\def\gsim{\mathrel{\raise.3ex\hbox{$>$\kern-.75em\lower1ex\hbox{$\sim$}}}}

\newcommand{ \slashchar }[1]{\setbox0=\hbox{$#1$}   
   \dimen0=\wd0                                     
   \setbox1=\hbox{/} \dimen1=\wd1                   
   \ifdim\dimen0>\dimen1                            
      \rlap{\hbox to \dimen0{\hfil/\hfil}}          
      #1                                            
   \else                                            
      \rlap{\hbox to \dimen1{\hfil$#1$\hfil}}       
      /                                             
   \fi}                                             %

\def\etmiss{\slashchar{E}_{T}}

\begin{document}
\preprint{\vbox {\vspace{2cm}
\hbox{MADPH--08--1526}
}}
\vspace*{2.7cm}

\title{ Collider Signatures for Heavy Lepton Triplet \\ in Type I+III  Seesaw }
\vspace*{-1cm}

\author{Abdesslam Arhrib} 
\affiliation{D\'epartement de Math\'ematiques, Facult\'e des Sciences et Techniques\\
  B.P 416 Tanger, Morocco}

\author{Borut Bajc}
\affiliation{J.~Stefan Institute, 1000 Ljubljana, Slovenia} 

\author{Dilip Kumar Ghosh}
\affiliation{
Department of Theoretical Physics and Centre for Theoretical Sciences\\
Indian Association for the Cultivation of Science\\
2A $\&$ 2B Raja S.C. Mullick Road, 
Kolkata 700 032, India }

\author{\\ Tao Han}
\affiliation{Department of Physics, University of Wisconsin,  Madison, WI 53706, USA} 

\author{Gui-Yu Huang}
\affiliation{Department of Physics, University of California, Davis, CA 95616, USA}

\author{Ivica Puljak}
\affiliation{FESB, University of Split, Split, Croatia}

\author{Goran Senjanovi\'c }
\affiliation{International Centre for Theoretical Physics,  Trieste, Italy }
 \vspace*{-0.5cm}

\begin{abstract}
The minimal SU(5) theory augmented by the fermionic adjoint representation 
restores the coupling constant unification and gives realistic neutrino masses and 
mixing through the hybrid Type I and Type III seesaw. The crucial prediction of the 
theory is an SU(2) lepton triplet with the mass below TeV. 
We study the signature of these heavy leptons and propose the strategy to
test this mechanism at the hadron and lepton colliders. The smoking gun evidence 
of the theory is $\Delta L=2$ lepton number violation through events of a pair of like-sign leptons 
plus four jets without significant missing energy
at hadron colliders. We find that via this unique channel, the heavy lepton can be searched
for up to a mass of 200 GeV at the Tevatron with 8 fb$^{-1}$, and  up to 450 (700) GeV at the LHC of 14 TeV 
C.M.~energy with 10 (100) fb$^{-1}$. 
The signal rate at the 10 TeV LHC is reduced to $60\% - 35\%$ for a mass of 200$-$700 GeV.
We also comment on how to distinguish this theory from other models with similar heavy leptons. 
Finally, we compare the production rates 
and angular distributions of heavy leptons in $e^+e^-$ collisions for various models.
\end{abstract}

\def\thepage{{}}
\maketitle
\def\thepage{\arabic{page}}

\section{Introduction}
\label{section1}

Within the context of the Standard Model (SM),  neutrino masses 
require the presence of the higher dimensional operator, symbolically written
as \cite{Weinberg:1979sa}
\begin{equation}
\label{weinberg}
{\cal L}_{eff}=y_{eff}\frac{LLHH}{M}\;,
\end{equation}
where $L$ is the usual leptonic doublet $L=(\nu ,e)_L$,  $H$ is 
the standard model Higgs doublet $H=(h^+,h^0)$, and $y_{eff}$ is an effective
Yukawa coupling. This operator breaks the lepton number by two units ($\Delta L=2$).
Once the electroweak symmetry is spontaneously broken by the Higgs vacuum 
expectation value $(v)$, neutrinos get Majorana masses 
\begin{equation}
m_\nu= y_{eff}\frac{v^2}{M}\,.
\end{equation}
The smallness of neutrino masses from the observation tells us that 
either $M$ is very large $M\gg v$, or $y_{eff}$ must be small, 
both cases equally natural from the technical point of view.

Observation of the lepton number violation would be the definitive test for the
generation of neutrino Majorana masses as in Eq.~(\ref{weinberg}). However, 
if we wish to understand the operator to a better level or to derive  $m_\nu$,
we need a completion of the SM which leads to Eq.~(\ref{weinberg}). 
There are three possible extensions  \cite{Ma:1998dn} to the SM which, 
after integrating out the
new heavy states, will lead to Eq.~(\ref{weinberg}): 
\begin{itemize}
\item[1.] fermionic singlets, called right-handed neutrinos; this is called 
Type I seesaw \cite{seesaw};

\item[2.] a scalar SU(2)$_L$ triplet ($Y=2$); this is called 
Type II seesaw \cite{Magg:1980ut,Lazarides:1980nt,Mohapatra:1980yp};

\item[3.] a fermionic SU(2)$_L$ triplet ($Y=0$) called Type III seesaw \cite{Foot:1988aq}.
\end{itemize}
The Type I  and Type II  seesaw scenarios have been studied in depth in the literature, 
while the Type III exploration is somewhat limited. This is quite 
natural since Types I and II are predicted in left-right symmetric theories, 
and thus can be incorporated in a predictive grand unified theory based 
on SO(10). The inclusion of right-handed neutrinos offers the simplest possibility
of gauging the anomaly free $B-L$ accidental symmetry of the SM, and the Higgs 
triplet provides the simplest way of generating their masses 
\cite{Mohapatra:1980yp} in L-R symmetric theories 
\cite{Pati:1974yy,Mohapatra:1974hk,Senjanovic:1975rk,Senjanovic:1978ev}.

It is tempting to address the issue of $M$ in 
the context of the minimal SU(5), which is known to have failed for two profound reasons: 
the gauge couplings do not unify and neutrinos are massless. A simple way of curing 
both is to add an adjoint $24_F$ fermionic representation \cite{Bajc:2006ia}. 
Neutrino masses result from the mixture of Type I and Type III seesaw since 
$24_F$ contains both the singlet $S$ and the triplet $T$, with an 
interesting prediction of one massless neutrino. Even more interestingly, 
the unification constraints force the triplet to be light: $M_T\lsim 1$ TeV. 
In short, we have a predictive grand unified theory that can be tested in future experiments. 
This motivates us to study the observability of the fermionic triplet with the predicted
characteristics in the collider experiments. 

We should stress that minimality 
plays a crucial role in the physics of the triplet. If, and only if minimality is assumed, 
the triplet is predicted to be light and to be a source of neutrino mass. For example, 
we are not allowed to add right-handed neutrinos as SU(5) singlets and any higher 
dimensional operator should be suppressed by a cutoff above $M_{GUT}$. As 
a consequence the theory does not allow for more generations, for otherwise 
the fourth generation neutral lepton is forced to be another light neutrino. 
The higher dimensional operator would give an infinitesimal mass, and since there 
is no room for seesaw, this particle would be effectively massless.

We calculate the production rate and subsequent decays of the triplet fermion 
responsible for Type III seesaw at hadron colliders. 
We focus on the gold-plated events: the $\Delta L=2$ 
channels, of same-sign dileptons  accompanied by multiple jets without significant
missing transverse energy. These events as 
originally discussed a long time ago \cite{Keung:1983uu} offer a possibility of directly 
observing the Majorana nature of the neutral triplet lepton and are characteristic of the 
seesaw mechanism that leads to Majorana light neutrinos. 

This was already studied for the LHC 
\cite{Ma:2002pf,Franceschini:2008pz,delAguila:2008cj,delAguila:2008hw} 
for the seesaw with three triplets. In our case we are dealing with a well defined 
theory with further constraints: a single triplet in the TeV region and one massless neutrino. 
In particular, we establish the correlations between the lepton flavors in the LHC 
signal and the neutrino mass patterns. The decays of the triplet are characterized by the 
same Yukawa couplings that enter into the neutrino mass matrix and so the collider 
signatures can be of great help in determining the rest of the neutrino mixings and phases.
 
It is worth emphasizing 
that the textbook example of $\Delta L=2$ violation, the neutrinoless double beta decay, 
depends on the nature of the seesaw completion of the standard model and thus 
by itself cannot point to the origin of the Majorana neutrino mass. 
 
The rest of this paper is organized as follows. In section \ref{section2} 
we describe the salient features of the SU(5) theory with an adjoint 
fermion. We briefly review the unification analysis which forces the triplet 
to be light, below TeV. 
In Sec.~\ref{sec2.5}, 
we then explore the constraints on the model parameters based on the
neutrino oscillation experiments. We find interesting 
correlations between the couplings and the neutrino mass patterns. 
In Sec.~\ref{section3}, we study the decays of the fermionic triplet. 
We once again present the correlations between the decay branching
fractions and the neutrino mass and the possible Majorana phases.
Section V is the core of our paper. 
Here we carefully discuss the smoking gun predictions of the 
theory: same-sign dilepton plus four jets without significant missing energy. 
We estimate the SM backgrounds 
and develop the strategies to separate the signal from  backgrounds. 
We discuss the discovery potential of 
both the Tevatron and LHC and demonstrate that the Tevatron could probe the 
lepton triplet up to a mass 200 GeV with a 8 fb$^{-1}$ integrated luminosity,
while the LHC could probe the triplets up to a mass 450 (700) GeV with 
a luminosity of 10 (100) fb$^{-1}$. 
We also calculate the signal rate at the 10 TeV LHC.
In particular, we 
provide some general remarks on how to distinguish the lepton triplet from 
other heavy leptonic states. As a complementary study in this regard, we
present in Sec.~VI 
the results for cross sections and distributions at an $e^+e^-$ Linear
collider for a few representative models.
Finally, in Sec.~VII, we conclude with a critical discussion and outlook.

\section{ SU(5) with an adjoint fermion }
\label{section2}

A practitioner of grand unification is faced with an embarrassing problem: 
after more than three decades we still do not have a prototype, minimal 
working model that is phenomenologically acceptable at low energies. 
We actually had one for a while: the original theory of Georgi and 
Glashow \cite{Georgi:1974sy}, the so called minimal SU(5). 
It worked great when neutrinos could have 
been massless and when the wrong measurement of $\sin^2{\theta_W}$ 
indicated the unification of gauge couplings of the SM. Today we know that the 
theory must be changed and if you are willing to make a brave assumption
of super split supersymmetry \cite{Fox:2005yp}, a particularly elegant and simple 
modification is to add an adjoint fermionic $24_F$.\footnote{Another possibility 
would be to add a scalar $15$-dimensional representation 
\cite{Dorsner:2005fq,Dorsner:2005ii}.} The resulting predictions are 
remarkable \cite{Bajc:2006ia,Bajc:2007zf}:
\begin{enumerate}
\item 
The triplet mass must be small: $M_T\lsim 1$ TeV, in order to keep the proton 
stable enough;
\item
For a very light triplet, close to its lower limit of about 100 GeV, the GUT scale reaches
its maximum of $10^{16}$ GeV, and falls off with increasing $M_T$. A non observation 
of the triplet would guarantee seeing the proton decay in the next generation of experiments;
\item
The triplet decays through the same lepton number violating Yukawa couplings 
that enters the seesaw formula;
\item
The theory is consistent if and only if there are three generations. As already discussed in 
the introduction, minimality arguments 
prevents the hypothetical fourth generation neutrino to get a mass bigger than $M_Z/2$; 
\item
One light neutrino remains massless;
\item 
The scalar triplet is also possibly light.
\end{enumerate}
This is an example of a predictive grand unified theory that offers a possibility
of observing the seesaw mechanism at colliders. The seesaw, in this case of both 
Type I and III, results from the Yukawa couplings of the singlet $S$ and triplet 
$T$ fermions from the adjoint $24_F$
\begin{equation}
\label{ynu}
{\cal L}_Y(\nu)=y_S^iL_iHS+y_T^iL_iHT
-\frac{M_S}{2}SS-\frac{M_T}{2}TT+h.c.
\end{equation}
where the index $i$ refers to the flavors defined with respect to the charged leptons.

After the SU(2)$_L\times$U(1) symmetry breaking $\langle H\rangle=v\approx 174 \gev$, 
one obtains in the usual manner the light neutrino mass matrix upon integrating out 
$S$ and $T$
\begin{equation}
\label{seesaw13}
m_\nu^{ij}=-v^2\left(\frac{y_T^iy_T^j}{M_T}+\frac{y_S^iy_S^j}{M_S}\right)
\end{equation}
with $M_T\lsim 1$ TeV and $M_S$ undetermined. For an alternative version of the 
electroweak seesaw see for example \cite{Hung:2006ap} with light doublets instead 
of a triplet.\footnote{Light doublets are predicted with unification arguments also in 
minimal walking technicolor \cite{Gudnason:2006mk}, while a light triplet in 
\cite{Ma:2005he}.}

The neutrino mass matrix in Eq.~(\ref{seesaw13}) is clearly rank 2, thus one massless 
neutrino. This is simply a reflection of the fact that there are only two heavy neutral leptons,  
$T^0$ and $S$. Of course, there may be a higher dimensional operator as in Eq.~(\ref{weinberg}) 
which will lift the zero mass eigenvalue, but for $M>M_{GUT}$ it is less than $10^{-4}$ eV, 
thus completely negligible. For simplicity we will keep referring to the lightest neutrino as being massless. 
The main point is that we have a hierarchical spectrum of neutrinos, and neutrino 
masses are known up to the normal (NH) versus inverted (IH) hierarchy ambiguity.

In short, the SU(5) theory augmented with an adjoint fermion can be said to keep the ugliness 
of the minimal theory: asymmetric fermionic representations and  the fine-tuning, but also its 
remarkable predictiveness. It offers serious hope for the observation of the proton decay in the 
next generation of experiments, but in contrast with the minimal theory it predicts a few oasis in 
the desert, the first one in the form of the fermion SU(2) triplet, at the energies accessible
by the current and next generation high-energy colliders.
We now turn to its implications for neutrino physics.

\section{Neutrino Parameter Constraints} 
\label{sec2.5}

To make a direct contact with observables, it is necessary to connect the model parameters
with the neutrino masses and the mixing angles. We first note the relations for the
neutrino mass parameters with the normal hierarchy (NH: $m^\nu_3>m^\nu_1=0$) 
\begin{equation}
m_1^\nu=0  , \ \ \ m_2^\nu=\sqrt{\Delta m_S^2}  , \ \ \ m_3^\nu=
\sqrt{\Delta m_A^2+\Delta m_S^2}  ,
\end{equation}
and the inverted hierarchy (IH: $0=m^\nu_3<m^\nu_1$) 
\begin{equation}
m_1^\nu=\sqrt{\Delta m_A^2-\Delta m_S^2}  , \ \ \ 
m_2^\nu=\sqrt{\Delta m_A^2} \ , \ m_3^\nu=0 ,
\end{equation}
where we take the neutrino mass parameters \cite{Schwetz:2007my}
as measured in the solar and atmospheric oscillation experiments 
\beq
\Delta m_S^2 \approx 7.7 \times 10^{-5}\ {\rm eV}^2,\quad 
\Delta m_A^2 \approx 2.4 \times 10^{-3}\ {\rm eV}^2.
\label{eq:mnu}
\eeq

Since one neutrino is massless in our model, the neutrino flavor mixing 
matrix has only one physical Majorana phase $\Phi$%
\beq
\label{pmns}
U=
\left(
\begin{array}{ccc}
 c_{12} c_{13} & s_{12} c_{13} & s_{13}e^{-\text{i$\delta $}} 
   \\
 - s_{12}c_{23} -c_{12} s_{23} s_{13} e^{\text{i$\delta $}}&
   c_{12} c_{23}-s_{12} s_{23} s_{13} e^{\text{i$\delta $}} &
    s_{23} c_{13} \\
 s_{12} s_{23}- c_{12} c_{23} s_{13} e^{\text{i$\delta $}}&
   -c_{12} s_{23} -s_{12} c_{23} s_{13} e^{\text{i$\delta $}}&
    c_{23} c_{13}
\end{array}
\right)\times \text{diag} (1,e^{i \Phi}, 1) .
\eeq
It is useful to invert the seesaw formula (\ref{seesaw13}) following 
\cite{Casas:2001sr,Ibarra:2003up} and to parameterize the Yukawa 
couplings by only one complex parameter.\footnote{For a review on 
the general rank two neutrino mass, see for example \cite{Guo:2006qa}.}
We thus have the relations 
\begin{equation}
\label{eq:tn}
vy_T^{i*}=
 \left\{ \begin{array} {cc}
 i\sqrt{M_T}\left( U_{i2}\sqrt{m^\nu_2}\cos{z}+U_{i3}\sqrt{m^\nu_3}\sin{z}\right), 
   &  {\rm NH}\quad (m^\nu_1=0), \\
i\sqrt{M_T}\left(U_{i1}\sqrt{m^\nu_1}\cos{z}+U_{i2}\sqrt{m^\nu_2}\sin{z}\right), 
   &  {\rm IH}\quad (m^\nu_3=0), 
\end{array} \right. 
\end{equation}
and
\begin{equation}
\label{eq:tns}
vy_S^{i*}=
 \left\{ \begin{array} {cc}
-i \sqrt{M_S}\left( U_{i2}\sqrt{m^\nu_2}\sin{z}-U_{i3}\sqrt{m^\nu_3}\cos{z}\right), 
   &  {\rm NH}\quad (m^\nu_1=0), \\
-i \sqrt{M_S}\left(U_{i1}\sqrt{m^\nu_1}\sin{z}-U_{i2}\sqrt{m^\nu_2}\cos{z}\right),
   &  {\rm IH}\quad (m^\nu_3=0) .
\end{array} \right. 
\end{equation}
There is another solution with the opposite sign of the second terms in (\ref{eq:tn}), 
(\ref{eq:tns}). The major advantage of this parameterization is to allow
us to untangle the contributions of $y_T$ and $y_S$, so that one can
separately explore the correlations between these Yukawa couplings and the neutrino
oscillation parameters, as we will demonstrate next. 

The singlet Yukawa couplings $y_S^{}$ are less useful for our consideration 
since one does not expect the singlet to be produced at the LHC. 
The Yukawa couplings $|y_T^i|$ are invariant  under the operations
\bea
y_T^i &\to & - y_T^i \quad {\rm under } \quad {\rm Re}(z)
 \to {\rm Re}(z)+\pi
\\
y_T^i &\to & - y_T^i \quad {\rm under } \quad \Phi
 \to \Phi+\pi,\ z \to -z
\\
y_T^i & \to &- (y_T^i)^* \ {\rm under } \quad \Phi
\to -\Phi,\quad \delta\to -\delta,\quad z \to z^* .
\eea
It suffices therefore to explore effects of the full range of neutrino parameters 
by restricting to 
\begin{equation}
{\rm Im}(z) \geq 0\,\,,\,\, \Phi \in [-\pi/2, \pi/2]\,\,,\,\,
{\rm Re}(z)\in [0,\pi]\,\,.
\end{equation}
If we scan over the whole parameter space, 
this range covers also the other solution to the seesaw equation,
Eq.~(\ref{seesaw13}), i.e. Eqs.~(\ref{eq:tn}) and (\ref{eq:tns}) with the opposite sign of 
the second term.

It is convenient to introduce a dimensionless parameter $a_T$, defined as,
\beq
a_T^i \equiv 
\left| y_T^{i} \right| \sqrt{\frac{v}{2M_T}}\  ,
\eeq
where $i=1,2,3$ refers to the lepton flavors $e,\mu,\tau$.
There exist significant constraints on the Yukawa couplings, or on $a_T^i$, 
from the low energy data, most notably by various flavor violating processes,  
such as $\mu\to 3e$ and lepton universality at the $Z$-pole 
\cite{Abada:2008ea, He:2009tf, Arhrib:2009xf, miha09}. 
These limits are intercorrelated and become 
of order $1\%$ when only one such coupling is nonzero. To be on the safe side 
hereafter we stick to 
\begin{equation}
a_T^i\le 10^{-2}\;.
\label{eq:yukawa}
\end{equation}
We take the experimental lower bound for the triplet mass to be 100 GeV from 
\cite{Amsler:2008zz} (although strictly speaking it is found for charged 
lepton doublets, a similar limit applies for charged triplet). 
The neutrino mass and mixing constraints are of essential importance in our
consideration. We adopt their numerical values from the $2\sigma$ fitting in a recent 
analysis~\cite{Schwetz:2007my}.  

\begin{figure}[tb]
\includegraphics[scale=1,width=8cm]{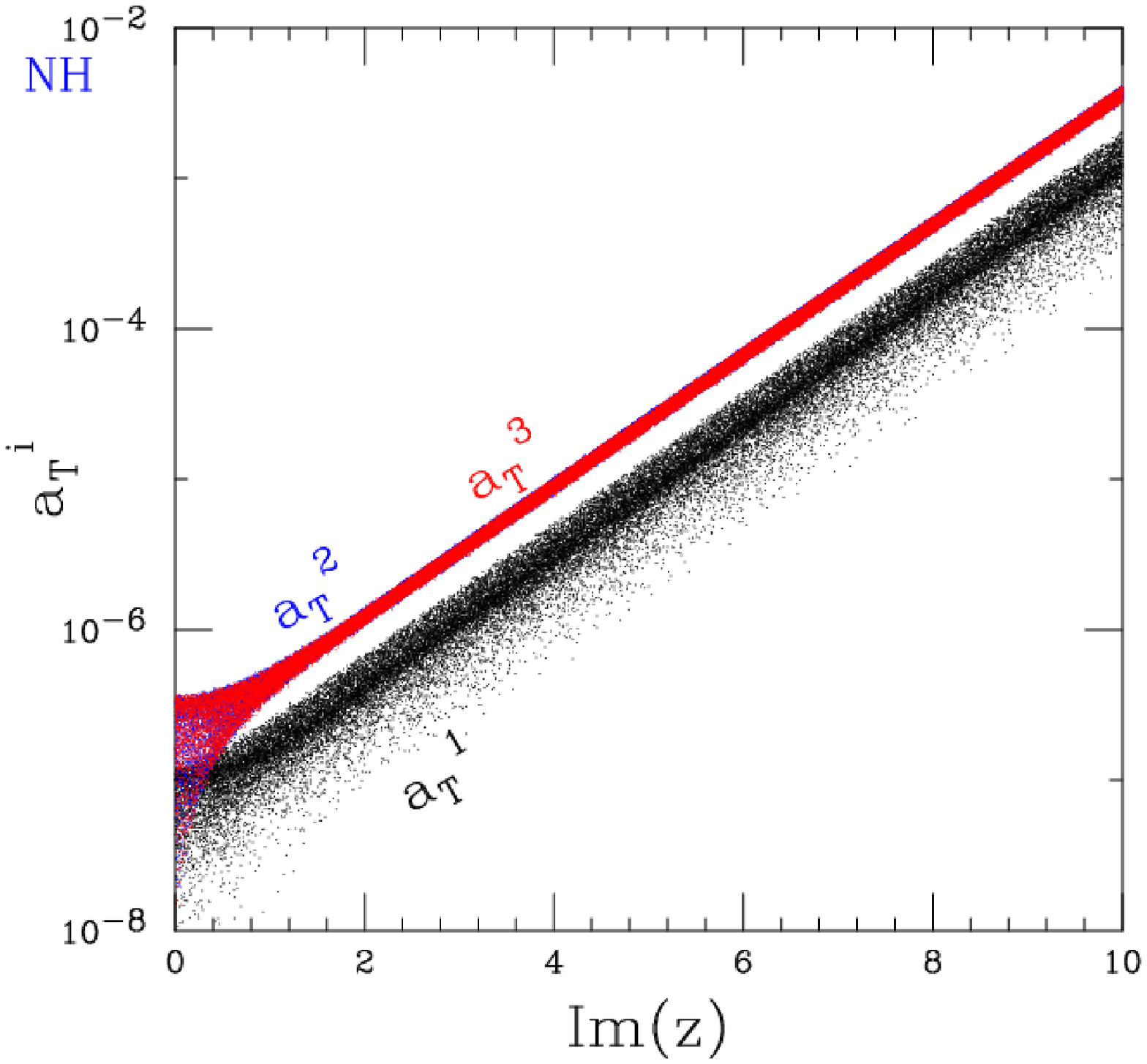}
\includegraphics[scale=1,width=8cm]{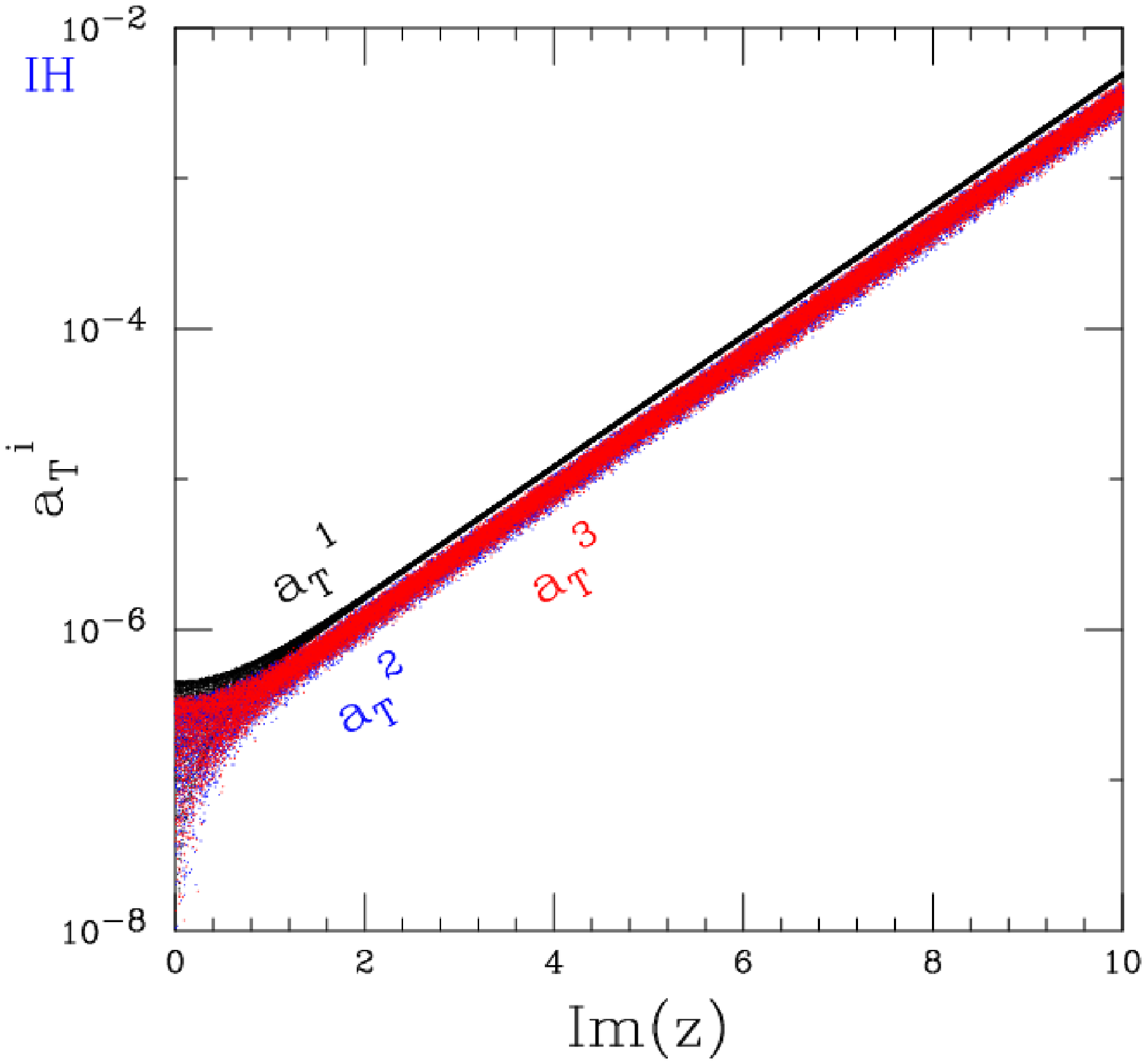}
\caption{$a_T^i$ versus ${\rm Im}(z)$ for NH (left) and IH (right). 
$\Phi=0$. 
}
\label{fig:zA00}
\end{figure}

\begin{figure}[tb]
\includegraphics[scale=1,width=8.0cm]{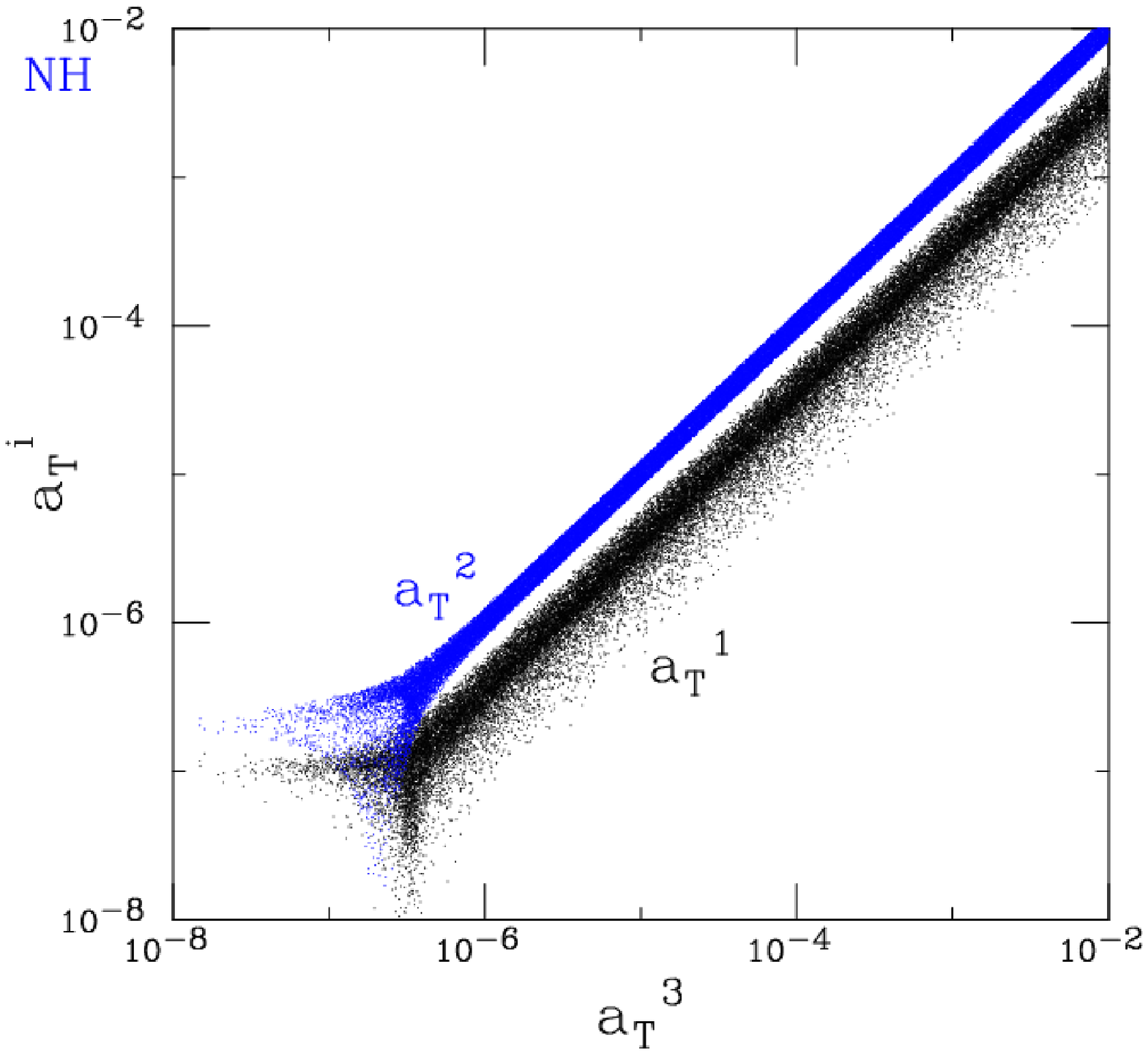}
\includegraphics[scale=1,width=8.0cm]{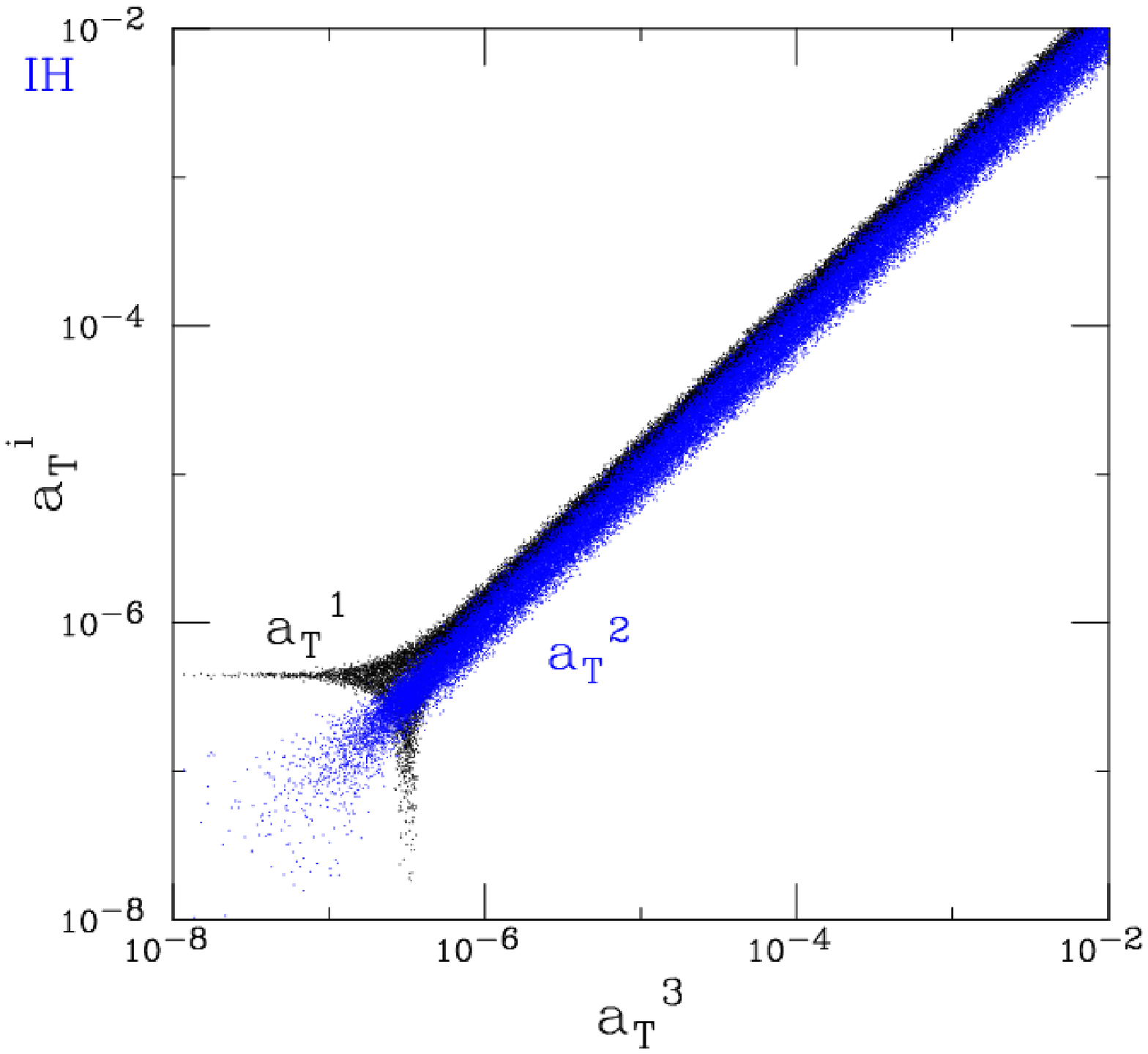}
\caption{Correlations between $a_T^i$'s for NH (left) and IH (right). 
$\Phi=0$. 
}
\label{fig:AA00}
\end{figure}

Based on the relations between our
model parameters and the oscillation data, one can predict the
size of the couplings in terms of the other parameters. We perform a comprehensive 
scan over the large parameter space, including both the neutrino oscillation parameters
and the theory parameters. 
The expression of Eq.~(\ref{eq:tn}) makes the Yukawa couplings transparent 
in terms of the complex parameter $z$.  The real part of $z$ leads to a simple 
oscillatory nature. It is the imaginary part of $z$ that affects the size of the 
coupling  $a_T$ most. 
This is plotted in  Fig.~\ref{fig:zA00} for the NH (left panel) and the IH (right panel),
when ignoring the influence of the Majorana phase $\Phi$. Several features are
transparent: (a). The Yukawa couplings $y_T^i$ 
are exponentially enhanced over Im($z$); 
(b). At large values of  ${\rm Im}(z)$, the couplings approximately factorize into a form
independent of ${\rm Re}(z)$
\begin{equation}
\label{eq:tnshort}
y_T^{i*} \propto
 \left\{ \begin{array} {cc}
e^{ {\rm Im}(z)} \left( \left(\Delta m_S^2\right)^{1/4} U_{i2} +i \,
\left(\Delta m_A^2\right)^{1/4}\ U_{i3} \right), 
   &  ({\rm NH}), \\
   e^{ {\rm Im}(z)} \ 
\left(\Delta  m_A^2\right)^{1/4} \left(U_{i1} +i\, U_{i2} \right), 
   &  ({\rm IH}).
\end{array} \right. 
\end{equation}
(c). We see the important correlations:  
due to the large mixing between $\mu-\tau$ from the atmospheric 
neutrino observation, the generic prediction is
\bea
\label{eq:nh}
&& a^1_T <  a^2_T \approx a^3_T \quad {\rm  for\ NH},\\
&& a^1_T >  a^2_T \approx a^3_T \quad {\rm  for\ IH}.
\label{eq:ih}
\eea
(d). For $|z| \lsim {\cal O}(1)$, the situation is less clear because of the dependence 
on ${\rm Re}(z)$.
The same feature is presented differently in Fig.~\ref{fig:AA00} for $a_T^{1,2}$ 
versus $a^3_T$. 
The band structure is due to the experimental uncertainties for the input parameters
of the neutrino masses and mixing angles. 

\begin{figure}[tb]
\includegraphics[scale=1,width=8.0cm]{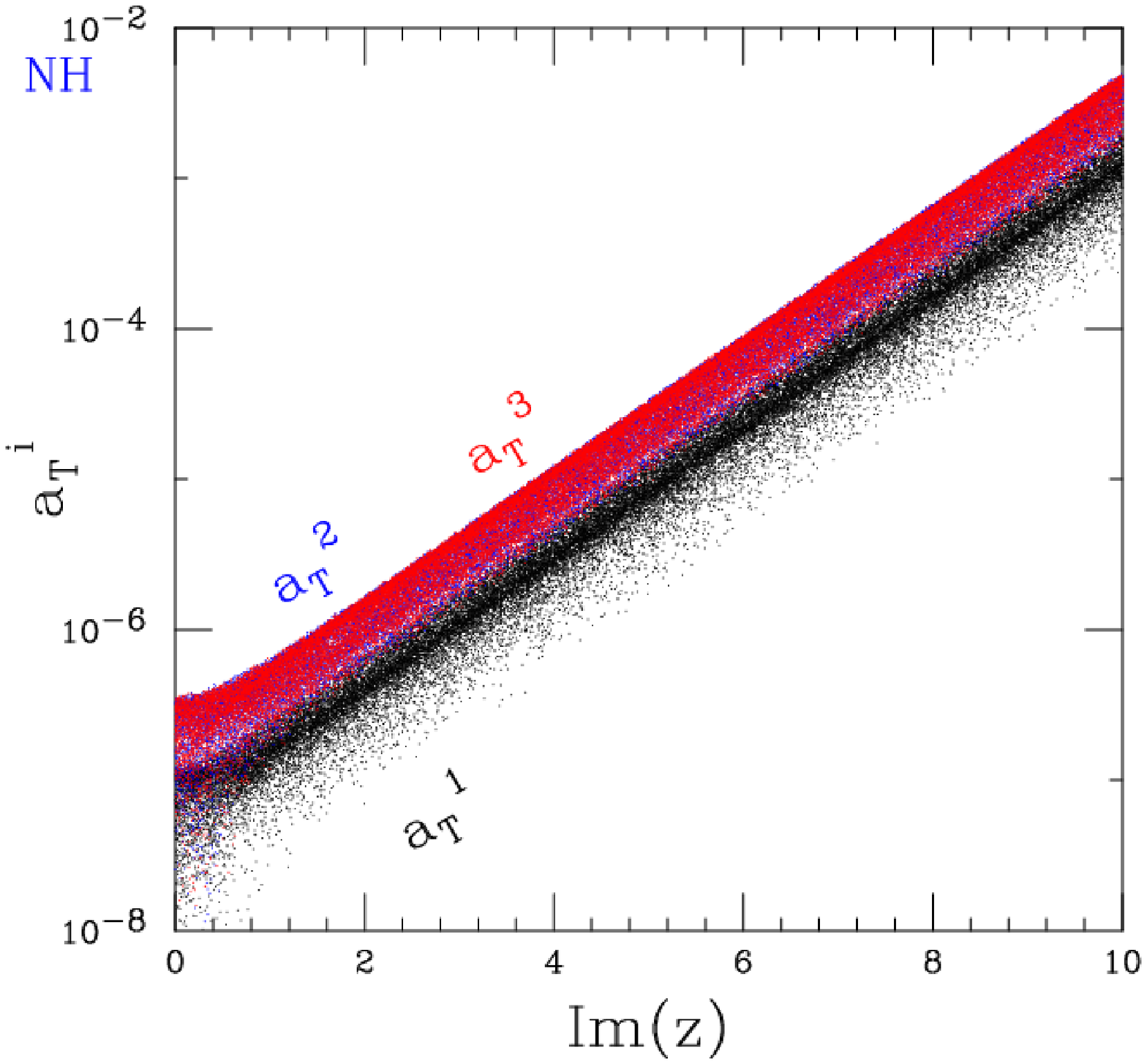}
\includegraphics[scale=1,width=8.0cm]{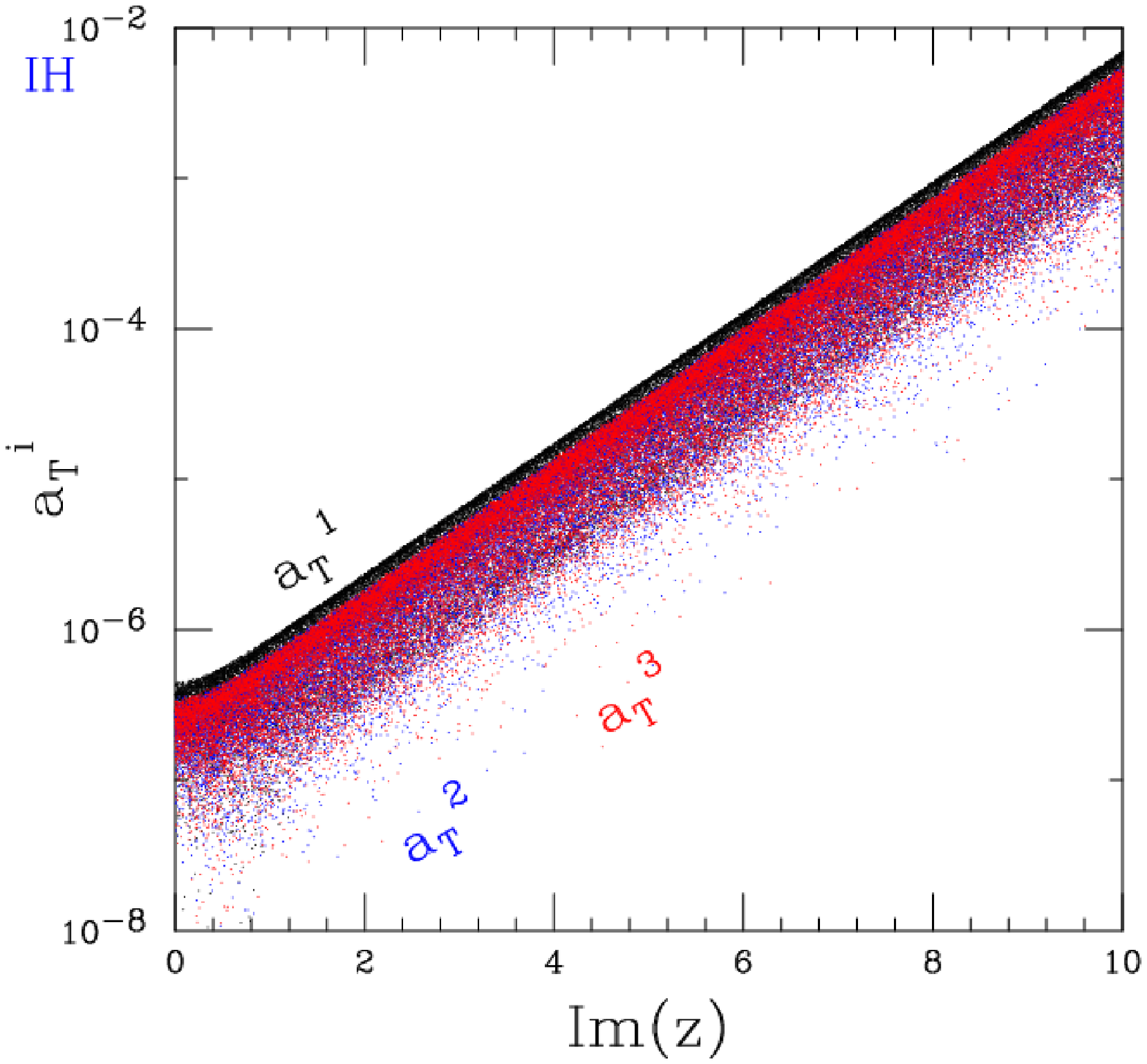}
\caption{$a_T^i$ versus ${\rm Im}(z)$ for NH (left) and IH (right). 
$\Phi \in [-\pi/2,\pi/2]$. 
}
\label{fig:zA}
\end{figure}

\begin{figure}[tb]
\includegraphics[scale=1,width=8.0cm]{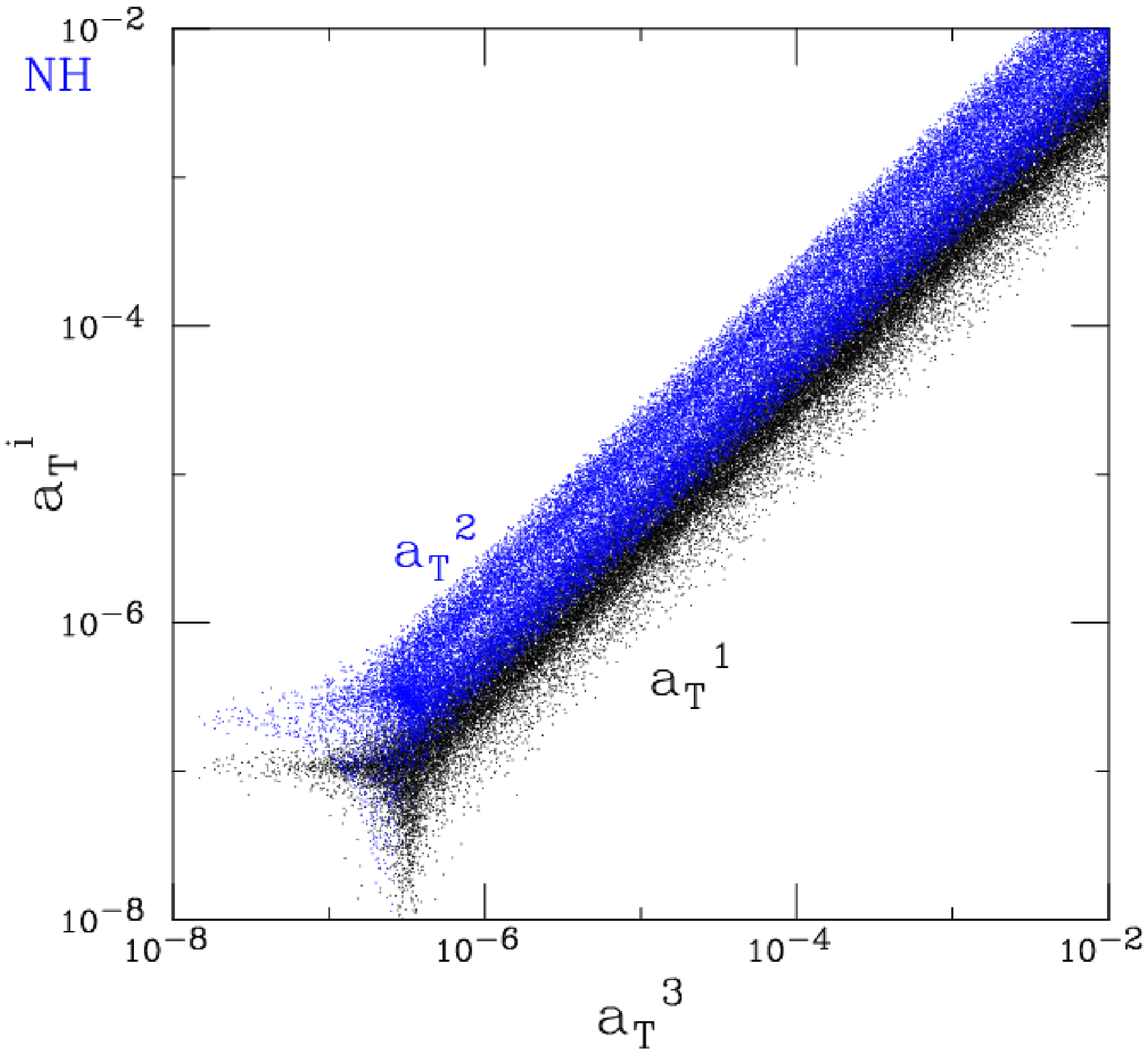}
\includegraphics[scale=1,width=8.0cm]{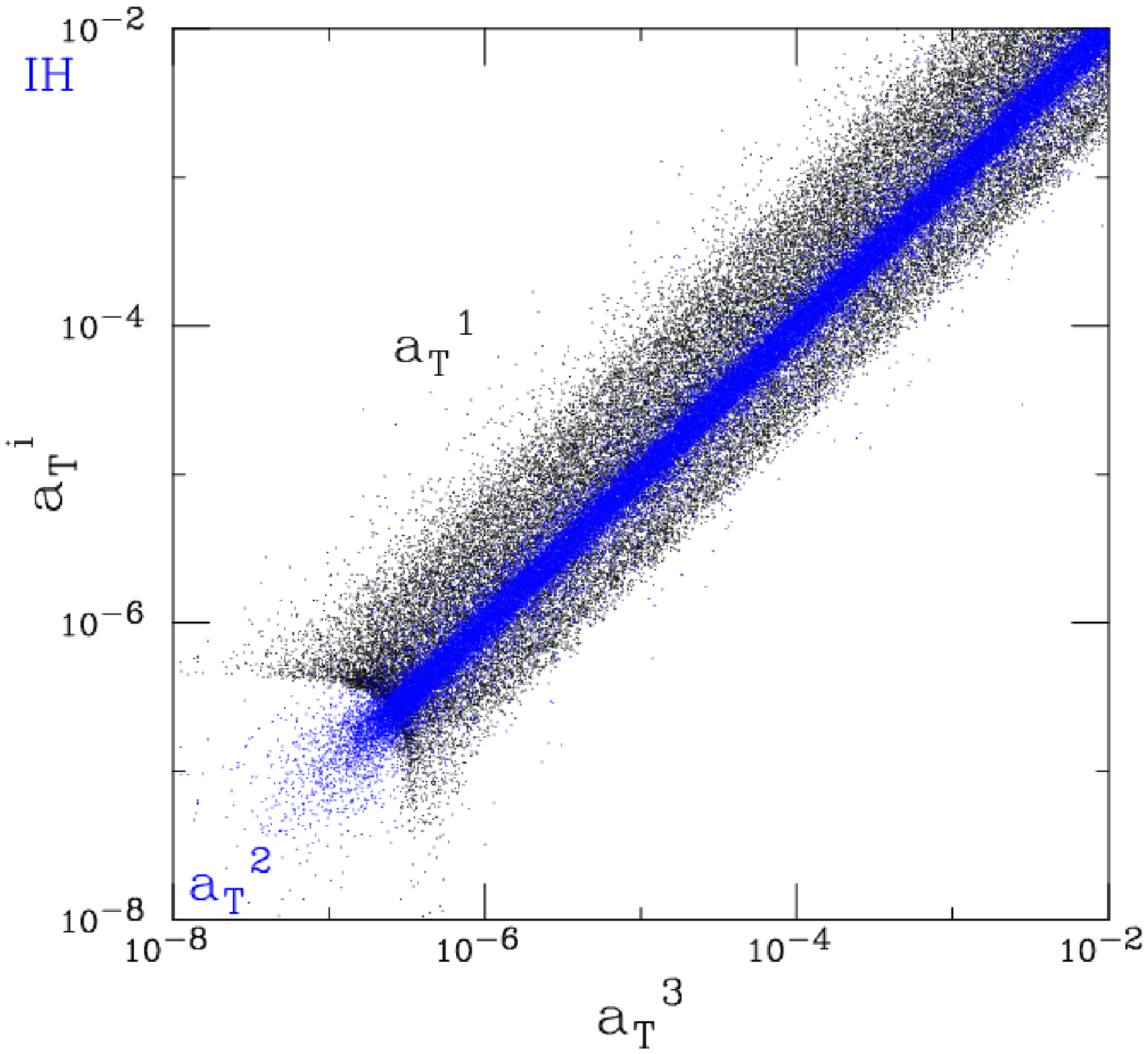}
\caption{Correlations between $a_T^i$'s for NH (left) and IH (right). 
$\Phi \in [-\pi/2,\pi/2]$. 
}
\label{fig:AA}
\end{figure}

Including the Majorana phase $\Phi$ as an additional unknown parameter 
greatly increases the spread of $a_T$ range as seen
in Figs.~\ref{fig:zA} and \ref{fig:AA}.  Although the generic features of 
Eqs.~(\ref{eq:nh}) and (\ref{eq:ih}) should still remain, the prediction is
less sharp due to the dilution of the phase parameters.

\section{Decays of $T$}
\label{section3}

For the leptonic triplet in our model, there are a number of decay channels as well as rich 
new physics  to explore. 

\subsection{$T\to W, Z, h + \;{\rm light \; lepton}$}

If kinematically accessible, 
the predominant decay modes of the triplet leptons are to a gauge boson (or a Higgs
boson) plus a SM lepton \cite{Bajc:2007zf}, 
whose coupling strength is dictated by the neutral Dirac Yukawa couplings. 
The expressions are given in Appendix B.
The partial width for each mode is proportional to the mass $M_T$.
From Eq.~(\ref{t0w}) one sees that the decays of $T^0$, just like a heavy  
Majorana  neutrino, violate lepton number by two units. Because of the
firm prediction for the leptonic triplets to be light near the TeV scale, 
it opens up the possibility to observe these Majorana states at hadron colliders 
via this fairly background-free signature as first discussed in \cite{Keung:1983uu}. 
These results are consistent with the decay formula of the gauge boson modes,
in accordance with the Goldstone boson equivalence theorem when $M_T \gg m_Z, m_h$.

\subsection{$T^\pm \rightarrow T^0\pi^\pm$ }
At the leading order, the heavy lepton triplet is mass-degenerate at $M_T$.
Electroweak radiative corrections at one loop lift the degeneracy by a shift 
\bea
\label{dmt}
&& \Delta M_T\equiv m_{T^+}-m_{T^0}=\frac{\alpha_2}{2\pi}\frac{m_W^2}{M_T}
\left[f\left(\frac{M_T}{m_Z}\right)-f\left(\frac{M_T}{m_W}\right)\right], \\
&& f\left(y\right)=\frac{1}{4y^2}\log{y^2}- \left(1+\frac{1}{2y^2}\right)
\sqrt{4y^2-1}\arctan{\sqrt{4y^2-1}} ,
\eea
which gives $\Delta M_T\approx 160$ MeV 
with $10\%$ accuracy in the whole range $100$ GeV$\le M_T\le\infty$. 

Given the rather small mass difference, the leading kinematically allowed 
 decay mode is 
$T^\pm\to T^0\pi^\pm$ with the decay width \cite{Ibe:2006de}
\begin{equation}
\Gamma_{\Delta M_T}=\frac{2G_F^2}{\pi}\cos^2{\theta_c}
f_\pi^2\Delta M_T^3\sqrt{1-\frac{m_\pi^2}{\Delta M_T^2}}.
\end{equation}
The corresponding lifetime is 
estimated to be ${\cal O}(10^{-10})$ sec $\approx$ 10 cm or more
(with $f_\pi=130$ MeV and $\cos^2{\theta_c}\approx 0.95$). This decay mode is thus 
negligible in comparison with the $W^\pm\nu$ or $Z l^\pm$ decay 
channels considered in the previous subsection (see below for details).

\subsection{Total widths and decay lengths}

\begin{figure}[tb]
\begin{center}
\scalebox{0.7}{\includegraphics[angle=0]{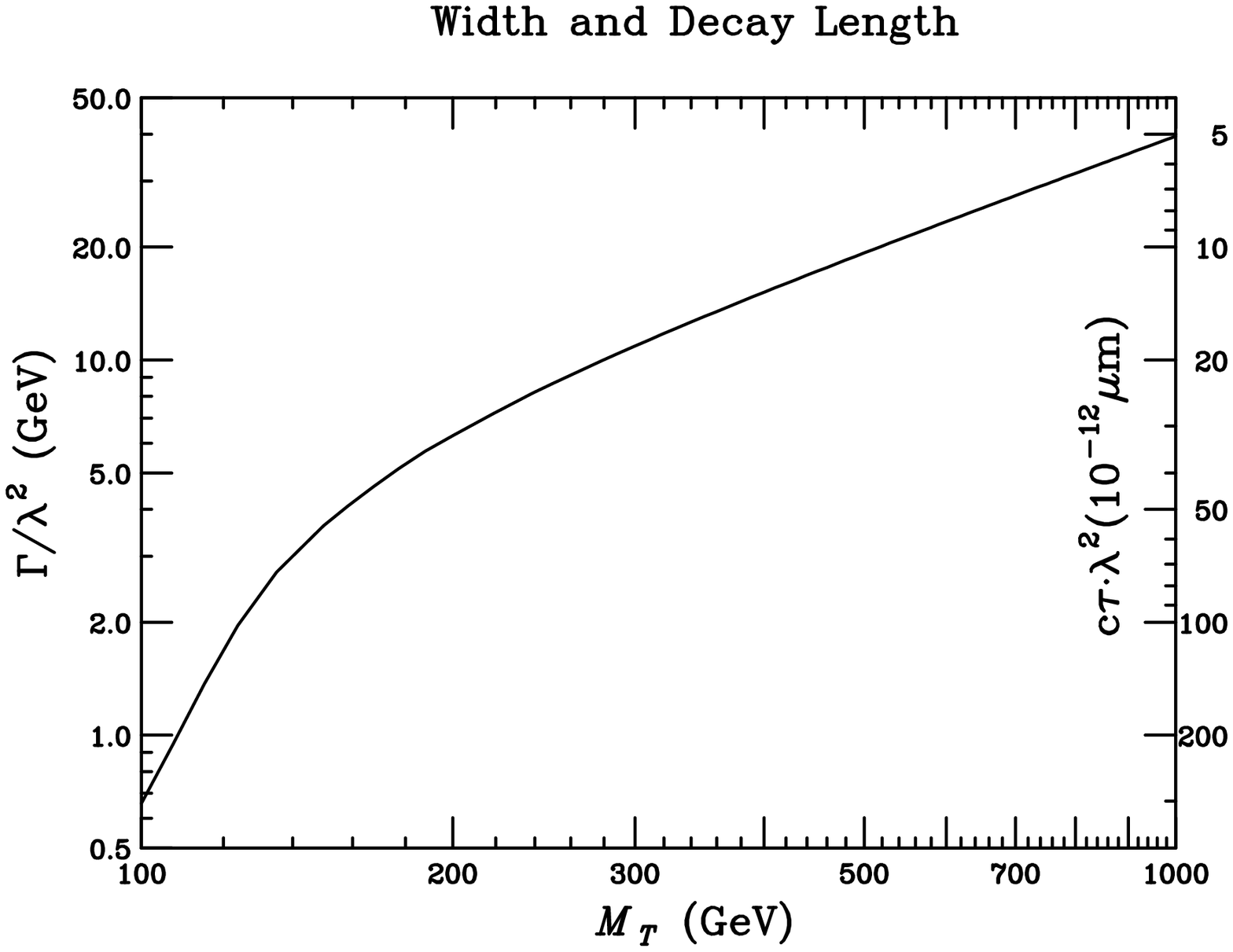}}
\caption{Total widths (left axis) and proper decay lengths (right axis) of $T^\pm/T^0$ 
 as functions of its mass. 
\label{fig:width}}
\end{center}
\end{figure}
\begin{figure}[tb]
\begin{center}
\scalebox{1.4}{\includegraphics{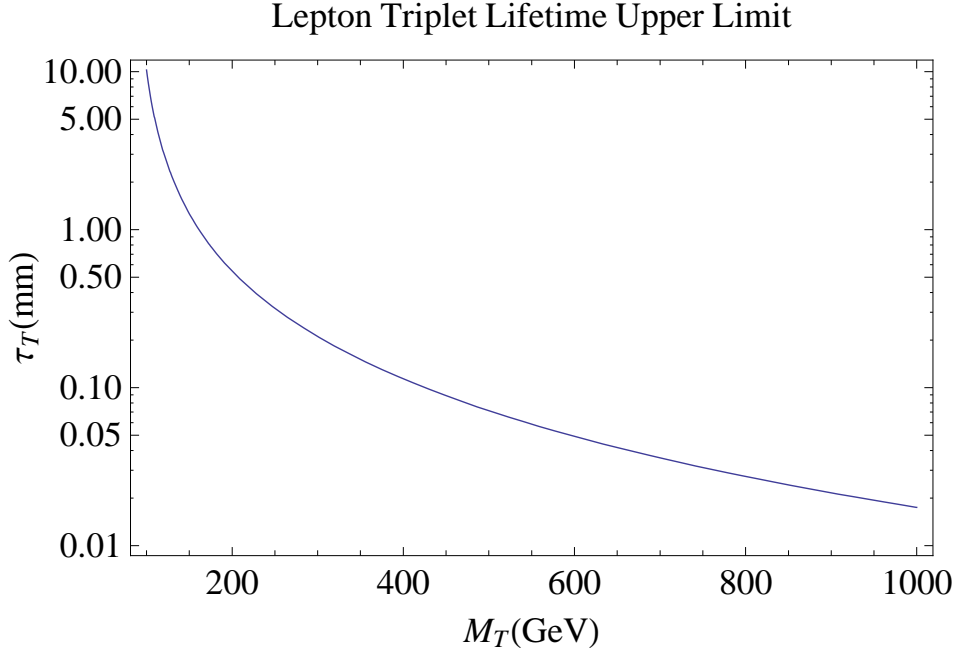}}
\caption{The upper limit on the triplet lifetime as a function of its mass for $m_h=120$ GeV.
\label{tripletlifetime2}}
\end{center}
\end{figure}

The total triplet decay width is given by the simple sum over the contributing channels
\bea
&& \Gamma_T=\frac{M_T}{32\pi}\left(\sum_k\left|y_T^k\right|^2\right)
\left[2f_1\left(\frac{m_W}{M_T}\right)+f_1\left(\frac{m_Z}{M_T}\right)
+f_0\left(\frac{m_h}{M_T}\right)\right]+\Gamma_{\Delta M_T}, \\
&& f_n(x)=(1-x^2)^2(1+2nx^2)\Theta(1-x) .
\eea
Given the couplings in our theory, 
the total width scales roughly linearly with the mass, particularly above the 
$W$, $Z$, and $h$ thresholds:
\beq
 \Gamma_T \approx 0.04 \lambda^2\  M_T, \quad 
 \lambda^2 \equiv \sum_k |y^k_T|^2 , 
\eeq
and the proper decay length 
\beq
 \tau_T = {1 \over \Gamma_T} \approx \dfrac{25}{\lambda^2 M_T}
  \approx \dfrac{1}{10^{13} \lambda^2} \left(\dfrac{100\, \gev}{M_T}\right) \, 0.5\  \rm{mm} 
\eeq
On the other hand, making the connection to the neutrino masses as
in Eq.~(\ref{eq:tn}), we have the lower  bound in complete generality 
\begin{equation}
\lambda^2 = \sum_k\left|y_T^k\right|^2 \ge
 \left\{ \begin{array} {cc}
\frac{M_T}{v^2}\sqrt{\Delta m_S^2}\ ,  &  {\rm NH}, \\
\frac{M_T}{v^2}\sqrt{\Delta m_A^2}\ ,  &  {\rm IH},
\end{array}\right.
\end{equation}
where in the IH case we approximated $\Delta m_A^2-\Delta m_S^2\approx\Delta m_A^2$. 
Using all the above we get an upper bound for the triplet lifetime (assuming for 
the moment that $\Gamma_{\Delta M_T}$ can be neglected)
\begin{equation}
\tau_T\le \frac{32\pi}{\sqrt{\Delta m_X^2}}\left(\frac{v}{M_T}\right)^2
\frac{1}{2f_1(m_W/M_T)+f_1(m_Z/M_T)+f_0(m_h/M_T)}
\end{equation}
with $X=S$ for NH and $X=A$ for IH. 
Taking the values of the neutrino masses as in Eq.~(\ref{eq:mnu}) 
and the Higgs mass $m_h=120$ GeV, one can easily calculate the upper 
limit for the triplet lifetime as a function of its mass. The limit in the NH case 
is shown on Fig. \ref{tripletlifetime2}. 
For $M_T\gsim 200$ GeV (enough to saturate $f_n\to 1$) the upper lifetime limit 
can be approximated by
\begin{equation}
\tau_T\lsim \left(\frac{200\gev}{M_T}\right)^2 \ 0.5\ {\rm mm}
\end{equation}
In the IH case one has to divide the lifetimes by 
$\sqrt{\Delta m_A^2/\Delta m_S^2}\approx 5.6$. So in no case the decay 
$T^\pm\to T^0\pi^\pm$ could be relevant \cite{Bajc:2007zf}. 
This is different compared to the rank three neutrino mass case considered in 
\cite{Franceschini:2008pz}, where the lightest triplet decay can have 
arbitrarily small Yukawa couplings with the light leptons. In that limiting case the 
charged triplet decays to the neutral one and a pion in approximate $10$ cm, 
while the neutral one does not have any upper limit, see also \cite{Ibe:2006de}.

This leads us to conclude that the heavy leptons produced at the LHC experiments
would decay inside the detector, possibly leaving displaced secondary vertices,
but not appearing as stable particles. At least for small enough 
triplet mass the total lifetime could be measured directly. But even if not the total lifetime, 
the branching fractions into different final lepton states could be determined if not too 
small. This we discuss in the next subsection.

\subsection{Branching fractions}

\begin{figure}[tb]
\begin{center}
\scalebox{0.7}{\includegraphics[angle=0]{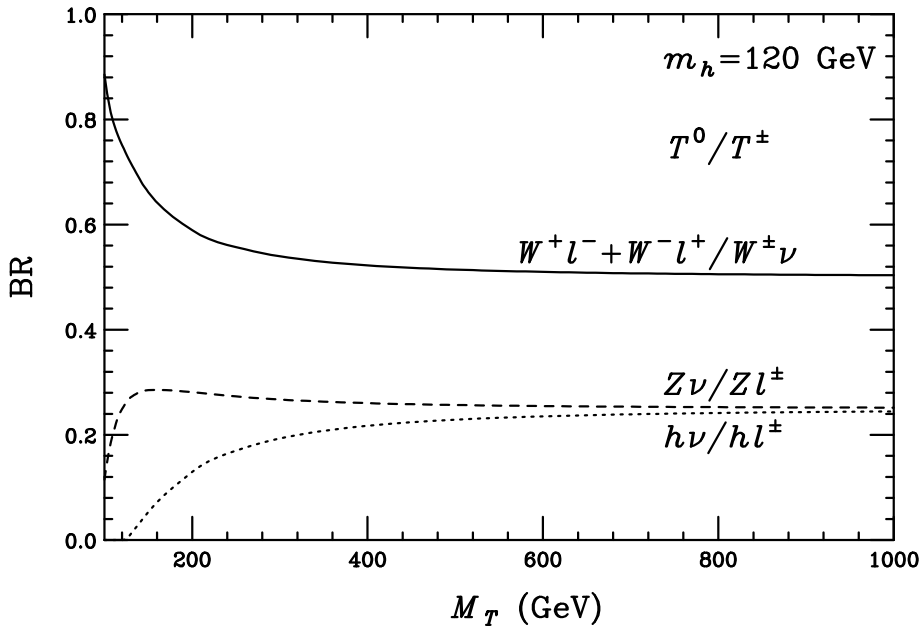}}
\caption{Branching fractions of $T^0/T^\pm$ as a function of its mass. A sum over
lepton final states has been assumed. 
\label{fig:br}}
\end{center}
\end{figure}

\begin{figure}[tb]
\includegraphics[scale=1,width=8.0cm]{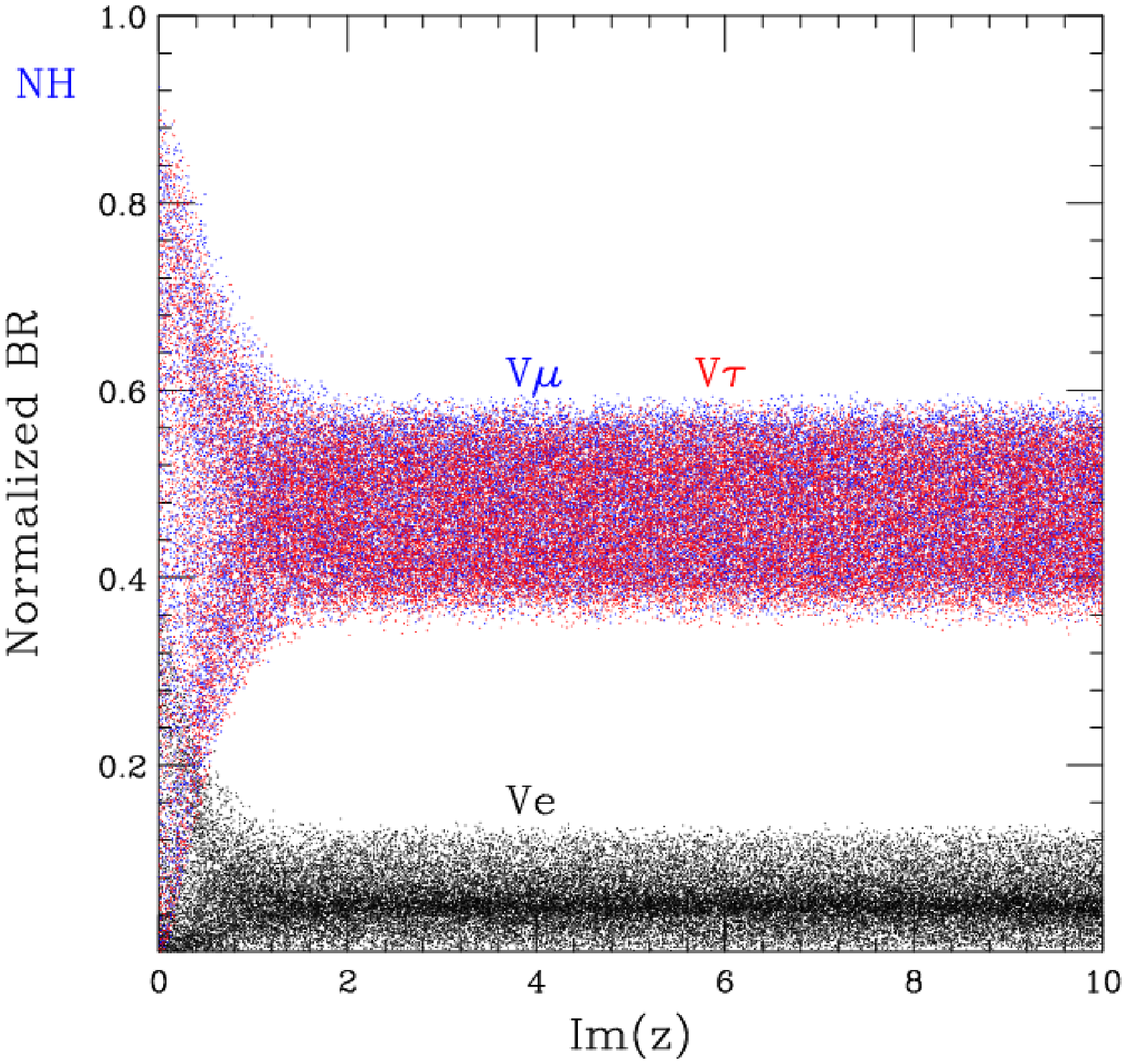}
\includegraphics[scale=1,width=8.0cm]{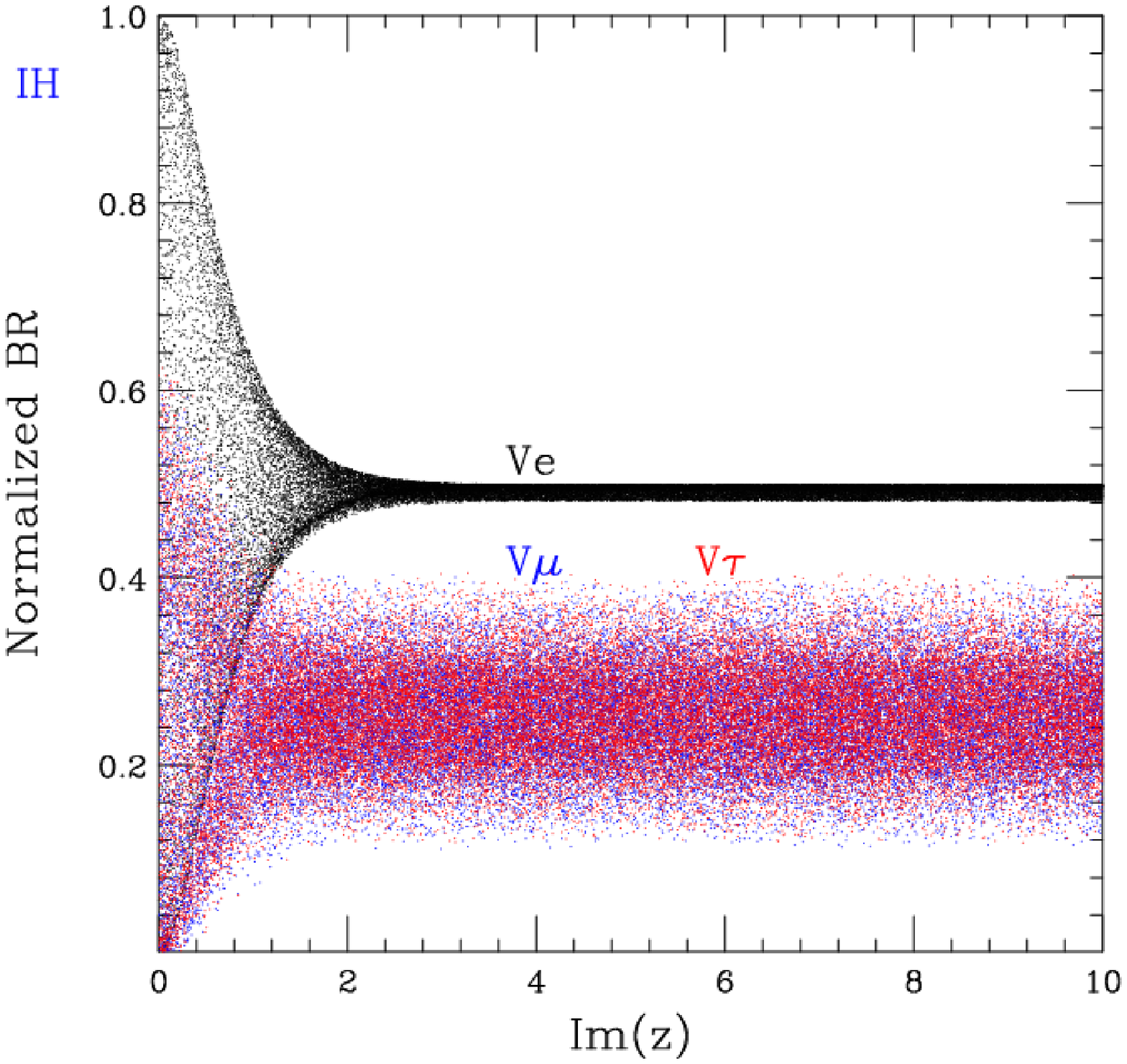}
\caption{Normalized branching of $T\to V e_i$ (NBR$_i$) 
versus ${\rm Im}(z)$ 
for NH (left) and IH (right); 
$\Phi= 0$. 
}
\label{fig:zB00}
\end{figure}

\begin{figure}[tb]
\includegraphics[scale=1,width=8.0cm]{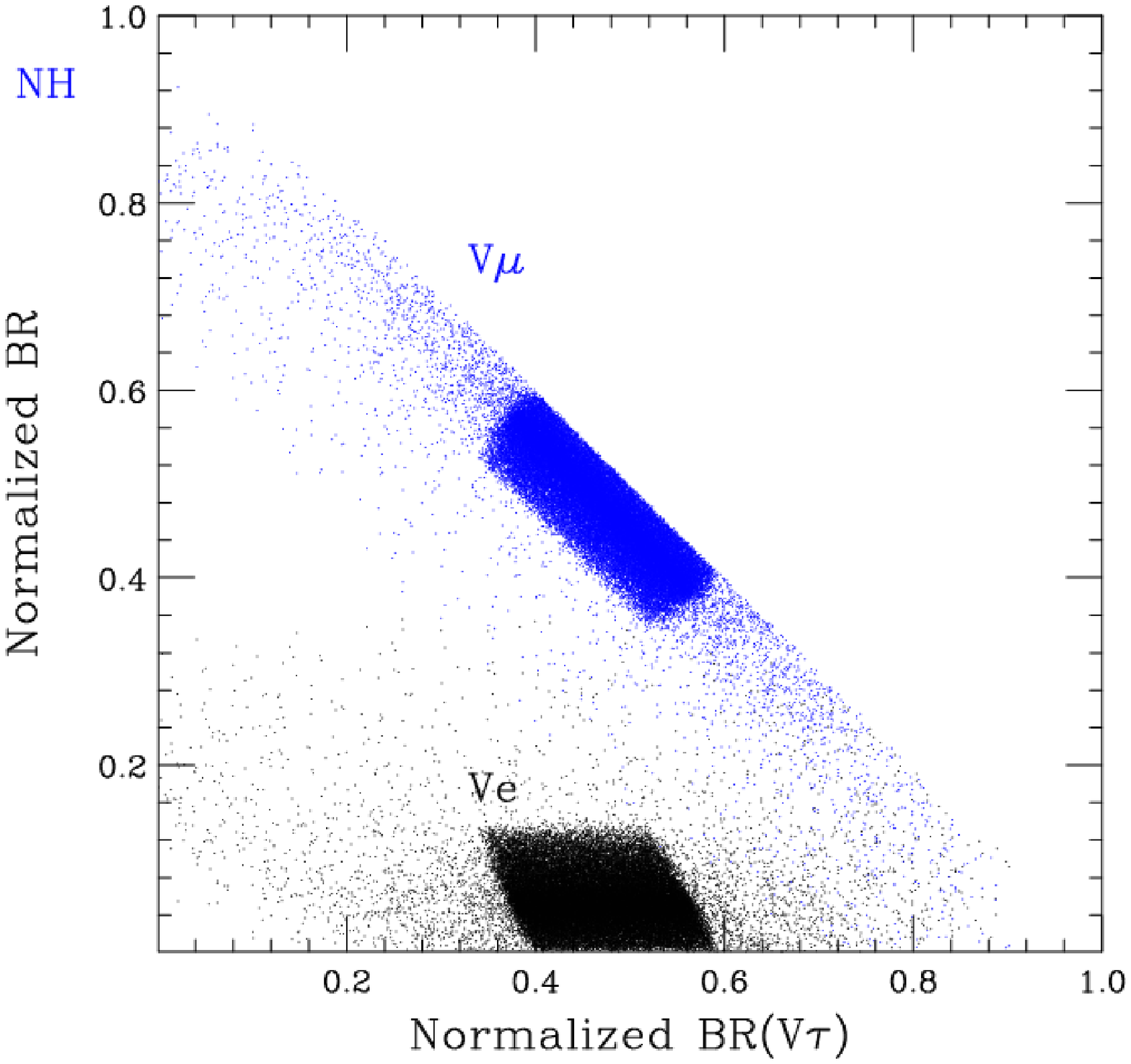}
\includegraphics[scale=1,width=8.0cm]{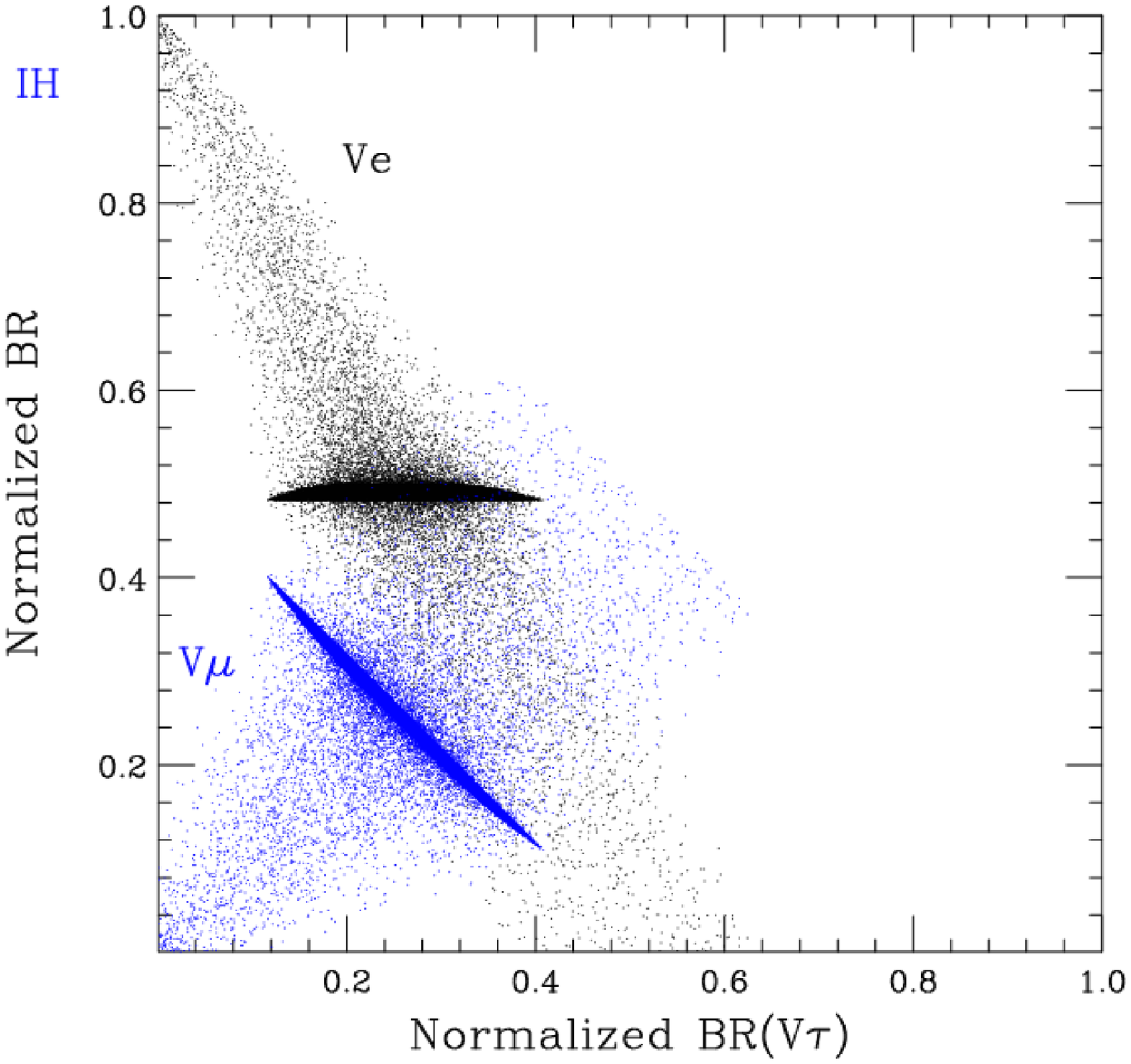}
\caption{Correlations between normalized branchings of $T\to V e_i$ 
(NBR$_i$) 
for NH (left) and IH (right); 
$\Phi = 0$. 
}
\label{fig:BB00}
\end{figure}

Decay branching fractions of $T$ to the three main decay channels 
involving $W$, $Z$ and $h$ are plotted in 
Fig.~\ref{fig:br}. Behavior in the low $M_T$ region is dominated by
threshold suppression. For sufficiently large $M_T$, these branching fractions
approach their asymptotic values of $1/2$, $1/4$ and $1/4$, respectively.
Due to the importance of charged leptons in the final state, 
we define the normalized branching fraction to a given charged lepton $e_i$
($e_i=e,\mu,\tau$ for $i=1,2,3$),
counted for the same final state gauge boson as 
\beq
 {\rm NBR_i}\equiv
 \frac{{\rm BR}(Ve_i)}{\sum_k{\rm BR}(Ve_k)}=
 \frac{|y_T^i|^2}{\sum_k|y_T^k|^2}.
\eeq
This quantity is universal for $V=W,Z,h$, and reflects the flavor structure of the final 
state leptons that is governed by the neutrino mass and mixing parameters. 
The ${\rm Im}(z)$ dependence of  ${\rm NBR}_i$ and their correlations are shown in
Figs.~\ref{fig:zB00}, and \ref{fig:BB00}, when ignoring the Majorana phase.

In most of the parameter space of NH (left panels), i.e. for ${\rm Im}(z)>1$, the 
normalized branching fraction for either $V\mu$ or $V\tau$ is about $0.35$ to $0.55$ and the 
normalized $Ve$ branching is less than $0.1$. We thus have the prediction 
\beq
{\rm BR}(V\mu) \approx  {\rm BR}(V\tau)  \gg {\rm BR}(Ve), 
\eeq
For the case of IH (right panels), we can establish
similarly the rough order of branchings and the combinations.
\beq
{\rm BR}(V\mu)  \approx  {\rm BR}(V\tau)  < {\rm BR}(Ve), 
\eeq

\begin{figure}[tb]
\includegraphics[scale=1,width=8.0cm]{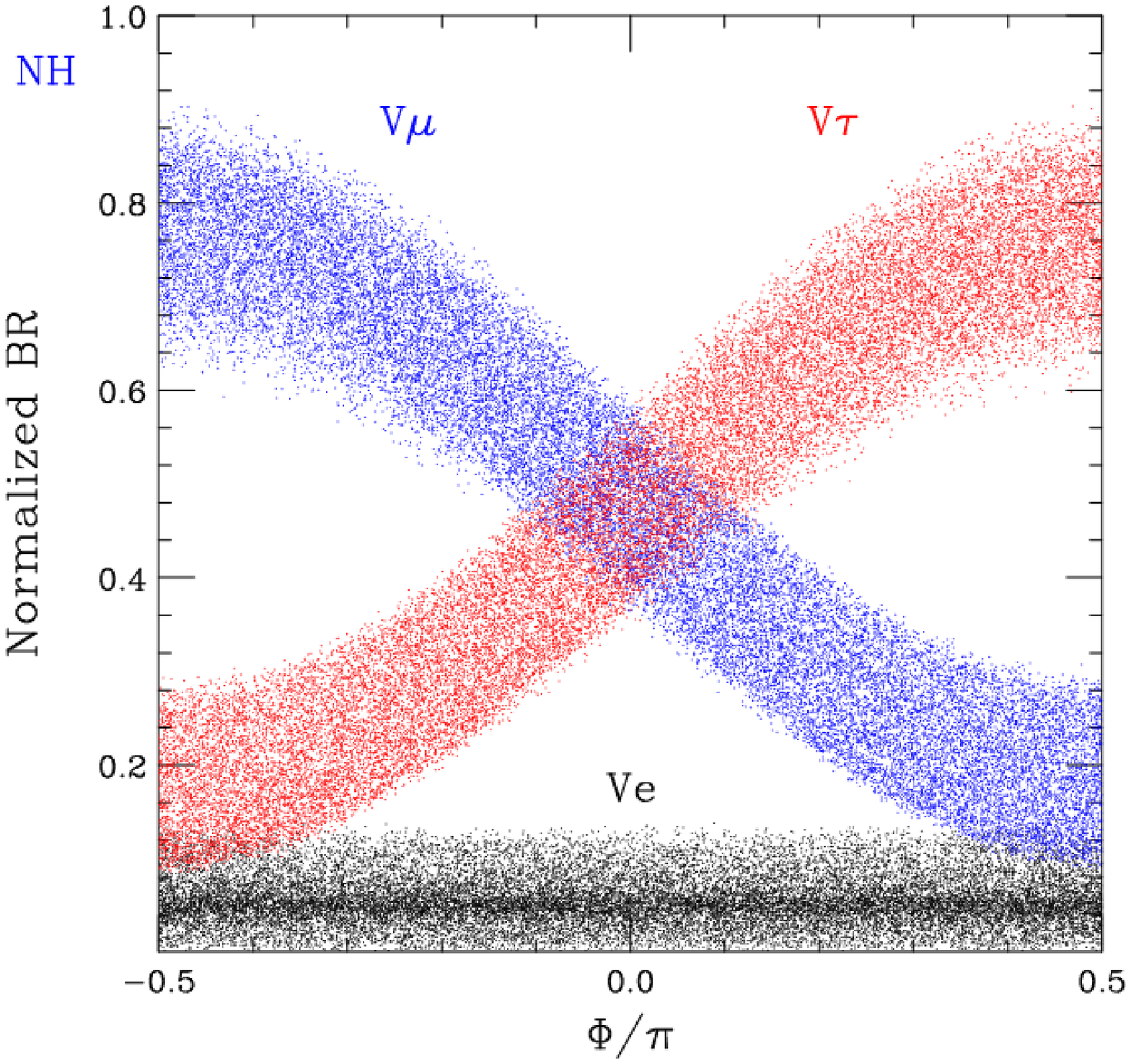}
\includegraphics[scale=1,width=8.0cm]{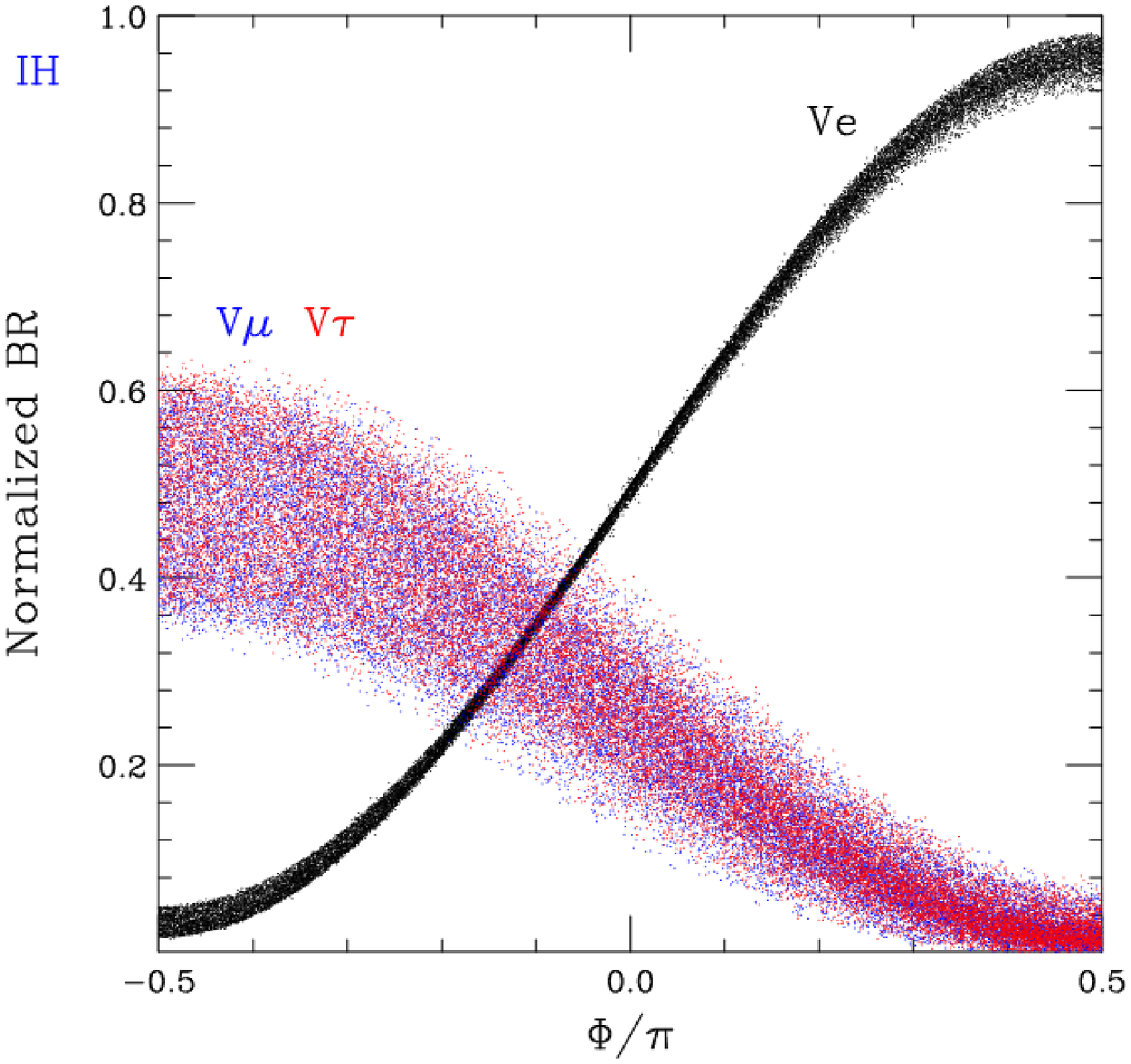}
\caption{Normalized branching versus Majorana phase for NH (left) and IH (right). 
${\rm Im}(z)\ge 2$.}
\label{fig:lXX}
\end{figure}

\begin{figure}[tb]
\includegraphics[scale=1,width=8.0cm]{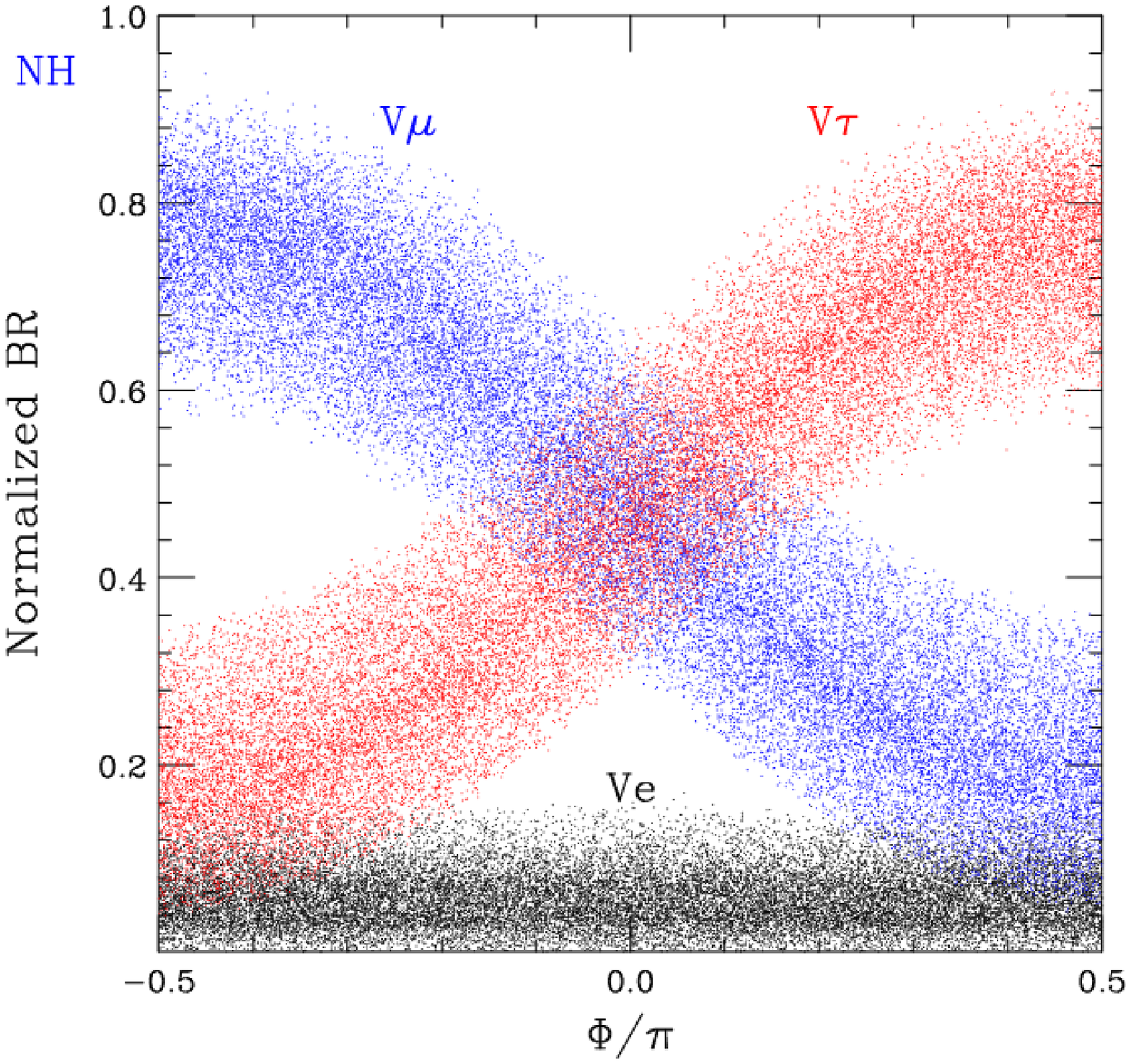}
\includegraphics[scale=1,width=8.0cm]{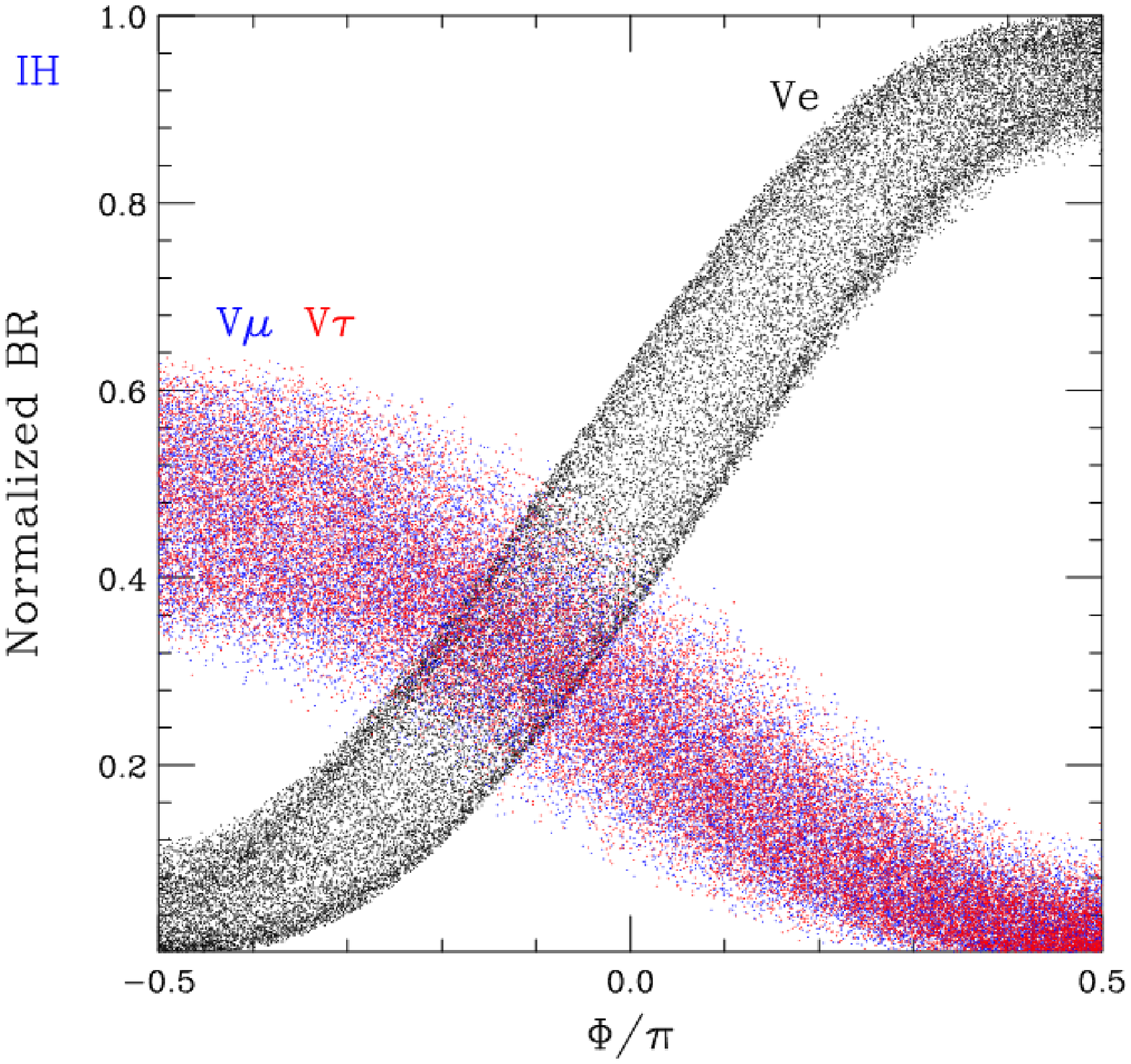}
\caption{Normalized branching versus Majorana phases for NH (left) and IH (right). 
${\rm Im}(z)=1$.}
\label{fig:sXX}
\end{figure}

Sizable  Majorana phases may dilute the flavor correlations. 
The dependence of the flavor branchings on Majorana phases 
is shown in Fig.~\ref{fig:lXX} for  ${\rm Im}(z) \ge 2$. 
The largest variations occur near $\Phi\approx \pm\pi/2$.
It is important to note that (for Im($z$)$\ge 2$):
\begin{itemize}

\item For NH, BR($V \mu$) is down (up) and BR($V \tau$) is up  (down) by an 
approximate factor of two for $\Phi \approx \pi/2$ ($-\pi/2$) with respect to 
$\Phi=0$, while BR($Ve$) is independent of the phase;
 
\item For IH, BR($V \mu$) $\approx)$ BR($V \tau$) in the whole $\Phi$ range 
and are highly suppressed at $\Phi\approx\pi/2$,  where BR($Ve$) is up by a factor 
of two with respect to $\Phi=0$.

\end{itemize}

We remind the reader again that one neutrino is massless in this set-up, 
a direct consequence of the underlying SU(5) symmetry.

\begin{figure}[tb]
\includegraphics[scale=1,width=8.0cm]{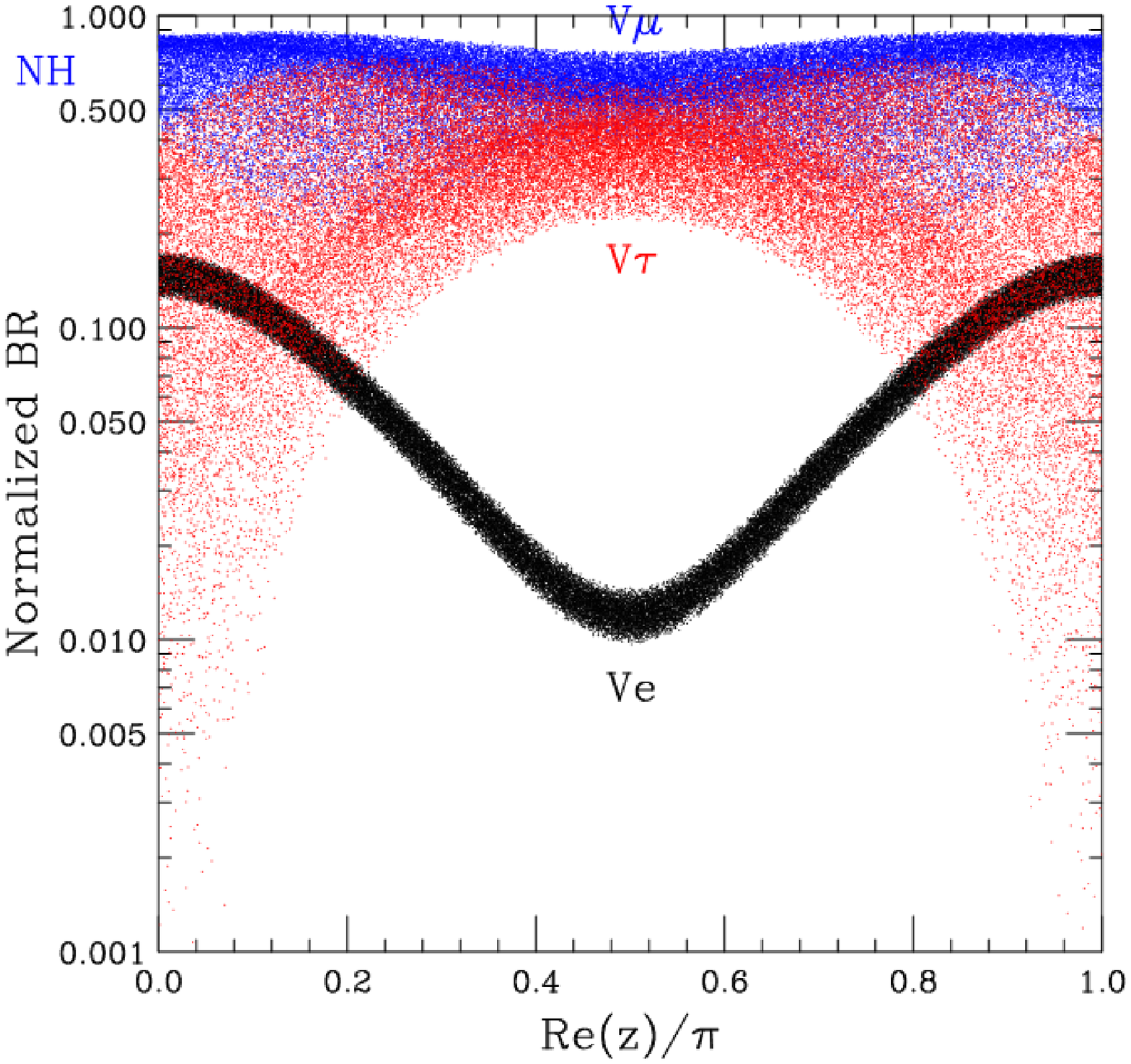}
\includegraphics[scale=1,width=8.0cm]{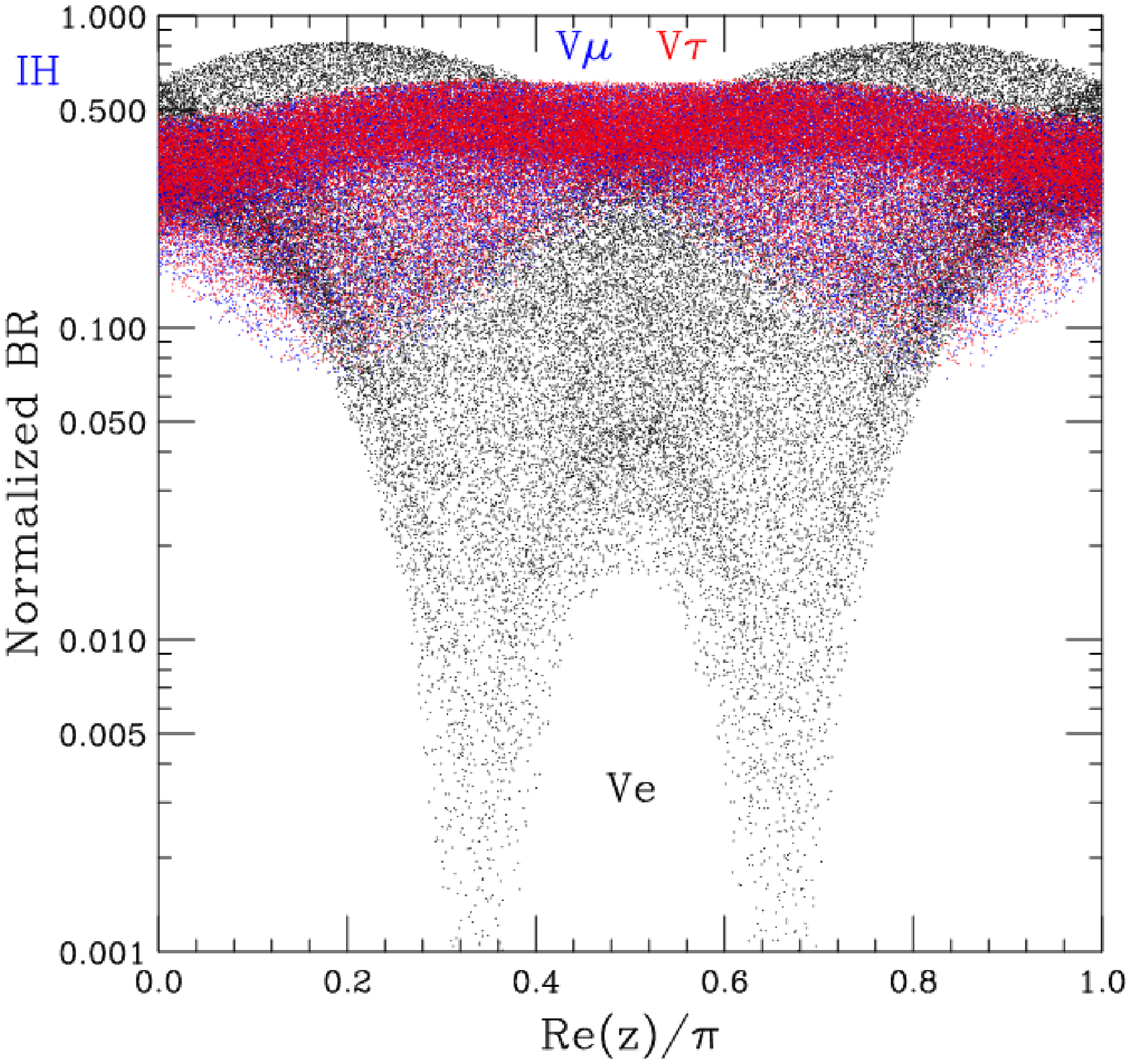}
\caption{Normalized branching versus ${\rm Re}(z)$ for NH (left) and IH (right). 
${\rm Im}(z)=0.5$ and $\Phi \in [-\pi/2,\pi/2]$. 
}
\label{fig:Rztxx}
\end{figure}

\begin{figure}[tb]
\includegraphics[scale=1,width=8.0cm]{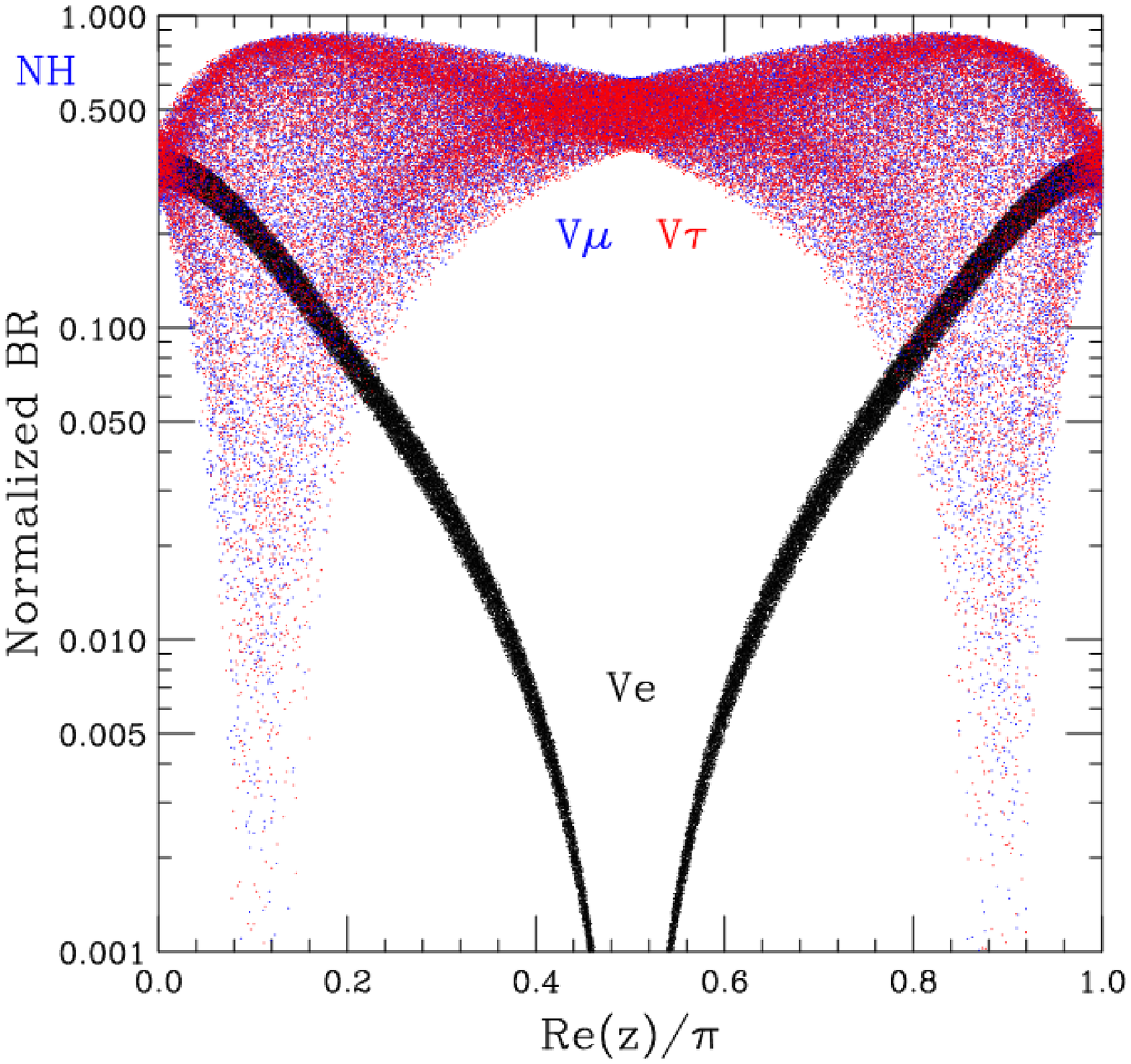}
\includegraphics[scale=1,width=8.0cm]{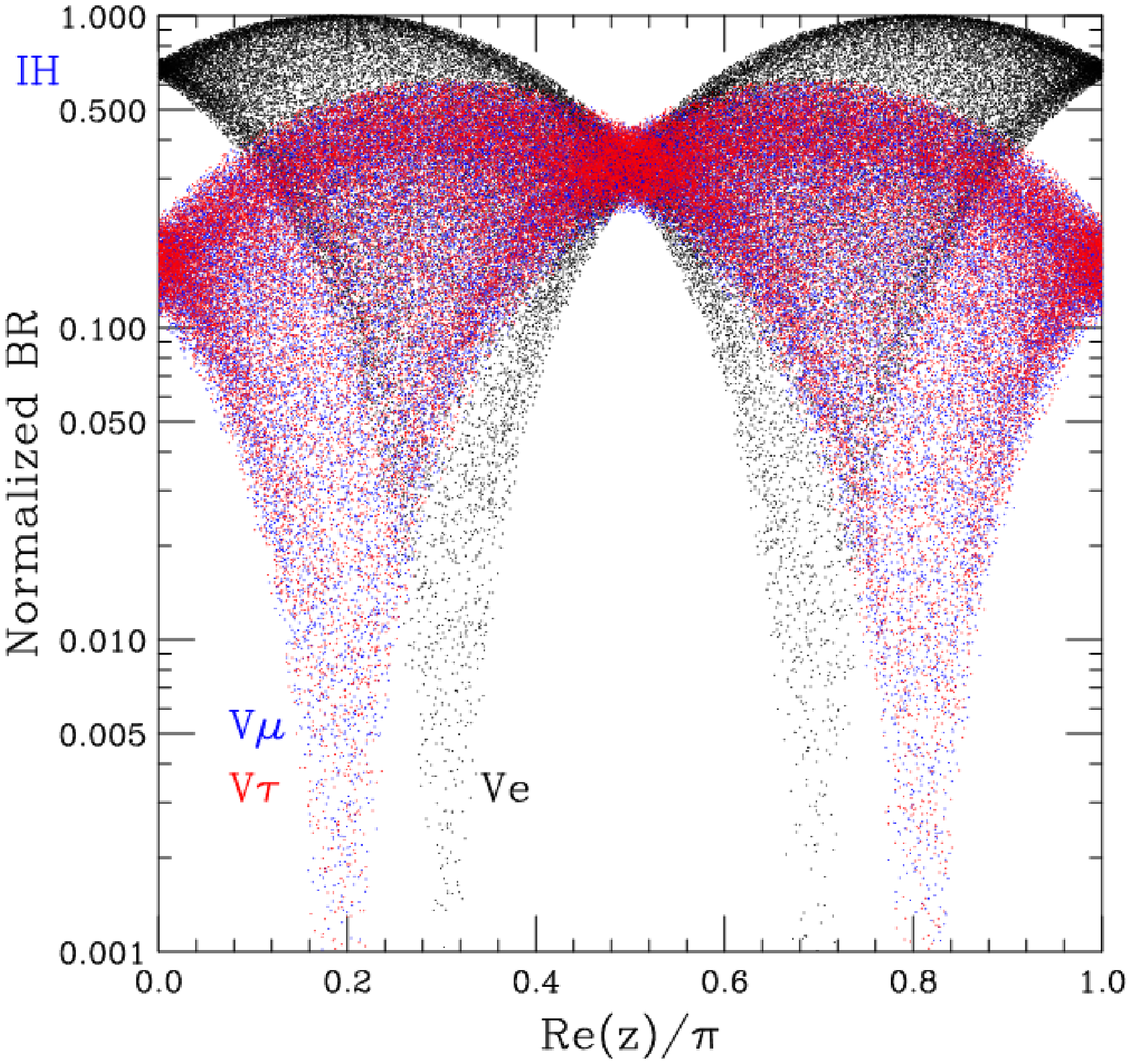}
\caption{Normalized branching versus ${\rm Re}(z)$ for NH (left) and IH (right). 
${\rm Im}(z)=0$ and $\Phi \in [-\pi/2,\pi/2]$. 
}
\label{fig:Rz0xx}
\end{figure}

For smaller ${\rm Im}(z)$, 
the branching fraction  dependence on $\Phi$ gets smeared up, 
as shown in Fig.~\ref{fig:sXX} for ${\rm Im}(z)=1$.
Instead, they have a clearer dependence on the real part of $z$, 
${\rm Re}(z)$,  another phase  with periodic behavior,  as seen in 
Figs.~\ref{fig:Rztxx} and \ref{fig:Rz0xx} for ${\rm Im}(z)=0.5$ and 0,
respectively. 
The reader should keep in mind that for large enough values of ${\rm Im}(z)$ the 
${\rm Re}(z)$ plays a much less significant role.

\subsection{Determining the leptonic mixing matrix}

The message of the above discussion is that while one cannot make 
predictions yet for the various branching fractions, the collider signatures can 
shed light on the lepton mixing parameters. The theory in question has only 
two real parameters on top of the $U_{PMNS}$: Re$(z)$ and Im$(z)$. Since we 
can measure in principle two branching fractions and the total lifetime, one can 
get information on the yet unknown mixing angle $\theta_{13}$ and the 
Dirac and Majorana phases $\delta$ and $\phi$. 

There is a good hope to probe 
$\theta_{13}$ in the near future through neutrino oscillation experiments. 
The large value of $\theta_{13}$ 
(close to the current experimental upper limit $\sin{\theta_{13}}\approx 0.18$) 
allows at least in principle for the determination of the Dirac phase $\delta$. 
Together with the observation of matter effects in oscillation, large $\theta_{13}$ 
allows also to distinguish the normal from the inverted hierarchy. On the other hand, 
for small values of $\theta_{13}$, neutrino physics may not be sufficient to clear 
the above issues. Here the information from the LHC may be particularly handy. 
For this reason we highlight the small $\theta_{13}$ case.
A few generic comments are warranted. 

1) In this model it turns out that $\mu\to 3e$ dominates by large over $\mu\to e\gamma$ 
\cite{Abada:2008ea}. Simply the observation of $\mu\to e\gamma$ by the MEG experiment 
would eliminate this theory. 

2) Since one neutrino is massless the neutrinoless double beta decay can play a 
clear role in distinguishing the normal from the inverted hierarchy. In particular, 
its observation in the next generation of experiments would rule out the normal 
hierarchy, as shown in Fig.~8.5 of \cite{Strumia:2006db} (see the far left limit of 
the far right plot).

3) The observation of $\mu\to 3e$ would imply large Yukawa couplings, which 
would in turn imply large Im($z$). In this limiting case the triplet lifetime would be 
too small to be measured. For normal hierarchy there is a clear prediction of 
less than $10\%$ triplet normalized branching  into electrons. So the final 
state with two electrons represents less than $1\%$ of the total two charged 
lepton final states in this case.

For negligible $\theta_{13}$ (that we set to zero), 
in both NH and IH cases clear predictions emerge. In the case of IH one finds 
\cite{Bajc:2007zf}

\begin{equation}
\frac{\rm NBR_{\tau}}{\rm NBR_{\mu}}=\tan^2{\theta_{23}}
\end{equation}

\begin{figure}[tb]
\includegraphics[scale=1.5]{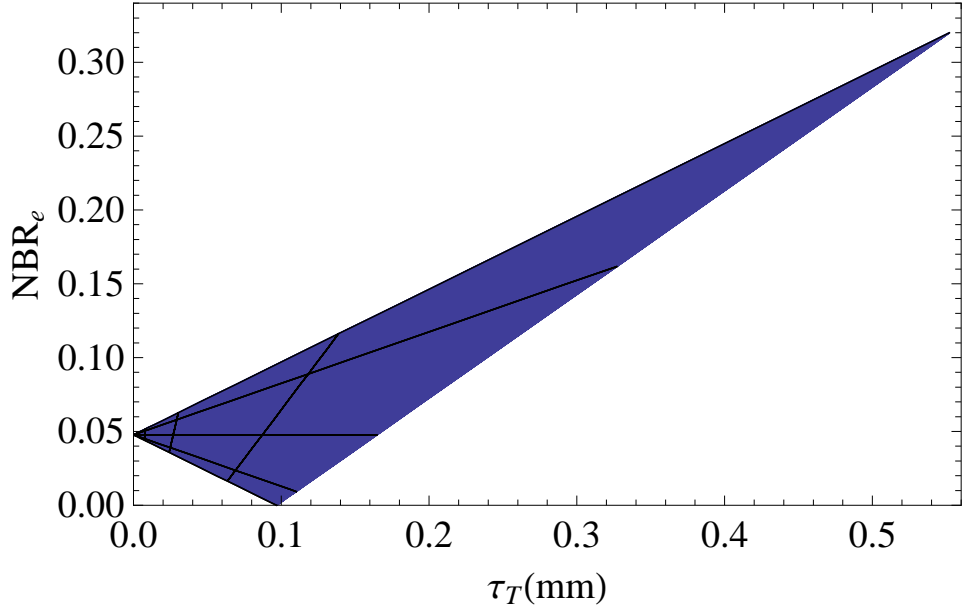}
\caption{The allowed values of the triplet lifetime and the electron 
normalized branching fraction in the NH case with $\theta_{13}=0$, $M_T=200$ GeV.}
\label{fig:nbr1tau}
\end{figure}

In the opposite case of NH the electron NBR and the total lifetime are 
functions of Re$[z]$ and 
Im$[z]$ only, which can thus be determined. The lifetime and the electron NBR 
cannot take arbitrary values, but are restricted, as can be seen 
from Fig.~\ref{fig:nbr1tau}.
The muonic NBR can then be used for 
the determination of the Majorana phase $\Phi$, since the Dirac phase $\delta$ 
disappears in this case. 

One could go on and on, but at this point it is not so useful due to the poor 
knowledge of the leptonic mixing matrix. As time goes on and knowledge 
of $U_{PMNS}$ grows, there will be ample opportunity to return to these issues.

\section{Heavy Lepton Signals at Hadron Colliders}
\label{section4}

Based on the model discussed in section \ref{section2}, the leading production 
of the heavy triplet leptons at a hadron collider is via the Drell-Yan type processes
\bea
&& q\bar q' \to W^{*\pm} \to T^\pm T^0,\quad  q\bar q' \to W^{*\pm} \to T^0 \ell^\pm,\\
&& q\bar q \to \gamma^*,Z^* \to T^+  T^-,\quad  q\bar q \to  Z^* \to T^\pm \ell^\mp.
\eea
Note that there is no tree-level process for $T^0\overline{T^0}$ production. 
In our numerical analysis, we use the CTEQ6L1 parton distribution function
\cite{cteq6l} with factorization scale $Q = \sqrt{s}/2$, where $s$ is the intermediate
gauge boson four-momentum squared (or the parton-level CM energy squared).

\subsection{Total cross sections}

\begin{figure}[tb]
\begin{center}
\scalebox{0.6}{\includegraphics[angle=0]{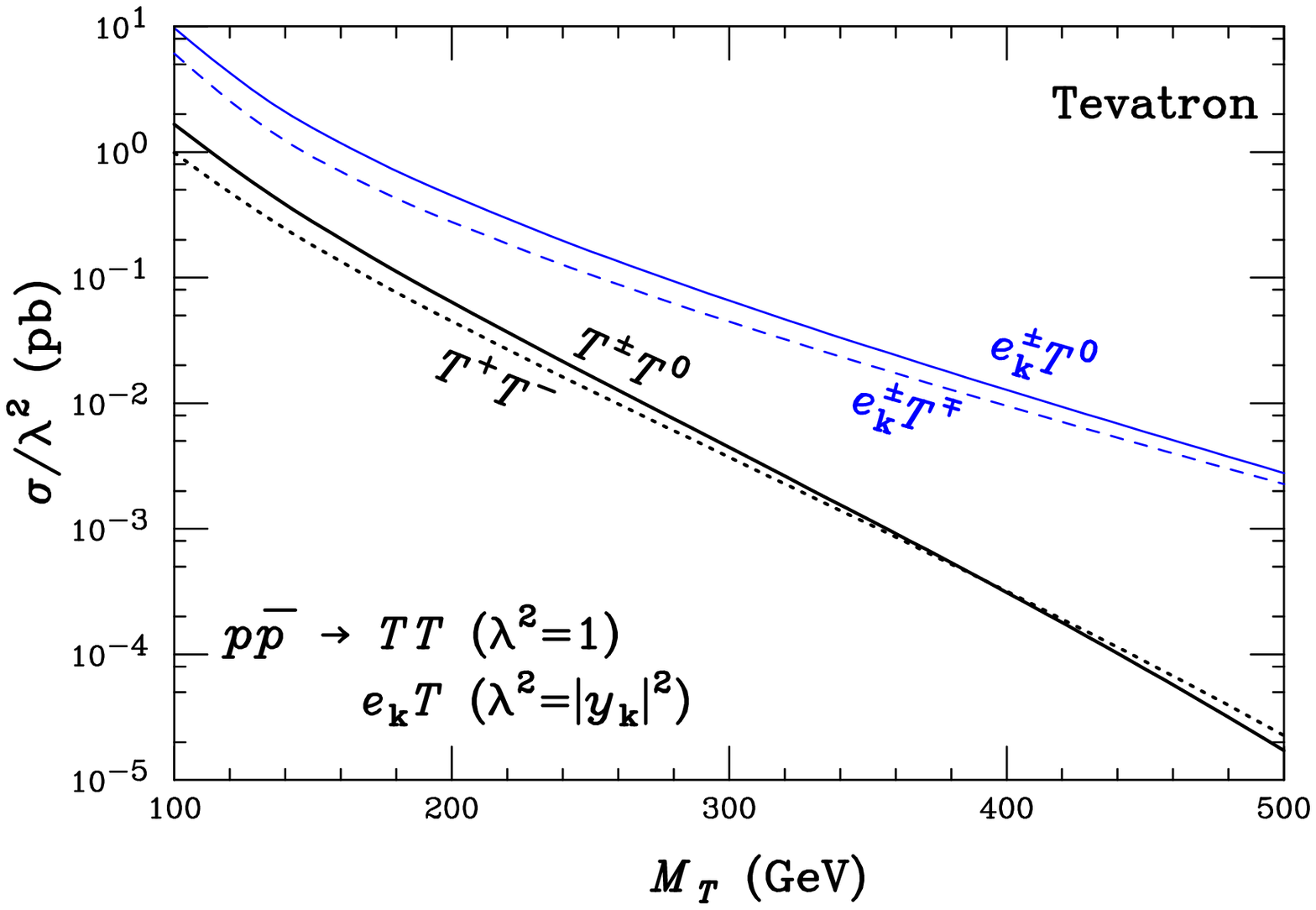}}
\scalebox{0.6}{\includegraphics[angle=0]{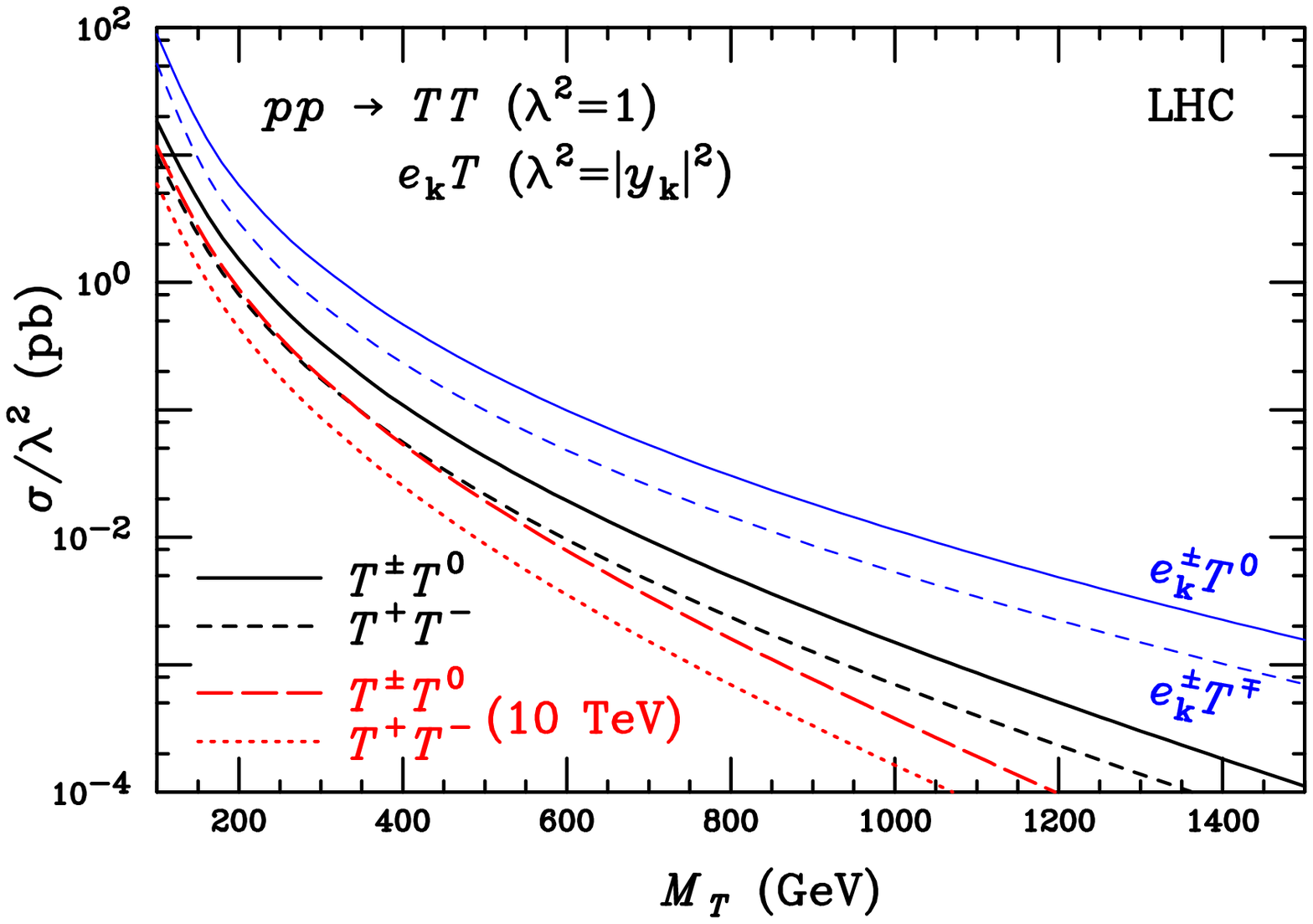}}
\caption{Cross sections of single and pair productions of
 $T^\pm/T^0$  as a function of its mass, (a) at the Tevatron (1.96 TeV) 
 and (b) at the LHC (14 TeV and 10 TeV). 
 The scaling constant $\lambda^2$ is 1 for $TT$, and $|y_k|^2$ for  $e_k T$. }
\label{fig:total1} 
\end{center}
\end{figure}

There are two main mechanisms to produce the heavy leptons in hadronic collisions:  
pair production of  $T\overline{T}$, and associated single production $Te_i$.
The cross sections for  heavy pair production are well predicted by the SM gauge
interactions. 
Those for  heavy-light associated production are governed by the Yukawa
coupling $y_T^i$.  In Fig.~\ref{fig:total1}, we present the total production cross
sections for those processes versus the heavy lepton mass  $M_T$, for 
(a) at the Tevatron ($p\bar p$ at $\sqrt S =1.96$ TeV) 
and (b) at the LHC ($pp$ at $\sqrt S =14$ TeV and 10 TeV). 
To view the generic feature, we have pulled out the effective couplings $\lambda^2$
 in the plots, which is normalized to unity for the pair production, 
 and to the Yukawa coupling squared for the single production. 

Given the consideration of Eq.~(\ref{eq:yukawa}) that leads to an upper bound on
the Yukawa couplings of $|y_k|^2 < {\cal O}(10^{-3})$,  we expect
the single production to be much smaller than the pair production of the
heavy triplet via gauge interactions. Thus the potential signal observation at 
hadron colliders 
can be significantly enhanced over the case of Type I seesaw 
\cite{Han:2006ip,del Aguila:2007em}, where the only weak signal comes
from the $T^0 e_i$ type due to mass mixing. 
Generically, the charged current process of $T^\pm T^0$ production has
a larger cross section than that of the neutral current for $T^+ T^-$ at the parton level.
This leads to the larger cross section rate by about a factor of 2 at the LHC as seen
in Fig.~\ref{fig:total1}(b) by the solid and short-dashed curves, 
while they are about equal at the Tevatron due to the compensation from the 
larger valence quark luminosity $u\bar u,\ d\bar d$ for the neutral
current. One sees that the cross section rate of the pair production can reach the
order of 20 fb for $M_T \sim$ 250 GeV at the Tevatron,
and the order of 2 fb for $M_T \sim$ 1000 GeV at the LHC.
The cross sections at a 10 TeV LHC will drop to $60\%-50\%$ at $M_T=200-400$ GeV,
and to $25\%$ at $M_T=1$ TeV, as given by the long-dashed and dotted curves 
in Fig.~\ref{fig:total1}(b).
We note that  the pair production of the triplets is the same as that 
of  the winos in Supersymmetric theories 
when we ignore the other superpartners \cite{Cheung:2005ba}. 
Our results are in accord with this case. 

\subsection {Final states from the pair production of $T\overline T$}

We focus on the pair production of the heavy triplets. 
The possible combinations of decay channels of pair-produced $T$'s are listed
in Table.~\ref{tab:PairModes}, where $\ell$ denotes the observable charged lepton
flavors $e$, $\mu$ and $\tau$.

\begin{table}[htb]
\begin{centering}
 \begin{tabular}[t]{| l || l | l | l|}
   \hline
    & $T^-\to \nu W^-$& $T^-\to \ell^- Z$& $T^-\to \ell^- h$ \\
   \hline \hline 
   $T^0\to\nu h$& $\nu\nu W^-h$ & $\nu\ell^- Zh$ & $\nu\ell^- hh$ \\
   \hline
   $T^0\to\nu Z$& $\nu\nu W^-Z$ & $\nu\ell^- ZZ$ & $\nu\ell^- Zh$ \\
   \hline
   $T^0\to\ell^- W^+$&$\nu\ell^- W^-W^+$ &$\ell^-\ell^- W^+Z$ &$\ell^-\ell^- W^+h$ \\
   \hline
   $T^0\to\ell^+ W^-$&$\nu\ell^+ W^-W^-$ &$\ell^+\ell^- W^-Z$ &$\ell^+\ell^- W^-h$ \\
   \hline \hline
   $T^+\to\nu W^+$& $\nu\nu W^-W^+$& $\nu\ell^- W^+Z$& $\nu \ell^- W^+h$\\
   \hline
   $T^+\to\ell^+ Z$& $\nu\ell^+ W^-Z$& $\ell^+\ell^- ZZ$& $\ell^+\ell^- Zh$\\
   \hline
   $T^+\to\ell^+ h$& $\nu\ell^+ W^-h$& $\ell^+\ell^- Zh$& $\ell^+\ell^- hh$\\
   \hline 
 \end{tabular}
 \caption{Triplet pair decay channels to the SM particles. $\ell=e,\mu,\tau$.
  \label{tab:PairModes}}
\end{centering}
\end{table}

We are set to search for the heavy leptons and to test the seesaw mechanism.
We wish to unambiguously identify the Majorana nature of $T^0$.
We are thus left with the unique channels with lepton number violation
\beq
  T^0 T^\pm \to (\ell^\pm W^\mp) (\ell^\pm Z/h), 
  \ {\rm or }\ \ell^\pm \ell^\pm W^\mp Z/h
\eeq
with $\ell = e, \mu, \tau$, which violate lepton number by two units $\Delta L=2$.
We must require the hadronic decay modes of $W, Z$ and $h$ in order 
to manifestly preserve the unique feature of the lepton number violation.
These particular final states are thus, without missing neutrinos, 
\beq
 \ell^\pm \ell^\pm\  j_1 j_1'\ j_2 j_2',
\label{LsJs}
\eeq
where $j_1 j_1'$ and $j_2 j_2'$ are from decays of $W$ and $Z/h$, respectively. 
From the observational point of view, these final states have unique kinematical
features and are quite clean. The
lepton number violation has no genuine  irreducible SM backgrounds. 

It is particularly important to identify the lepton flavors for those final state processes.
Given the predicted relations between the neutrino mass and oscillation
parameters and the couplings, we tabulate the predicted 
leading  channels of $\Delta L=2$ processes along with the indicative ranges
of their branching fractions in Table \ref{TabV}, where the first two factors
$(1/2) (1/4)$ come from the decay branching to a given boson $T^+ \to Z/h \ell^+,\ T^0\to W^- \ell^+$
in Fig.~\ref{fig:br}, respectively, and the next factors are from the branching to specific flavor of
charged leptons as in Figs.~\ref{fig:zB00} and \ref{fig:BB00}.
We see a hierarchical order among the event rates of their flavor combinations.
\bea
&&\mu\mu, \mu \tau, \tau\tau  \gg ee , \quad {\rm for\ NH,}\\
&& \mu\mu, \mu \tau, \tau\tau <  ee, \quad {\rm for\ IH.}
\eea
The observation of these qualitative features should serve as direct test  of our 
Type I + Type III Seesaw mechanism. 
For simplicity from the observational point of view,  we will mainly consider $e,\ \mu$ final states,
and only comment on the tau final state briefly. 
The final decay branching fraction of  $T^\pm, T^0, W^\pm, Z$ and $h$ to the final state of Eq.~(\ref{LsJs}) 
based on Table \ref{TabV} is thus 
\begin{equation}
BR = 
 \left\{ \begin{array} {cc}
 {1\over 2}\ {1\over 4} \cdot ({1\over 2})^2\ (70\%)^2 \approx {1 \over 64}, &  \quad {\rm only}\ \mu^\pm \mu^\pm\ {\rm for\ NH}, \\
 \\
{1\over 2}\ {1\over 4} \cdot ({1\over 2})^2\ 2\ (70\%)^2 \approx {1 \over 32}, & \quad {\rm both}\ e^\pm\ {\rm and}\ \mu^\pm\ {\rm for\ IH}.
\end{array}\right.
\label{eq:BRs}
\end{equation}

\begin{center}
\begin{table}[tb]
\begin{tabular}[t]{|c|c|c|}
\hline
Signal channels & Leading modes and BR & Leading modes and BR\\
& Normal Hierarchy & Inverted Hierarchy \\ 
\hline
    $T^\pm\  T^0 $ 
 &  &  \\
    $\Phi \approx 0$ 
 & $\mu^\pm\mu^\pm  \ \ \ 
     {1\over 2}\ {1\over 4} \cdot ({1\over 2})^2$ 
 & $e^\pm e^\pm \ \ \ 
     {1\over 2}\  {1\over 4} \cdot ({1\over 2})^2$\\
 &     $\tau^\pm \tau^\pm  \ \ \ 
    {1\over 2}\  {1\over 4} \cdot ({1\over 2})^2$ 
 & $e^\pm \mu^\pm \ \ \ 
     {1\over 2}\ {1\over 4} \cdot  {1\over 2}\ {1\over 4}\ 2$ 
\\
 & $\mu^\pm\tau^\pm  \ \ \ 
    {1\over 2}\ {1\over 4} \cdot  ({1\over 2})^2\ 2$ 
 & $e^\pm \tau^\pm\ \ \ 
     {1\over 2}\ {1\over 4} \cdot\  {1\over 2}\ {1\over 4}\ 2$ 
\\
   & 
   & $\mu^\pm \mu^\pm \ ({\rm or}\ \tau^\pm\tau^\pm) \ \ \ 
        {1\over 2}\ {1\over 4} \cdot\ ({1\over 4})^2$   
\\
$\Phi \approx\pi/2 $ & $\mu:\times 1/2;\ \tau:\times 2$ & BR$(Ve) \to 1$ 
\\
\hline
\end{tabular}
\caption{Leading  channels of $\Delta L=2$ and the indicative ranges
of their branching fractions, as discussed in the text for both cases of 
the NH and IH. Another factor of $(70\%)^2$ should be included to count
for the 4-jet final state decays from $W,Z$ and approximately from $h$ as well.}
\label{TabV}
\end{table}
\end{center}

\subsection{Tevatron}

We first explore the signal observability at the Tevatron. We define the signal
identification with two charged leptons and four jets 
\bea
\nonumber
&& p_T(\ell) > 18\ {\gev}, \  |\eta_\ell | < 2;\quad 
p_T(j) > 15\ {\gev}, \  |\eta_j| < 3; \\
&& \Delta R(jj) > 0.4,\ \Delta R(j\ell) > 0.4,\ \Delta R(\ell\ell) > 0.3,
\label{eq:cuts}
\eea
where the particle separation is 
{$\Delta R({\alpha\beta}) \equiv \sqrt{ \left
  (\Delta \phi_{\alpha\beta} \right)^2 + \left (\Delta
    \eta_{\alpha\beta} \right )^2}$} 
    with $\Delta \phi$ and $\Delta \eta$
      being the azimuthal angular separation and rapidity difference
        between two particles.
         We further choose to look for $e,\mu$ events only, and 
         demand there be no significant missing transverse energy
\beq
 \etmiss < 15\ {\gev}.
\label{ETmissTeV}
\eeq
In  our parton-level simulation, we smear the lepton (electron here) and 
jet energies with a Gaussian distribution according to 
\beq
\frac{\delta E}{E} = \frac{a}{\sqrt{E/{\rm GeV}}} \oplus b ,
\label{resol}
\eeq
where  $a_{e} = 13.5\% ,\ b_{e} = 2\%$ and $a_{j} = 75\% ,\ b_{j} = 3\%$ 
($\oplus$ denotes a sum in quadrature) \cite{Abazov:2007ev}.

\begin{figure}[tb]
\begin{center}
\includegraphics[scale=0.7]{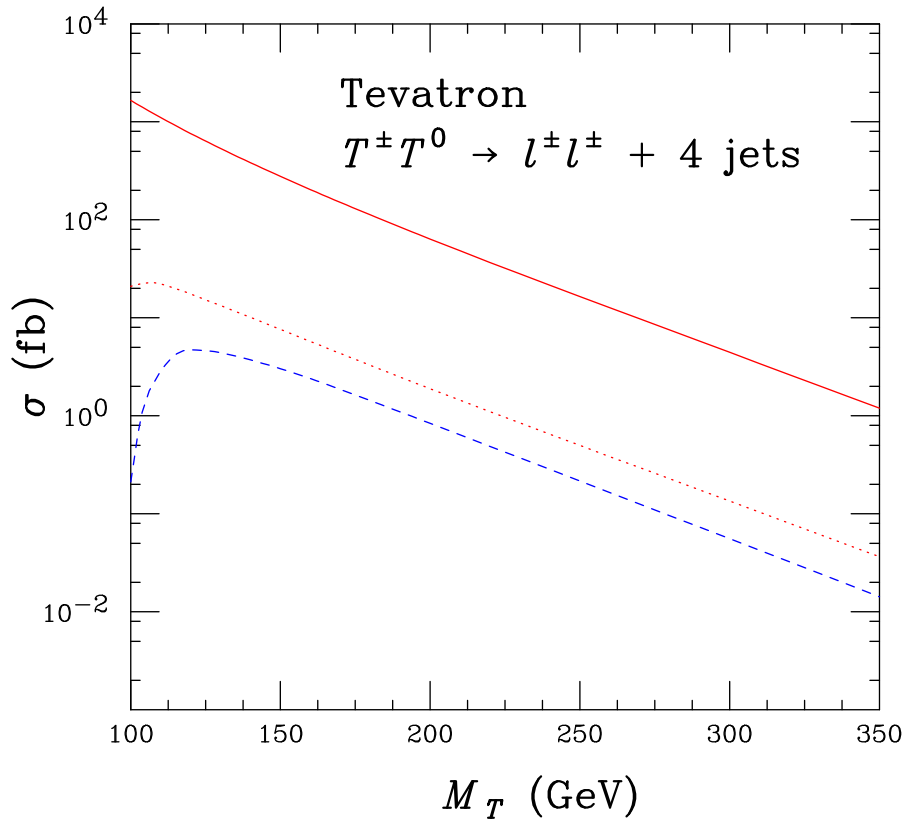}
\end{center}
\caption{
Total cross section for $p\bar p \to T^\pm  T^0$ production and decay 
at the Tevatron energy
$\sqrt S = 1.96$ TeV as a function of the heavy lepton mass. The solid curve (top) is
for the  production rate of $T^+ T^0 + T^- T^0 $ before any decay or kinematical cuts.  
The dotted (middle) curve represents production cross section including 
appropriate branching fraction of Eq.~(\ref{eq:BRs}), for the case of IH for illustration, 
with $\ell=e,\mu$ taken from the leading channels in Table \ref{TabV}.
The dashed (lower) curve shows variation of signal cross section after taking
into account the cuts in Eqs.~(\ref{eq:cuts}) and (\ref{ETmissTeV}). } 
\label{fig:cs_sigTeV}
\end{figure}

In Fig.~\ref{fig:cs_sigTeV},  we show the total cross section in units of fb, 
for $p\bar p \to T^\pm  T^0$ production and their decays at the Tevatron energy
$\sqrt S = 1.96$ TeV as a function of the heavy lepton mass. The solid curve (top) is
for the  production rate of $T^+ T^0 + T^- T^0 $ before any decay or kinematical cuts.  
The dotted (middle) curve represents production cross section including 
 appropriate branching fraction  of Eq.~(\ref{eq:BRs}) for the case of IH 
with $\ell=e,\mu$ taken from the leading channels in Table \ref{TabV}.
The dashed (lower) curve shows variation of signal cross section after taking
into account all the kinematical cuts as in Eqs.~(\ref{eq:cuts}) and (\ref{ETmissTeV}).
As a result of the cuts, the cross section is reduced by about a factor of 3.
As mentioned above, these final states with lepton number violation 
have no genuine  irreducible SM backgrounds. The other fake backgrounds
from multiple $W,Z$ production leading to the final state of Eq.~(\ref{LsJs})
the Tevatron energies are negligibly small.
Assuming an integrated luminosity of 8 fb$^{-1}$ is available
in the near future, a $99\%$ Confidence Level (CL) signal would require 5$-$7 events,
that would lead to mass reach $M_T\sim 200$ GeV or higher.
For the case of NH, the electron mode is absent while the tau mode shows up with
a larger branching fraction, as a distinctive feature of this neutrino mass pattern. 

Before ending this section, two remarks are in order. First, as argued in \cite{Perez:2008ha}, 
the additional channels $e^\pm \tau^\pm,\ \mu^\pm \tau^\pm$ and $\tau^\pm\tau^\pm$
may be fully reconstructable kinematically. This  will significantly enhance our signal
observability, as well as the discrimination power for the NH and IH mass patterns
as outlined in Table \ref{TabV}. The $\tau$ identification would be particularly crucial
if the effects of the CP phase are present. 
Further studies will be needed to incorporate the $\tau^\pm$ modes.
Second, our results are based on parton level
simulations although we have implemented the detector acceptance and smeared
the energy and momenta. We realize that there will be additional detection 
efficiencies associated with the final state particle identification and construction. 
Even with high efficiencies of over $90\% $ for each object \cite{TeVe}, 
the complex final state of two leptons and four jets will
result in about a factor of two reduction in rate. More realistic simulations are needed
for further conclusions, that are beyond the scope of the current work.

\subsection{LHC}

The LHC signatures of the Type III seesaw and their background were already 
studied in the case of three lepton 
triplets~\cite{Ma:2002pf,Franceschini:2008pz,delAguila:2008cj,delAguila:2008hw},
while there is only one light triplet in our theoretical setting. 
Our results below are compatible with 
their findings whenever the comparison is possible. 

At the LHC energies, we follow a similar approach for the signal search to the above. 
We select the events  with two charged leptons and four jets 
by the following basic kinematical acceptance \cite{LHCe}
\bea
\nonumber
&& p_T(\ell) > 15\ {\gev}, \  |\eta_\ell | < 2.5 ;\quad 
p_T(j) > 20\ {\gev}, \  |\eta_j| < 3; \\
&& \Delta R(jj) > 0.4,\ \Delta R(j\ell) > 0.4,\ \Delta R(\ell\ell) > 0.3.
\label{eq:cuts2}
\eea
Once again, we look for clean $e, \mu$ events and 
demand there be no significant missing transverse energy
\beq
 \etmiss < 25\ {\gev}.
 \label{ETmissLHC}
\eeq
As for the energy smearing of the leptons and jets, we adopt the same form
of Eq.~(\ref{resol}), 
with the CMS parameterization  
$a_{e} = 5\% ,\ b_{e} = 0.55\%$ and  $a_{j} = 100\% ,\ b_{j} = 5\%$.
For simplicity, we did not separately smear the muon momenta by tracking, 
which would result in a better resolution at lower energies 
and become worse at higher energies, typically when $M_T \gsim 500$ GeV.

\begin{figure}[tb]
\begin{center}
\includegraphics[scale=0.7]{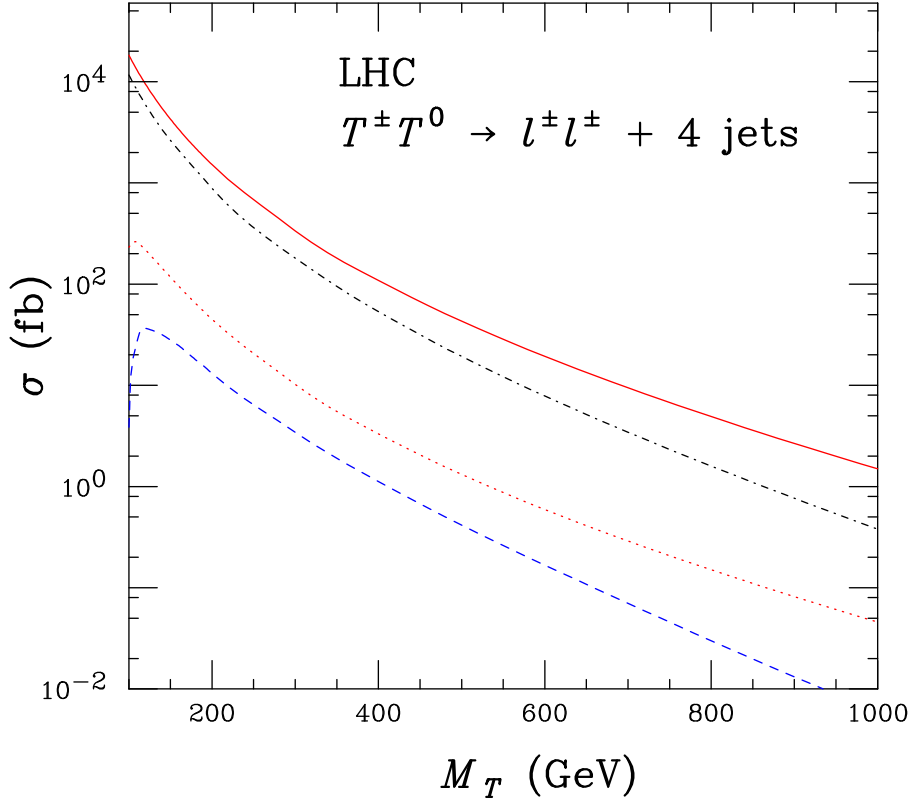}
\end{center}
\caption{ 
Total cross section for $p p \to T^\pm  T^0$ production and decay at the LHC energy
 at $\sqrt S = 14$ TeV  as a function of the heavy lepton mass. The solid curve (top) is
for the  production rate of $T^+ T^0 + T^- T^0 $ before any decay or kinematical cuts.  
The cross section at the 10 TeV LHC is also plotted (the curve right below) for comparison.
The dotted (middle) curve represents production cross section including 
appropriate branching fraction of Eq.~(\ref{eq:BRs}), for the case of IH for illustration,
with $\ell=e,\ \mu$ taken from the leading channels in Table \ref{TabV}. 
The dashed (lower) curve shows variation of signal cross section after taking
into account the cuts in Eqs.~(\ref{eq:cuts2}$-$ \ref{eq:MT}). }
\label{fig:cs_sig}
\end{figure}

 \begin{figure}[tb]
 \begin{center}
  \includegraphics[scale=0.7]{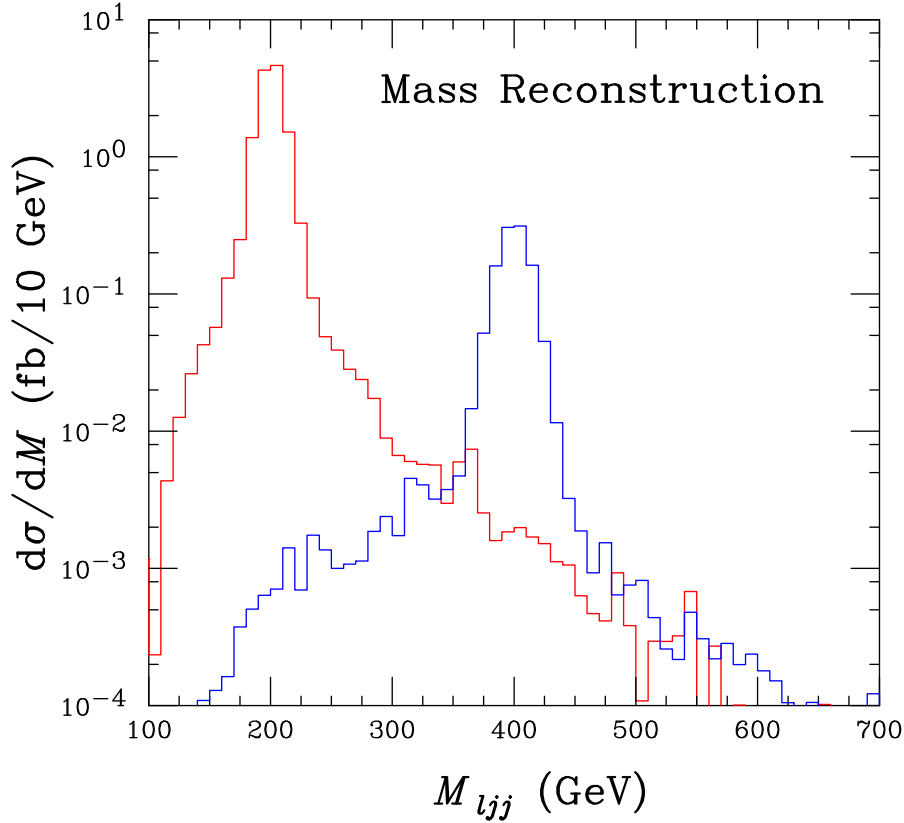}
 \end{center}
 \caption{Differential distribution of the reconstructed mass $M_{\ell jj}$ at the LHC for two 
 representative 
values of heavy triplet lepton mass $M_T=200,\ 400$ GeV. The energy bin size is $10$ GeV. }
 \label{fig:mljj_dist}
 \end{figure}

In Fig.~\ref{fig:cs_sig} we show the total cross section for $p p \to T^\pm  T^0$ production 
at the LHC as a function of the heavy lepton mass. The solid curve (top) is
for the  production rate of $T^+ T^0 + T^- T^0 $ at $\sqrt S = 14$ TeV 
before any decay or kinematical cuts.  
The cross section at 10 TeV is also plotted (the dot-dashed curve right below). In comparison
with LHC at 14 TeV, the rate is scaled down to $60\% - 50\%$ at $M_T=200-400$ GeV, and to 
$25\%$ at $M_T=1$ TeV. 
The dotted (middle) curve represents production cross section including 
appropriate branching fraction of Eq.~(\ref{eq:BRs}), for the case of IH for illustration,
with $\ell=e,\ \mu$ taken from the leading channels in Table \ref{TabV}. 
The dashed (lower) curve shows variation of signal cross section after taking
into account all the kinematical cuts in Eqs.~(\ref{eq:cuts2}$-$ \ref{eq:MT}). 
As a result of the cuts, similar to the discussions in the previous section, 
the cross section is reduced by about a  factor of 4
for a modest lepton mass, but the reduction becomes more severe due to the
fact that the decay products are more collimated for a much heavier lepton
and that the faked missing energy increases.

\begin{table}[tb]
\begin{tabular}{| c || c | c |c | c |  }
    \hline
     $\sigma$ (fb) & Basic & $ \etmiss $ &  $m_{jj}$ cut  & $M_{\ell jj}$   \\
Process   & Cuts  &  $< 25~{\rm GeV}$ & $65 - 105$ GeV & $200 \pm 50$ GeV \\
       \hline
\hline
      $T^\pm T^0,\ M_T=200$ GeV & 15.0 & 13.5 & 13.4  & 13.4 \\
\hline
      $T^\pm T^0,\ M_T=400$ GeV & 1.49 & 1.13 & 1.12 & 1.12 ($400 \pm 50$ GeV) \\
\hline
\hline
  $W^+W^+ (W^-W^-) \ 4$ jets & 1.26 (0.51) & 0.038 (0.024)& 7.6 (5.6)$\times 10^{-3}$ & 1.3 (1.0)$\times 10^{-3}$ \\
\hline
          $W^+W^+W^- (W^-W^-W^+)\ 2$ jets & 3.2 (1.8) & 0.14 (0.08) & 0.087  (0.046) & 0.032 (0.014) \\
\hline
\hline
   Total backgrounds &  6.8 & 0.28 & 0.15 & 0.048 \\
\hline
  \end{tabular}
  \caption{Effects of the kinematical cuts on the production cross sections (in fb) 
at the LHC for  the signal  
$pp\rightarrow T^\pm T^0 \rightarrow \ell^\pm \ell^\pm j_1 j^\prime_1 j_2 j^\prime_2
(\ell = e, \mu ) $ as in the case of IH, and their leading SM backgrounds.
    We take $M_T=200,\ 400$ GeV for illustration. The final state branching fractions
have been included as given in Table \ref{TabV} (IH). The background results in the last
column are with a cut $M_{\ell jj} = 200\pm 50$ GeV. }
\label{Tab:III}
\end{table}
%

To further purify the signal sample, we can consider constraining the
di-jet mass $m(jj) \approx m_W,\ m_Z$ or $m_h$, where we will assume that the Higgs
mass is already known. To break the combinatorial degeneracy of the three possible
parings for the four jets, we first pick the one that fits well with 
\beq
65~{\rm GeV} < m(jj) < 105~{\rm GeV},\quad {\rm or}\quad 
100~{\rm GeV} < m(jj) < 140~{\rm GeV}~{\rm for}\ h.
\label{eq:mv}
\eeq
The most important feature of our signal events is the effective  
reconstruction of  the heavy lepton mass from the final state leptons and jets
 $M_{T^\pm} \approx m(\ell jj)_1 \approx M_{T^0} \approx m(\ell jj)_2$. 
The complication again is the combinatorial, with two choices of paring for
$\ell_{1,2}$ and $(jj)_{1,2}$. 
The best reconstruction is the one which has the least 
difference between the two reconstructed invariant masses 
$m_1$ and $m_2$ and the best reconstructed heavy 
lepton mass is the corresponding mean value $M_{\ell jj} = (m_1+m_2)/2$.
In Fig.~\ref{fig:mljj_dist}  we show the differential distribution of  the reconstructed
mass $M_{\ell jj}$ 
for two representative values of heavy lepton masses, 200 and 400 GeV.
It is evident from the shape of the distribution and location of the
peak that the reconstruction of the heavy lepton mass can be quite effective 
using the technique mentioned above. Since the physical width of the
heavy lepton is very narrow, the broad distribution at  
the peak is largely due to the detector resolution of the leptons and jets.
Moreover, we do not expect formation
of such a peak at those particular values from SM backgrounds. The 
background can at most contribute to a continuum distribution.  
We thus propose to examine a wide window for the reconstructed mass peak
\beq
M_T \pm 50\ {\rm GeV},
\label{eq:MT}
\eeq
when estimating the signal statistical significance. 
In Fig.~\ref{fig:cs_sig}, 
the short dot-dashed (lower) curve shows variation of signal cross section after taking
into account all cuts as discussed in the text. 
The signal efficiency for the cuts are as high as
$25\%$ for $M_T\approx 200-600$ GeV.

Although there is no intrinsic SM background to the lepton-number violating
processes, there are always some fake backgrounds that lead to some
similar final states to our signal events. We have estimated the different contributions 
using Madgraph/Madevent \cite{Alwall:2007st}. 
The immediate background to the
$\ell^\pm \ell^\pm 4j$ signal that comes in mind will be 
$$W^\pm W^\pm+4\ {\rm QCD\ jets.}$$ 
This background with our basic acceptance cuts has a cross section of 1.8 fb. 
After the selective cuts for the mass reconstructions, it is reduced to a 
negligible level, about three orders of magnitude down. A larger background is from
$$W^\pm W^\pm W^\mp +2\ {\rm jets} \to  W^\pm W^\pm  +4\ {\rm jets ,}$$ 
in which about $90\%$ of the events are actually from 
$t\bar t W^\pm \to W^\pm W^\pm b\bar b +2\ {\rm jets}$.
With the appropriate  branching fraction $BR(W\to l\nu)^2 BR(W\to jj) \approx 0.028$, 
the total cross section is about 15 fb and is reduced to about 5 fb with the basic acceptance cuts. 
After the selective cuts for the mass reconstructions, 
this background can be reduced by two orders  of  magnitude, to about 0.05 fb
at $M_T=200\pm50$ GeV. It becomes negligible at a higher mass window. 

Other backgrounds include $W^\pm W^\pm Z+2$ jets and $W^\pm W^\pm VV$ ($V=W, Z$).
These processes are rather small in production rate, typically less than 1 fb to begin with.
We will not consider them further. 

Another potentially large background is from the $b$ decays that give a charged
lepton. However, it is known that the stringent requirement of the lepton isolation 
would effectively separate the heavy quark backgrounds. Quantitatively, the
suppression efficiency is difficult to estimate reliably and perhaps will be better
understood once the real data become available.

In Table \ref{Tab:III},  we show the incremental effect of different cuts on the signal 
cross section for two different values of heavy triplet lepton masses,
200 GeV and 400 GeV, respectively,  along with the leading SM backgrounds.
Again, the signal events are essentially background free after all of the kinematical 
reconstructions.  To reach a $99\%$ CL, the signal would require 5$-$7 events,
that would lead to a mass reach $M_T\sim$ 450$-$480 (700$-$740) GeV with an integrated luminosity of 
10 (100) fb$^{-1}$.

As already noted in the previous section, we reiterate that 
the additional channels $e^\pm \tau^\pm,\ \mu^\pm \tau^\pm$ and $\tau^\pm\tau^\pm$
may be fully reconstructable kinematically \cite{Perez:2008ha}. 
This  will significantly enhance our signal observability 
as well as the discriminating power between the NH and IH mass patterns. 
As seen in Table \ref{TabV}, $\tau$ identification would be particularly crucial
if the effects of the CP phase are present. 
On the other hand,  we have not performed detailed detector simulations in our studies. 
Even with high reconstruction efficiencies of over $90\% $ for each object \cite{LHCe}, 
the complex final state of two leptons and four jets will
result in about a factor of two reduction in rate. More realistic simulations are needed
for further conclusions, which are beyond the scope of the current work.

\subsection{The conspiracy problem: how to distinguish triplets from doublets}

The main point  of our paper is the search for a light fermionic triplet with $\Delta L=2$ 
lepton number violating signatures such as  same-sign dileptons plus jets 
without significant missing energy at the LHC. 
Although we have established the signal observability for the lepton number violating 
processes in a large region of the parameter space, 
it is important to ask if we can confirm the triplet nature of the signal that does not follow 
from other sources of different new physics in the leptonic sector. 
The existence of the charged heavy leptons could be from
either a gauge doublet or triplet. 
Can the doublet of leptons leads to similar or, worse, same predictions? 
The answer is possibly yes. 
We explore the means to distinguish them both qualitatively and quantitatively.
The extra doublet of leptons can be either a part of a chiral sequential fourth generation 
or a vector-like particle, that we discuss next in turn.

\subsubsection{Sequential fourth generation heavy leptons}

The immediate example would be the fourth family sequential heavy leptons. 
There are a number of features that will distinguish the fourth family case from the 
(three family) triplet case. 

\begin{itemize}
\item[1] {\bf The $4^{th}$ family quarks:}
In an anomaly-free formulation, there will be a new family of quarks.
Due to the strong production, the new heavy quarks should be much easier
to observe  at the LHC up to a mass about 700 GeV \cite{LHCe} or higher. 

\item[2]  {\bf Gauge coupling strength in charged currents:}
An obvious difference between the doublet and the triplet is their
gauge couplings, as summarized in the Appendix. 
It turns out that the pair production
cross section for a triplet is larger than that for a doublet by a factor of two. 
Although it would be non-trivial to determine the cross section normalization
at the LHC for some processes with a complex final state, we are
optimistic to assume that our channel is clean and the cross section can
be determined with sufficient statistics when $M_T\sim 500$ GeV.

\item[3]  {\bf Charged-neutral mass difference:}
In the triplet model under consideration, there is a 
high degree of degeneracy of the triplet, $\Delta M_T\lsim 160$ MeV.
This is  in great contrast to a sequential doublet  model, where the lepton masses
are not predicted. In fact, the fit to the electroweak oblique parameter $\Delta S$  indicates 
a preference for a significant mass splitting, on the order of $30-60$ GeV,   
between the charged and the neutral leptons \cite{Kribs:2007nz}.  
Due to this the dominant channel for the decay of the charged heavy lepton becomes 
a neutral one ($N$, a Majorana neutrino) plus an off-shell $W$. This means that the $E^\pm N$ 
production  may lead to two same-sign leptons plus six jets (instead of four in the triplet case 
$T^\pm T^0$). For the same reason the $E^+E^-$ production in the doublet 
case will even lead to lepton number violating signatures of $\ell^\pm\ell^\pm+8$ jets. 
A word of caution is needed: it is not impossible that future studies find yet 
another possibility for extra sequential generation even with degenerate leptons. For this 
reason we study carefully the other distinguishable features at the international linear 
collider in the next section.

\item[4]  {\bf Neutral current coupling:} 
The new feature of the doublet is that the neutral lepton couples to the $Z$ boson, 
which distinguishes it crucially from the triplet, as summarized in the Appendix.
The production of a pair of neutral heavy leptons is absent in the triplet model,
and is negligibly small for a gauge singlet  $N$ as well.  Any clear signal for $N\overline N$ 
production would indicate a lepton doublet.

However, if the signal $N\overline N$ yields the final state $\ell^\pm \ell^\pm +4$ jets, 
it would be difficult to tell them apart from the triplet signal as we discussed earlier,
due to the fact that the $W$ and $Z$ in their hadronic decays are indistinguishable 
in the LHC environment. 
It is only hopeful  through the four charged lepton mode
 with missing energy $\ell^+ \ell^- \nu \ \ell^+ \ell^- \bar\nu$. 
Such events will not have a resonant structure for $N\to \ell^\pm W^\mp,\ \nu Z$;
while they will reveal the mass peak for $T^\pm \to \ell^\pm Z\to  \ell^\pm \ell^+ \ell^-$.

Let us remind the reader that in general there will be a non-negligible non-diagonal 
heavy-light leptonic coupling to $Z$. In the standard model 
the off diagonal lepton couplings to $Z$ are suppressed by tiny neutrino masses and thus 
completely negligible. Since the fourth neutral lepton has to be heavy, it is easy to generate 
the above mentioned vertex through the GIM at one loop, even if the fourth generation 
mixes very little with the first three. 

\item[5] {\bf Chiral couplings:} 
Perhaps the most conclusive test for the lepton doublet is to establish the 
charge forward-backward asymmetry, due to the chiral feature of their couplings.
However, since we are unable to distinguish $W$ and $Z$ in their hadronic decay,
we may have to rely on the pure leptonic decay of the $Z$, such as 
$T^+T^- \to  \ell^+ \ell^+ \ell^-,\   \ell^- \ell^+ \ell^-$, that would suffer from low statistics
and the potential ambiguity in identifying the incoming quark direction. 
On the other hand, this type of measurements would be straightforward at an
$e^+e^-$  linear collider, as we will comment on later. 
\end{itemize}
Obviously, there is no guarantee that either one of the above should be readily
observable at the LHC. 
However, the confirmation of any of the above signatures would be
nearly convincing to establish a triplet or a doublet model. 

\subsubsection{Vector-like heavy lepton doublet}

Another prominent example would be a vector-like heavy lepton. 
The characteristic features discussed  in points 2, 3 and 4 in the previous section are
still valid. Let us elaborate them.

\begin{itemize}
\item[1]  {\bf Gauge coupling strength:}
As summarized in the Appendix, due to the stronger gauge coupling for the triplet. 
the pair production cross section for a triplet is larger than that for a doublet 
by about a factor of two. 

\item[2]  {\bf Charged-neutral mass difference: } 
The crucial new feature in this case is that one can not have the degeneracy and 
appreciable lepton number violation. The vector-like nature of the doublets 
$L_L$ and $L_R$
\begin{eqnarray}
L_{L,R}=
\begin{pmatrix}
N
\cr
E
\end{pmatrix}_{L,R}
\end{eqnarray}
implies a gauge invariant mass term 
\begin{equation}
-{\cal L}_{Dirac}=M_D\bar L_LL_R+h.c.=M_D(\bar EE+\bar NN)
\end{equation}
where as usual $E\equiv E_L+E_R$ and in the same manner $N\equiv N_L+N_R$. 
At first glance there is a conspiracy because of the complete degeneracy between 
the charged and the neutral lepton, just as in the triplet case. However, as opposed to the 
triplet case, at this 
point the neutral lepton is a Dirac particle, which implies no lepton number violating 
signatures as the ones discussed throughout our paper.

Lepton number violation requires breaking the degeneracy through the Majorana masses 
\begin{equation}
-{\cal L}_{Majorana}=\delta M_LN_LN_L+
\delta M_RN_RN_R+h.c.\;.
\end{equation}
These mass terms can simply emerge due to Weinberg-like dimension 5 operators 
and they will lead to lepton number violation proportional to $(\delta M_{L,R}/M_D)^2$. 
Clearly in order not to be too small, the degeneracy between $E$ and $N$ must 
be substantially broken through $\delta M_{L,R}$. This strongly broken degeneracy will surely 
discriminate between the triplet and vector-like doublet. 

\item[3]  {\bf Neutral current coupling:} 
Also, as in the case of sequential fourth generation heavy leptons, 
the neutral lepton $N$ couples to the $Z$, so that 
the above discussion follows here as well.

\end{itemize}

\section{Heavy Leptons at a Linear Collider}
\label{sectionILC}

It has been demonstrated in Ref.~\cite{ILCLHC} that physics at the 
LHC and the $e^+e^-$ International Linear Collider (ILC) will be 
complementary to each other in many respects.  

In this section we study the production of a pair of heavy leptons $(E^\pm)$ 
at a linear collider. As discussed above, those heavy leptons can
originate either from a triplet model with the Type III Seesaw mechanism, 
a sequential fourth generation leptons model, or a vector-like doublet model. 
In those three scenarios, the heavy leptons couple differently to the $Z$ boson. 
For triplet fermions and vector-like leptons, the coupling to the $Z$
boson are purely vectorial, while a sequential fourth generation lepton 
has also an axial coupling which will lead to a non vanishing forward-backward
asymmetry as will demonstrate later. We summarize these couplings in Table \ref{Tab:IV}
of Appendix A.

\begin{figure}[tb]
\begin{center}
\vskip -4cm
\scalebox{0.51}{\includegraphics[angle=0]{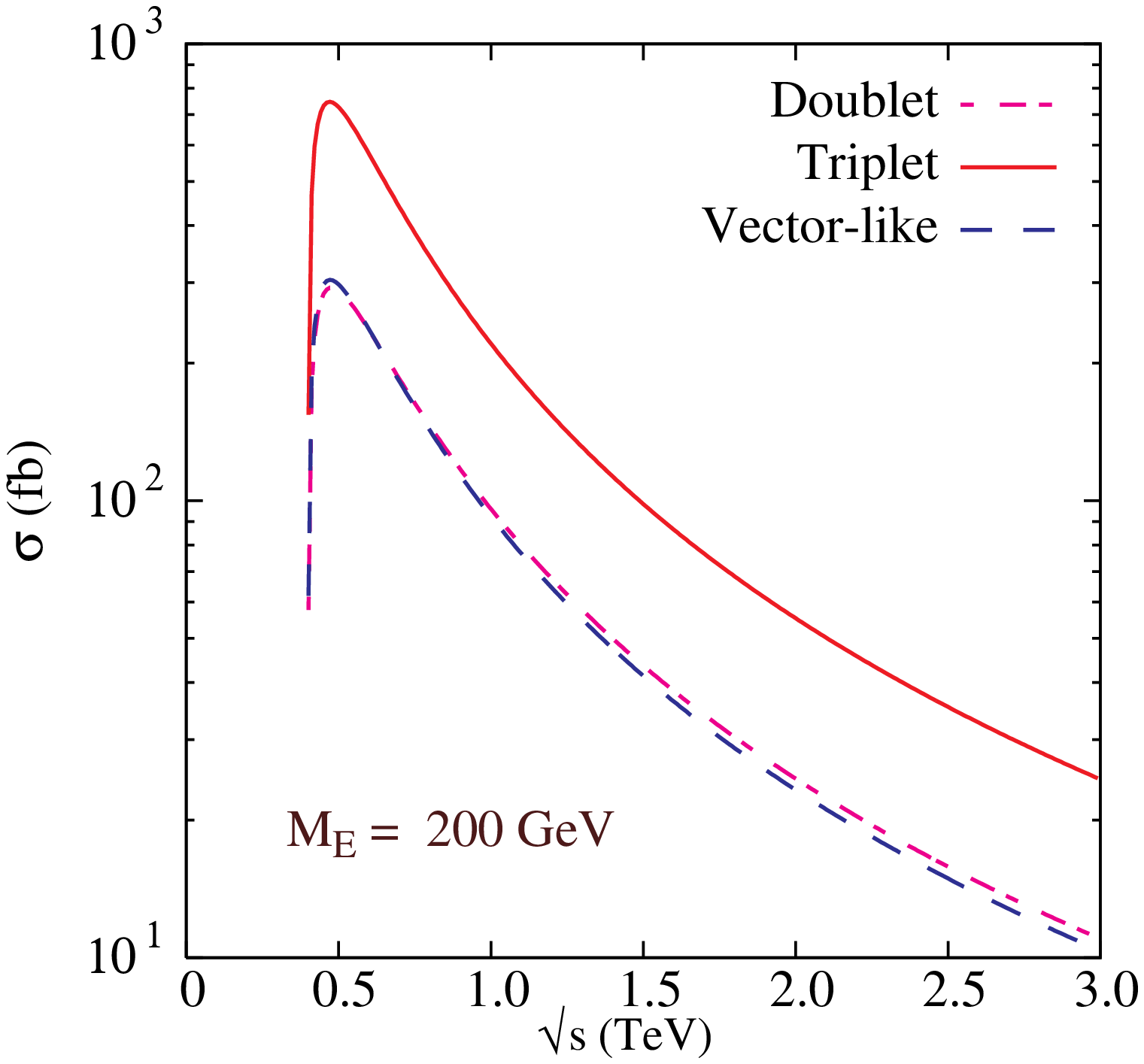}} 
\scalebox{0.51}{\includegraphics[angle=0]{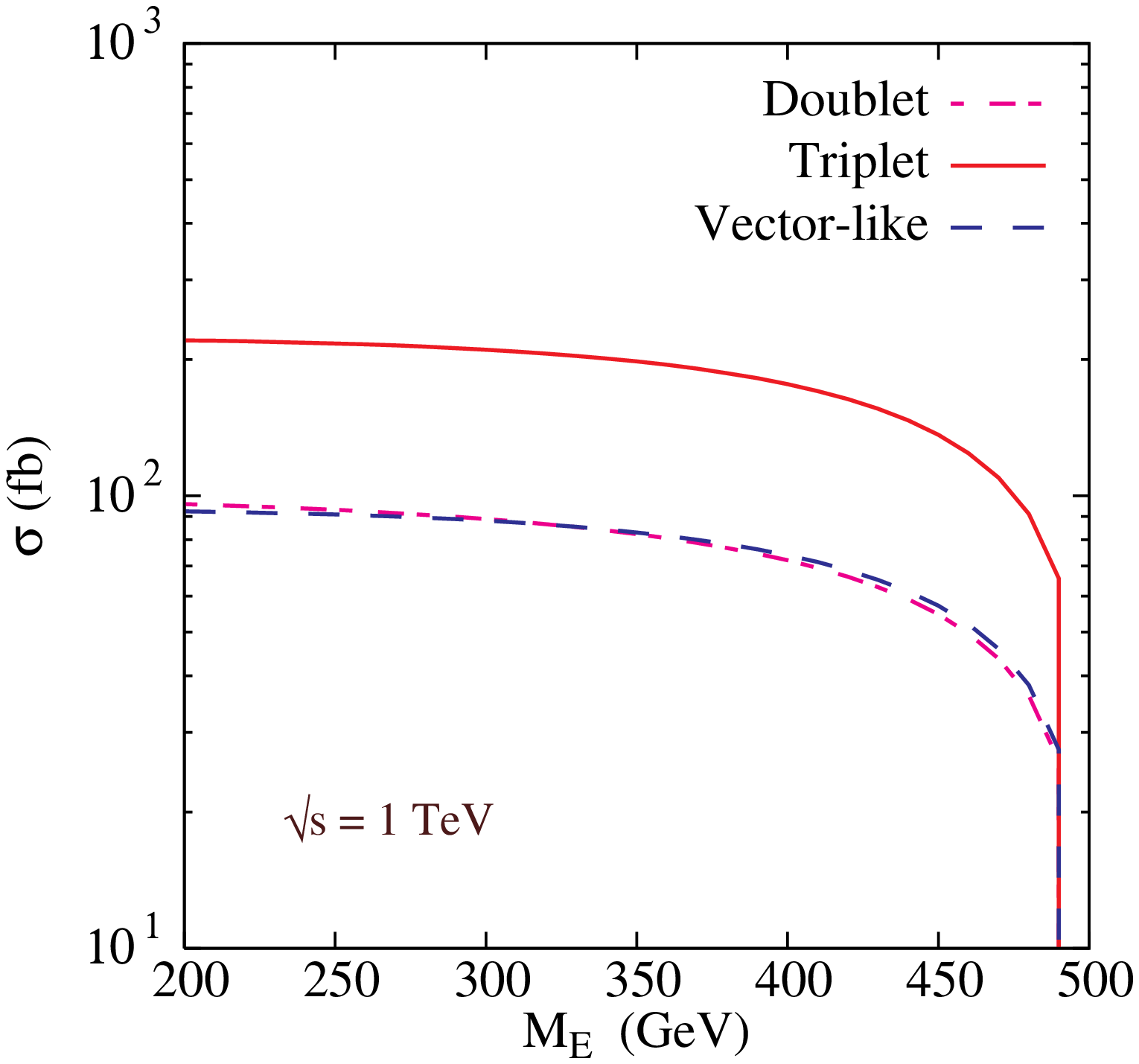}}\\
\end{center}
\caption {$\sigma(e^+ e^- \to E^+E^-)$ in units of fb as a function
  of center-of-mass energy $\sqrt s$ 
  with $M_E=200 \gev$ (left), and as a function of $M_E$
with $\sqrt{s}=1 \tev$ for a triplet, a sequential fourth generation and vector-like doublet models.}
\label{ilc1}
\end{figure}

At an $e^+e^-$ collider, the only Feynman diagrams 
that contribute to a pair production of heavy leptons are 
the photon and $Z$ boson $s$-channel exchange. The corresponding 
cross section formulae  are given in Appendix B.
The total cross sections for the three models under consideration 
are illustrated in Fig.~\ref{ilc1}, as a function of (a) the  
center-of-mass energy $\sqrt{s}$ for $M_E=200$ GeV 
(left panel) and (b) $M_E$ for $\sqrt{s}=1$ TeV (right panel).
It is clear from the plots, that the sequential doublet and vector-like model
gives a similar cross  section while the triplet model give a cross section
which is more than twice as large. The reason is that in the case of triplet model
the purely vectorial coupling of $-2 \cos^2\theta_W$ is much larger than
the corresponding couplings for doublet and vector-like, as listed in  
Table \ref{Tab:IV}.

As one can see from Fig.~\ref{ilc1}(a), the 
total cross section above the threshold 
can reach 700 fb for $\sqrt{s}=500 \gev$ and a triplet mass
of $200 \gev$. For an integrated luminosity of 
$500\ \rm{fb}^{-1}$, this cross section would yield a couple of hundred thousand
events before the detector acceptance. 
For center of mass energy around 1 TeV, the cross
section for a triplet is still about 200 fb for  masses in the range of
200$-$400 GeV.

\begin{figure}
\begin{tabular}{cc}
\hspace{-1.cm}\resizebox{120mm}{!}{\includegraphics{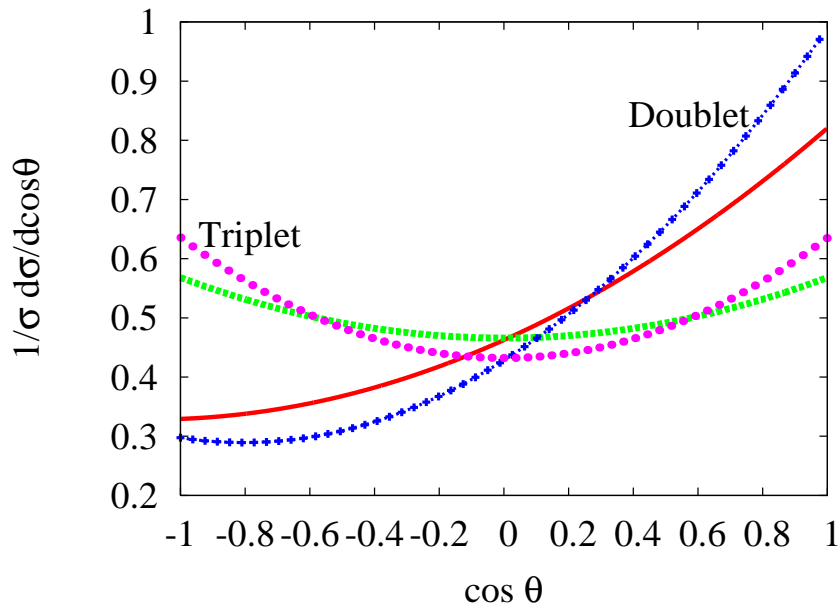}} 
\end{tabular}
\caption {Normalized angular  distributions for a pair production of 
lepton doublet (forward) and  triplet (central) with $M_E=200 \gev$ 
and $\sqrt{s}=500 \gev$ (solid curves) and $M_E=300 \gev$ and 
$\sqrt{s}=1 \tev$ (dotted curves).}
\label{ilc2}
\end{figure}

The process for triplet or vector-like leptons with 
$Z \bar EE$  coupling purely vectorial ($a_E=0$)
has an angular distribution of the generic form
\beq
{d\sigma\over d\cos\theta} \propto a(1+\cos^2\theta) + b(1-\cos^2\theta),
\eeq
with $\theta$ the scattering angle between the $e^-$ beam and $E^+$. 
This angular distribution gives a vanishing forward backward asymmetry due to parity conservation.
This would distinguish between the triplet and a sequential leptonic doublet. 
In Fig.~\ref{ilc2} we illustrate the angular distributions normalized to the
total cross section for $M_E=200\ \gev,\ \sqrt{s}=500 \gev$ (solid curves) 
and $M_E=300\ \gev,\ \sqrt{s}=1 \tev$ (dotted curves).
As expected from the above discussion, the angular
distribution for the triplet and vector-like fermion is completely symmetric
and would give vanishing forward-backward cross section. While in the case of
sequential fourth generation leptons we get an asymmetric angular distribution
which would give a forward-backward asymmetry of the order of 50\% (70\%)
for $\sqrt{s}=500 \gev\ (1 \tev)$. Evidently, the asymmetry variables would be the most
conclusive test of the chiral nature of the couplings. Even with a clear signal observation of heavy
lepton production at the LHC,
the fact that it is hard to determine the absolute cross section and to measure an asymmetry in an angular
distribution makes an $e^+e^-$ linear collider an ideal machine for detailed property studies. 

\section{Conclusions}

The minimal Georgi-Glashow SU(5) model is a remarkably predictive grand 
unified theory. It leads to massless neutrinos and it fails to unify the SM gauge
couplings with the current accuracy from the electroweak precision measurements. 
In a sense the failure of the minimal theory may be for the best, for it
predicted the desert,
with no new physics beyond the electro-weak scale, all the way up to the $M_{GUT}$
around
$10^{16}$ GeV or so. Even if correct, it would leave us with nothing new to
observe  at the next generation of  colliders such as the LHC,  
and we would have only indirect tests, albeit exciting, such as proton
decay. The desert could have an oasis of supersymmetry, but its motivation is mainly
inspired by the hierarchy issue and not a physical, phenomenological need.
And, strictly
speaking, the supersymmetry scale could be above the reach of LHC energies,
leaving us with
a mild hierarchy issue, far less dramatic than the original one of the
doublet-triplet splitting.

Instead, one can ask what is the minimal extension that can give realistic
neutrino masses and mixings
and the unification of gauge couplings.  It turns out that augmented 
by an adjoint fermion, it remains equally remarkably predictive but now in accord 
with the experiment. In the popular seesaw language it is a hybrid of Type I and 
Type III and automatically predicts one massless neutrino. The Type III triplet fermion 
must lie below TeV and should be looked for at the LHC. Its decays probe the 
Yukawa couplings responsible for the neutrino mass matrix. It is noteworthy that the 
term predictive is used here in a narrow strict sense: no further assumption or principle 
has been used beyond the one of grand unification.

In summary, in the Georgi-Glashow SU(5)  the desert was intimately related with 
the desire of massless neutrinos. Non vanishing
neutrino masses lead a number of oasis in the desert, and mostly important offer
new physics for the 
LHC.  If anything, this can be viewed as an example of how a phenomenological
motivation of
neutrino masses and mixings can lead to new phenomena at the LHC, and provides a
counter example
to a prejudice that the seesaw mechanism in the context of grand unification 
needs new large scales. Just as
right handed neutrinos of the Type I  seesaw tend to lie close to $M_{GUT}$ in
SO(10), here the mass of the analogous SU(2) triplet fermion must lie below TeV.

In this paper we have offered an in-depth study of the collider signatures of this model
at colliders such as the Tevatron, the LHC, and an $e^+e^-$ linear collider. 
The smoking gun is the production of lepton number violating same-sign dileptons 
plus four jets without significant missing energy. 
Our analysis shows that for an integrated luminosity of 8 fb$^{-1}$,
the Tevatron could probe the lepton triplet up to a mass 200 GeV.
At the  LHC with a luminosity of 10 (100) fb$^{-1}$, one may 
probe the triplet up to a mass 450 (700) GeV. 
We also provide some general remarks on how to distinguish the lepton triplet from 
other heavy leptonic states. As a complementary study in this regard, we
present the results for cross sections and distributions at an $e^+e^-$ linear
collider for a few representative models.

\subsection*{Acknowledgment}
We thank Ilja Dor\v sner, Pavel Fileviez Perez, Srubabati Goswami, Michel Herquet, 
Anjan Joshipura, Borut Ker\v sevan, Rajko Krivec, Miha Nemev\v sek, 
Kerim Suruliz and Enkhbat Tsedenbaljir for discussion and correspondence. 
A.A., D.K.G. and I.P.~would like to thank the ICTP for hospitality during 
their stay where this work was started.
The work of B.B. has been supported by the Slovenian Research Agency.
The work of D.K.G. has been partially supported by the Department of
Science and Technology, India under grant SR/S2/HEP-12/2006. 
The work of T.H.~is supported by the U.S.~Department of Energy under grant 
No.~DE-FG02-95ER40896. 
The work of G.-Y.H.~is supported by the U.S.~Department of Energy under
grant No.~DE-FG02-91ER40674 and by the U.C.~Davis HEFTI program.
The work of I.P.~is supported by the Croatian Ministry of Science,
Education and Sport under grant No. 023-0982887-3064.
The work of G.S.~is partially supported by the EU FP6 Marie Curie
Research $\&$ Training Network ``UniverseNet" (MRTN-CT-2006-035863). 
\appendix

\section{Heavy leptons in Type III Seesaw model }
\subsection{The Lagrangian}
We start with the Lagrangian of the SM leptons, i.e the three generations of 
left-handed doublet $L_i$ and right-handed singlet $e^c_i$
\begin{eqnarray}
L_i=\begin{pmatrix}
\nu_i \cr
e_i\cr
\end{pmatrix}, \qquad
e^c_i  , \nonumber
\end{eqnarray}
and an SU(2)$_L$ vector-like triplet $T$ with zero hypercharge, plus a leptonic singlet $S$:
\begin{equation}
T^k=(T^1,T^2,T^3)\equiv\left(\frac{T^-+T^+}{\sqrt{2}},\frac{T^--T^+}{i\sqrt{2}},T^0\right),
\qquad S. \nonumber
\end{equation}
In the two-component notation, we write 
\begin{eqnarray}
\label{kin}
{\cal L}_{kin}&=&\overline{L}_ji\bar\sigma^\mu\left(\partial_\mu
+ig'\frac{1}{2}B_\mu-ig\frac{\tau^a}{2}A_\mu^a\right)L_j
+ \overline{e^c}_ji\bar\sigma^\mu\left(\partial_\mu-ig'B_\mu\right)e^c_j 
\nonumber\\
&+&\overline{T^k}i\bar\sigma^\mu\left(\partial_\mu T^k-
g\epsilon^{akj}A_\mu^aT^j\right)+\overline{S}i\bar\sigma^\mu\partial_\mu S.
\end{eqnarray}
with 
$\bar\sigma^\mu=\left(1,\sigma^i\right),$ 
where $\sigma^i$ are the Pauli matrices.

One can add to this Lagrangian Majorana  mass terms for the triplet
and singlet 
\begin{equation}
{\cal L}_{M}=-\frac{M_T}{2}\left(2T^+T^-+T^0T^0\right)
-\frac{M_S}{2}SS+h.c.
\end{equation}
 With properly defined $T^k$ and $S$ the masses 
$M_T$ and $M_S$ can be made real and positive. 

The Yukawa terms are written as
\begin{equation}
\label{yukawa}
{\cal L}_Y=
-y_E^iH^\dagger e^c_iL_i
+y_T^iH^Ti\tau^2\tau^aT^aL_i
+y_S^iH^Ti\tau^2SL_i+h.c.
\end{equation}
where $y_E^i$ are the physical (real and diagonal) charged lepton Yukawas. 
After the spontaneous symmetry breaking of SU(2)$_L\times$U(1)$_Y$, 
working in the Unitary gauge, 
\begin{eqnarray}
H\to
\begin{pmatrix}
  0
\cr
  v+h/\sqrt{2}
\cr
\end{pmatrix} ,
\end{eqnarray}
with $v\approx 174$ GeV, we have the Yukawa terms (\ref{yukawa}) written as 
\begin{equation}
\label{yukawaafter}
{\cal L}_Y\to -
\left(v+\frac{h}{\sqrt{2}}\right)\left(y_E^ie^c_ie_i
+y_T^i\left(\sqrt{2}T^+e_i+T^0\nu_i\right)
+y_S^iS\nu_i\right)+h.c.
\end{equation}

The electroweak gauge interactions of the fermions in (\ref{kin})
are thus written as
\begin{eqnarray}
\label{gauge}
{\cal L}_{gauge}&=&
eA_\mu\left(-\bar e_j\bar\sigma^\mu e_j+
\bar{e^c}_j\bar\sigma^\mu e^c_j+\overline{T^+}\bar\sigma^\mu T^+
-\overline{T^-}\bar\sigma^\mu T^-\right)\nonumber\\
&+&\frac{eZ_\mu}{2s_wc_w}\left(
\bar\nu_j\bar\sigma^\mu\nu_j+
\left(2s_w^2-1\right)\bar e_j\bar\sigma^\mu e_j
-2s_w^2\bar{e^c}_j\bar\sigma^\mu e^c_j\right.\nonumber\\
&&{\hskip 1.2cm}+\left.2c_w^2\overline{T^+}\bar\sigma^\mu T^+
-2c_w^2\overline{T^-}\bar\sigma^\mu T^-\right)\nonumber\\
&+&\frac{e}{s_w}W_\mu^+\left(
\frac{1}{\sqrt{2}}\bar\nu_j\bar\sigma^\mu e_j
+\overline{T^0}\bar\sigma^\mu T^-
-\overline{T^+} \bar\sigma^\mu T^0
\right)\nonumber\\
&+&\frac{e}{s_w}W_\mu^-\left(
\frac{1}{\sqrt{2}}\bar e_j \bar\sigma^\mu \nu_j
+\overline{T^-} \bar\sigma^\mu T^0
-\overline{T^0} \bar\sigma^\mu T^+
\right), 
\end{eqnarray}
where
\begin{eqnarray}
B_\mu=c_wA_\mu-s_wZ_\mu&,&A_\mu^1=\frac{W_\mu^-+W_\mu^+}{\sqrt{2}}\;,\\
A^3_\mu=s_wA_\mu+c_wZ_\mu&,&A_\mu^2=\frac{W_\mu^--W_\mu^+}{i\sqrt{2}}\; ,
\end{eqnarray}
and $c_w\equiv \cos{\theta_w}$, $s_w\equiv \sin{\theta_w}$, $e=gs_w=g'c_w$.

\subsection{Fermion mass eigenstates}

The mass terms for the neutral fermions are 
\begin{eqnarray}
-\frac{1}{2}
\begin{pmatrix}
  \nu_i
& 
  T^0
&
  S
  \end{pmatrix}
\begin{pmatrix}
  0
& 
  vy_T^i
&
  vy_S^i
\cr
  vy_T^j
& 
  M_T
&
  0
\cr
  vy_S^j
&
  0
&
  M_S
\cr
\end{pmatrix}
\begin{pmatrix}
  \nu_j
\cr
  T^0
\cr
  S
\cr
\end{pmatrix}
+h.c.
\end{eqnarray}
The symmetric complex $5\times 5$ mass matrix can be 
diagonalized by a unitary transformation
\begin{eqnarray}
\label{nic}
\begin{pmatrix}
  \nu_j
\cr
  T^0
\cr
  S
\cr
\end{pmatrix}
\to U_0
\begin{pmatrix}
  \nu_j
\cr
  T^0
\cr
  S
\cr
\end{pmatrix}\;.
\end{eqnarray}
In the leading order for $\left|vy_{T,S}^i\right|\ll M_{T,S}$, 
the unitary matrix is 
\begin{eqnarray}
\label{unic}
U_0 \approx 
\begin{pmatrix}
  1_{3\times 3} & \epsilon_T^* & \epsilon_S^*
\cr
  -\epsilon_T^T  & 1          & 0 
\cr
  -\epsilon_S^T  & 0          & 1
\cr
\end{pmatrix},\quad
\epsilon_X^i= \frac{vy_X^i}{M_X}\;.
\end{eqnarray}
After this transformation the mass terms of the neutral 
fields become approximately
\begin{equation}
-\frac{1}{2}M_TT^0T^0
-\frac{1}{2}M_SSS
-\frac{1}{2}m^\nu_{ij}\nu_i\nu_j\;,
\end{equation}
with
\begin{equation}
\label{neutrinomass}
m^\nu_{ij}=-M_T\epsilon_T^i\epsilon_T^j-
M_S\epsilon_S^i\epsilon_S^j\;.
\end{equation}
The neutrino mass matrix gets diagonalized with 
the final transformation
\begin{equation}
\nu\to U_{PMNS}\nu\;.
\end{equation}

The mass matrix for charged fermions is 
\begin{eqnarray}
-
\begin{pmatrix}
  e^c_i
& 
  T^+
  \end{pmatrix}
\begin{pmatrix}
  m_E^i\delta_{ij}
& 
  0
\cr
  vy_T^j
& 
  M_T
\cr
\end{pmatrix}
\begin{pmatrix}
  e_j
\cr
  T^-
\cr
\end{pmatrix}
+h.c.\ \  , \quad 
m_E^i\equiv vy_E^i \ .
\end{eqnarray}
This $4\times 4$ complex mass matrix can be diagonalized by two 
unitary matrices:

\begin{eqnarray}
\label{minus}
\begin{pmatrix}
  e_j
\cr
  T^-
\cr
\end{pmatrix}
\to U_-
\begin{pmatrix}
  e_j
\cr
  T^-
\cr
\end{pmatrix}\;,\quad
\begin{pmatrix}
  e^c_j
\cr
  T^+
\cr
\end{pmatrix}
\to U_+
\begin{pmatrix}
  e^c_j
\cr
  T^+
\cr
\end{pmatrix}\;.
\end{eqnarray}
In the same approximation as before and for $|m_E^i|\ll M_T$
\begin{eqnarray}
\label{upm}
U_- \approx 
\begin{pmatrix}
  1_{3\times 3} & \sqrt{2}\epsilon_T^* 
\cr
  -\sqrt{2}\epsilon_T^T  & 1
\cr
\end{pmatrix}\;,\quad
U_+\approx 1\;.
\end{eqnarray}

\subsection{Interaction in the mass eigenbasis}

We will eventually be interested in the decay rates of the 
triplets into a light lepton and a gauge boson. 
These go through one power of the small Dirac 
Yukawa couplings $y_T^i$. It is thus sufficient at leading order to 
make the following substitutions in (\ref{gauge}), coming from 
(\ref{nic})$-$(\ref{unic})
\begin{eqnarray}
\nu_j \to \nu_j+\epsilon_T^{j*}T^0+
\epsilon_S^{j*}S\;,\quad
T^0 \to T^0-\epsilon_T^k\nu_k \;,\quad
S \to S-\epsilon_S^k\nu_k\;,
\end{eqnarray}
and (\ref{minus})$-$(\ref{upm})
\begin{eqnarray}
e_j \to e_j+\sqrt{2}\epsilon_T^{j*}T^- \;,\quad
T^- \to T^--\sqrt{2}\epsilon_T^ke_k\;.
\end{eqnarray}
All the fields on the right-hand sides except the light neutrinos are the 
mass eigenstates, i.e. the asymptotic states in scattering and decays. 
Neutrinos in final states will be always summed over all three 
generations, so their true basis is irrelevant. 
Equation (\ref{gauge}) gets some new terms, that come from mixings:
\begin{eqnarray}
\label{gaugeafter}
\delta{\cal L}_{gauge}&=&
\frac{eZ_\mu}{2s_wc_w}\left(
\epsilon_T^j \overline{T^0} \bar\sigma^\mu \nu_j +
\sqrt{2}\epsilon_T^j\overline{T^-} \bar\sigma^\mu e_j + 
\epsilon_S^j \overline{S} \bar\sigma^\mu \nu_j
\right)\nonumber\\
&+&
\frac{e}{s_w}W_\mu^+\left(
\epsilon_T^j \overline{T^+} \bar\sigma^\mu \nu_j
-\frac{1}{\sqrt{2}} \epsilon_T^j \overline{T^0} \bar\sigma^\mu e_j
+\frac{1}{\sqrt{2}} \epsilon_S^j \overline{S} \bar\sigma^\mu e_j 
\right)
+h.c.
\end{eqnarray}

\subsection{Interaction in the four component notation}

For some convenience, we rewrite the Lagrangian in four-component notation, in which 
the Dirac fields are
\begin{eqnarray}
e_j=
\begin{pmatrix}
  e_j
\cr
  \bar{e^c}_j
\cr
\end{pmatrix}\;,\quad
T^-=
\begin{pmatrix}
  T^-
\cr
  \overline{T^+}
\cr
\end{pmatrix}\;,
\end{eqnarray}
and the Majorana fields are 
\begin{eqnarray}
\nu_j=
\begin{pmatrix}
  \nu_j
\cr
  \bar\nu_j
\cr
\end{pmatrix}\;,\quad
T^0=
\begin{pmatrix}
  T^0
\cr
  \overline{T^0}
\cr
\end{pmatrix}\;.
\end{eqnarray}
Keeping in mind our convention \cite{Derendinger:1990tj} 
(see also \cite{Dreiner:2008tw})
\begin{eqnarray}
\gamma^\mu=
\begin{pmatrix}
  0 & \sigma^\mu
\cr
  \bar\sigma^\mu & 0
\cr
\end{pmatrix},\quad
\gamma^5=
\begin{pmatrix}
  1 & 0
\cr
  0 & -1
\cr
\end{pmatrix},\quad 
\sigma^\mu=\left(1,-\sigma^i\right), 
\end{eqnarray}
and using the relation
\begin{equation}
\bar\psi\bar\sigma^\mu\chi=
-\chi\sigma^\mu\bar\psi
\end{equation}
with $\psi$ and $\chi$ two-component spinors, 
the quadratic terms are rewritten in the four-component notation as 
\begin{eqnarray}
{\cal L}_{kin}
=\bar e_ji\gamma^\mu\partial_\mu e_j
+\overline{T^-}\left(i\gamma^\mu\partial_\mu-M_T\right)T^- 
+\frac{1}{2}\bar\nu_ji\gamma^\mu\partial_\mu \nu_j
+\frac{1}{2}\overline{T^0}\left(i\gamma^\mu\partial_\mu-M_T\right)T^0,\quad
\end{eqnarray}
where we assume that all SM light charged leptons and neutrinos are massless. 
Before mixing the gauge interactions (\ref{gauge}) become
\begin{eqnarray}
\label{gaugefour}
{\cal L}_{gauge} &=&- eA_\mu \left(\bar e_j\gamma^\mu e_j+
\overline{T^-}\gamma^\mu T^-\right)\nonumber\\
&+& \frac{e}{2s_wc_w}Z_\mu \left( \bar\nu_j\gamma^\mu P_+\nu_j+
 (2s_w^2-1 )\bar e_j\gamma^\mu P_+e_j
+2s_w^2\bar e_j\gamma^\mu P_-e_j 
- 2c_w^2\overline{T^-}\gamma^\mu T^- \right) \nonumber\\
&+& \frac{e}{s_w} W_\mu^+ \left( \frac{1}{\sqrt{2}}\bar\nu_j\gamma^\mu P_+e_j
+\overline{T^0}\gamma^\mu T^- \right)  
+ \frac{e}{s_w} W_\mu^- \left( \frac{1}{\sqrt{2}} \bar e_j\gamma^\mu P_+\nu_j
+\overline{T^-}\gamma^\mu T^0 \right),~~~~~
\end{eqnarray}
with the projection operators defined as $P_\pm=(1\pm\gamma^5 )/2$
for the left-handed (+) and right-handed ($-$) chiralities. 

The extra terms (\ref{gaugeafter}) are put in the form
\begin{eqnarray}
\label{deltagaugefour}
\delta{\cal L}_{gauge}&=& 
\frac{e}{2s_wc_w}Z_\mu \left(
\epsilon_T^j \overline{T^0} \gamma^\mu P_+ \nu_j
+\sqrt{2} \epsilon_T^j \overline{T^-} \gamma^\mu P_+ e_j 
+\epsilon_S^j \overline{S} \gamma^\mu P_+ \nu_j
\right)
\nonumber\\
&+&
\frac{e}{s_w}W_\mu^+ \left(
- \epsilon_T^j \bar\nu_j \gamma^\mu P_- T^- 
- \frac{1}{\sqrt{2}} \epsilon_T^j \overline{T^0} \gamma^\mu P_+ e_j  
+ \frac{1}{\sqrt{2}} \epsilon_S^j \bar S \gamma^\mu P_+ e 
\right)
+h.c.
\end{eqnarray}

One can transform to the four component notation also the 
Yukawa interactions (\ref{yukawaafter}). The pieces new with respect 
to the SM are 
\begin{eqnarray}
\delta{\cal L}_{Y}= - \frac{h}{\sqrt{2}} \left(
\sqrt{2} y_T^j \overline{T^-} P_+ e_j 
+ y_T^j \bar\nu_j P_+ T^0 
+ y_S^j \bar\nu_j P_+ S 
\right) + h.c.
\end{eqnarray}

\subsection{Comparison of heavy leptons gauge couplings}

In this appendix, we list the  heavy lepton couplings to the SM gauge bosons. 
We denote a generic charged lepton by $E$ (with charge -1), and a neutral lepton by $N$. 
We first neglect the small mixings with the SM leptons 
and write the gauge couplings in a  form
\begin{eqnarray}
{\cal{L}}
\nonumber
&=& \frac{g}{\sqrt 2 } W^{+\mu} \ 
             \overline{N} \gamma_\mu (v_{C}+a_C\gamma_5 ) E + h.c. \\
&+& \frac{g}{2 \cos\theta_W}Z^\mu \ 
             (\overline{E} \gamma_\mu (v_E+a_E\gamma_5)E +
             \overline{N} \gamma_\mu (v_N+a_N\gamma_5)N)
\end{eqnarray}
The vector and axial couplings are summarized in Table.~\ref{Tab:IV}.

\begin{table}[tb]
\begin{tabular}{| c || c | c || c | c || c | c|  }
\hline
      & $v_C$ & $a_C$ & $v_E \qquad$ & $a_E$ & $v_N$ & $a_N$  \\
\hline
\hline
      triplet& $\sqrt{2}$ & $0$ & $-2\cos^2{\theta_W}$ & $0 $ & $0 $ & $0 $   \\
\hline
      vector-like doublet& $1$ & $0$ & $-1+2\sin^2{\theta_W}$ & $0 $ & $1 $ & $0 $   \\
\hline
      sequential doublet& $1/2$ & $1/2$ & $-1/2+2\sin^2{\theta_W}$ & $-1/2 $ & $1/2 $ & $1/2$   \\
\hline
  \end{tabular}
  \caption{Electroweak couplings of the extra leptons.}
\label{Tab:IV}
\end{table}

For off-diagonal couplings between a heavy lepton and a SM lepton, 
simply include a mixing such as $\epsilon_T^i$. 

\section{Production cross sections and decay rates}

\subsection{ Production cross sections}
\subsubsection*{\underline{$f\bar{f}\to E^+E^-$}}
The Drell-Yan  mechanism $f\bar{f}\to E^+E^-$, 
with $E^\pm$ is a heavy lepton, proceed through photon and $Z$ 
boson $s$-channel exchange. 
The corresponding differential cross section is found to be:
\begin{eqnarray}
\frac{d\sigma}{d\Omega} = 
\frac{\beta}{64\pi^2 s}\frac{1}{4}\frac{1}{N_c} (|M_\gamma|^2
+ |M_Z|^2 + 2 \Re e(M_\gamma^* M_Z))
\end{eqnarray}
with
\begin{eqnarray}
|M_\gamma|^2&=& 4Q_f^2 Q_E^2 e^4   [
(1+\cos^2\theta) +
(1-\cos^2\theta) \gamma^{-2}\  ]\nonumber\\
|M_Z|^2&=&  \frac{g^4}{4 \cos^4\theta_W} \ {s^2\over (s -m_Z^2)^2 }\ 
[8 a_f v_f a_E v_E \beta \cos \theta  \nonumber\\ &+&  
(a_f^2+v_f^2) \{ (a_E^2 \beta^2 + v_E^2) ( 1+\cos^2\theta) 
 + v_E^2 (1-\cos^2\theta) \gamma^{-2}\}  ] \\
\Re e(M_\gamma^* M_Z)
&=& Q_f Q_E \frac{e^2g^2}{ \cos^2\theta_W} 
{s \over s - m_Z^2}   
[ 2 a_f a_E  \beta \cos \theta + v_f v_E \{ 
( 1+\cos^2\theta) + (1-\cos^2\theta) \gamma^{-2} \} ] 
 \nonumber
\label{diff}
\end{eqnarray}
where $N_c$ is a color factor and is 1 (3) for $f$ is a lepton (quark).
$\theta$ is the scattering angle of 
$E^+$ with respect to the ${f}$ beam direction, 
$\beta=\sqrt{1-4 M_E^2/s}$ is the speed of the outgoing
particle in the CM frame, with $\gamma^{-2}=4 M_E^2/s$. 
The coupling of the heavy lepton to Z boson are listed 
in Table \ref{Tab:IV} while the SM coupling of $Z$ to initial states 
fermions are given by:
\begin{eqnarray}
&& v_e=-1/2-2Q_e\sin^2{\theta_W} \ , \ v_d=-1/2-2Q_d\sin^2{\theta_W} \ , \ 
v_u=1/2-2 Q_u \sin^2{\theta_W} \nonumber\\
&&a_e=-1/2\ , \ a_d=-1/2 \ , \ 
a_d=1/2 \nonumber
\end{eqnarray}
with $Q_e=-1$, $Q_d=-1/3$ and $Q_u=2/3$

\subsubsection*{\underline{$f \bar{f'}\to W^*\to E^+N^0$}}

The production mechanism $f\bar{f' }\to E^+N$, 
with $E^\pm$ is a heavy charged lepton and $N$ a neutral one, 
proceed through $W^\pm$ boson $s$-channel exchange. 
The differential cross section is:
\begin{eqnarray}
\frac{d\sigma}{d\Omega} &=& 
\frac{\beta}{64\pi^2 s}\frac{1}{4}\frac{1}{N_c} 
g^4 \ {s^2\over (s -m_W^2)^2 }\ 
[8 v_f^2  v_C a_C  \beta \cos \theta  \nonumber\\ &+&  
2 v_f^2 \{ (a_C^2 \beta^2 + v_C^2) ( 1+\cos^2\theta) 
 +  v_C^2 (1-\cos^2\theta) (1-\beta^2)\}  ]
\end{eqnarray}
with $v_f=a_f=1/2$, $V_C$ and $a_C$ are given in Table \ref{Tab:IV}.

\subsubsection*{\underline{$f \bar{f'}\to W^* \to N^0 l_j^\pm$}}

The production mechanism $f'\bar{f}\to N^0 l_j^\pm$, 
with $N^0$ is a heavy neutral lepton 
proceed through $W^\pm$ boson $s$-channel exchange. 
The differential cross section is:
\begin{eqnarray}
\frac{d\sigma}{d\Omega} &= &
\frac{\beta}{64\pi^2 s}\frac{1}{4}\frac{1}{N_c} g^4 {s^2\over (s -m_W^2)^2 }
 [ 8 a_f a_C v_f v_C \cos\theta 
\nonumber \\ 
&&
+\beta (v_f^2+a_f^2) 
 (2 + \cos^2\theta - \beta^2) (a_C^2 + v_C^2) ] 
 \label{prod:nl}
\end{eqnarray}
with $v_f=a_f=1/2$, $V_C=a_C=1/2 \epsilon_T^j$, and
$\theta$ is the scattering angle of 
$N^0$ with respect to the ${f}$ beam direction.

\subsubsection*{\underline{$f\bar{f}\to Z^*\to E^\mp l_j^\pm$}}

The production mechanism $f\bar{f}\to E^\mp l_j^\pm$, 
with $E^\pm$ is a heavy charged lepton proceed through $Z$ boson $s$-channel 
exchange. 
The differential cross section is the same as Eq.~(\ref{prod:nl})
with the following replacements:
$g\to g/(\sqrt{2}\cos\theta_W)$, $m_W\to m_Z$, $M_N\to M_E$, 
$v_f\to v_{u,d}$, $a_f\to a_{u,d}$
and $a_C=v_C=1/\sqrt{2} \epsilon_T^j$

\subsection{Decay widths for $T\to W, Z, h + \;{\rm light \; lepton}$}

If kinematically accessible, 
the predominant decay modes of the triplet leptons will be to the gauge bosons
(or a Higgs boson) plus a SM light lepton, 
whose coupling strength is dictated by the neutral Dirac Yukawa couplings. 
Those decay widths are listed below \cite{Bajc:2007zf}:
\begin{eqnarray} 
\label{t-w}
\sum_k\Gamma(T^-\to W^-\nu_k)&=&
\frac{M_T}{16\pi}\left(\sum_k\left|y_T^k\right|^2\right)
\left(1-\frac{m_W^2}{M_T^2}\right)^2\left(1+2\frac{m_W^2}{M_T^2}\right)\;,\\
\label{t-z}
\Gamma(T^-\to Ze_k^-)&=&
\frac{M_T}{32\pi}\left|y_T^k\right|^2 
\left(1-\frac{m_Z^2}{M_T^2}\right)^2\left(1+2\frac{m_Z^2}{M_T^2}\right)\;,\\
\Gamma\left(T^-\to h e_k^-\right)&=&\frac{M_T}{32\pi}\left|y_T^k\right|^2
\left(1-\frac{ m_h^2  }{ M_T^2  }\right)^2\;,\\
\label{t0w}
\Gamma(T^0\to W^+e_k^-)&=&
\Gamma(T^0\to W^-e_k^+)=\nonumber\\
&=&\frac{M_T}{32\pi}\left|y_T^k\right|^2
\left(1-\frac{m_W^2}{M_T^2}\right)^2\left(1+2\frac{m_W^2}{M_T^2}\right)\;,\\
\label{t0z}
\sum_k\Gamma(T^0\to Z\nu_k)&=&
\frac{M_T}{32\pi}\left(\sum_k\left|y_T^k\right|^2\right)
\left(1-\frac{m_Z^2}{M_T^2}\right)^2\left(1+2\frac{m_Z^2}{M_T^2}\right)\;, \\
\label{t0h}
\sum_k\Gamma\left(T^0\to h \nu_k\right)&=&\frac{M_T}{32\pi}
\left(\sum_k\left|y_T^k\right|^2\right)
\left(1-\frac{ m_h^2  }{ M_T^2  }\right)^2\; ,
\end{eqnarray}
where we averaged over initial polarizations and summed over final ones.



\begin{thebibliography}{99}
\bibitem{Weinberg:1979sa}
  S.~Weinberg,
  Phys.\ Rev.\ Lett.\  {\bf 43} (1979) 1566.

\bibitem{Ma:1998dn}
  E.~Ma,
  Phys.\ Rev.\ Lett.\  {\bf 81} (1998) 1171
  [arXiv:hep-ph/9805219].

\bibitem{seesaw}
P.~Minkowski,
Phys.\ Lett.\ B {\bf 67} (1977) 421;
T.~Yanagida, proceedings of the {\em Workshop on Unified Theories
and Baryon Number in the Universe}, Tsukuba, 1979, eds.
A. Sawada, A. Sugamoto;
S.~Glashow, in {\em Cargese 1979, Proceedings, Quarks and
Leptons}
(1979) ;
M.~Gell-Mann, P.~Ramond, R.~Slansky, proceedings of the
{\em Supergravity Stony Brook Workshop}, New York, 1979,
eds. P. Van Niewenhuizen, D. Freeman; 
R.~Mohapatra, G.~Senjanovi\' c,
Phys.~Rev.~Lett.~{\bf 44} (1980) 912.

\bibitem{Magg:1980ut}
  M.~Magg and C.~Wetterich,
  Phys.\ Lett.\ B {\bf 94} (1980) 61;

\bibitem{Lazarides:1980nt}  
  G.~Lazarides, Q.~Shafi and C.~Wetterich,
  Nucl.\ Phys.\ B {\bf 181} (1981) 287;
 
\bibitem{Mohapatra:1980yp}
  R.~N.~Mohapatra and G.~Senjanovi\' c,
  Phys.\ Rev.\ D {\bf 23} (1981) 165.

\bibitem{Foot:1988aq}
  R.~Foot, H.~Lew, X.~G.~He and G.~C.~Joshi,
  Z.\ Phys.\ C {\bf 44} (1989) 441.

\bibitem{Pati:1974yy}
  J.~C.~Pati and A.~Salam,
  Phys.\ Rev.\  D {\bf 10}, 275 (1974)
  [Erratum-ibid.\  D {\bf 11}, 703 (1975)].

\bibitem{Mohapatra:1974hk}
  R.~N.~Mohapatra and J.~C.~Pati,
  Phys.\ Rev.\  D {\bf 11}, 566 (1975).

\bibitem{Senjanovic:1975rk}
  G.~Senjanovi\' c and R.~N.~Mohapatra,
  Phys.\ Rev.\  D {\bf 12}, 1502 (1975).

\bibitem{Senjanovic:1978ev}
  G.~Senjanovi\' c,
  Nucl.\ Phys.\  B {\bf 153}, 334 (1979).

\bibitem{Bajc:2006ia}
  B.~Bajc and G.~Senjanovi\' c,
  JHEP {\bf 0708} (2007) 014
  [arXiv:hep-ph/0612029].

\bibitem{Keung:1983uu}
  W.~Y.~Keung and G.~Senjanovi\' c,
  Phys.\ Rev.\ Lett.\  {\bf 50}, 1427 (1983).

\bibitem{Ma:2002pf}
  E.~Ma and D.~P.~Roy,
  Nucl.\ Phys.\  B {\bf 644} (2002) 290
  [arXiv:hep-ph/0206150].

\bibitem{Franceschini:2008pz}
  R.~Franceschini, T.~Hambye and A.~Strumia,
  Phys.\ Rev.\  D {\bf 78} (2008) 033002
  [arXiv:0805.1613 [hep-ph]].

\bibitem{delAguila:2008cj}
  F.~del Aguila and J.~A.~Aguilar-Saavedra,
  arXiv:0808.2468 [hep-ph].

\bibitem{delAguila:2008hw}
  F.~del Aguila and J.~A.~Aguilar-Saavedra,
  arXiv:0809.2096 [hep-ph].

\bibitem{Georgi:1974sy}
  H.~Georgi and S.~L.~Glashow,
  Phys.\ Rev.\ Lett.\  {\bf 32} (1974) 438.

\bibitem{Fox:2005yp}
  P.~J.~Fox {\it et al.},
  arXiv:hep-th/0503249.

\bibitem{Dorsner:2005fq}
  I.~Dor\v sner and P.~Fileviez Perez,
  Nucl.\ Phys.\  B {\bf 723} (2005) 53
  [arXiv:hep-ph/0504276].

\bibitem{Dorsner:2005ii}
  I.~Dor\v sner, P.~Fileviez Perez and R.~Gonzalez Felipe,
  Nucl.\ Phys.\  B {\bf 747} (2006) 312
  [arXiv:hep-ph/0512068].

\bibitem{Bajc:2007zf}
  B.~Bajc, M.~Nemev\v sek and G.~Senjanovi\' c,
  Phys.\ Rev.\  D {\bf 76}, 055011 (2007)
  [arXiv:hep-ph/0703080].

\bibitem{Hung:2006ap}
  P.~Q.~Hung,
  Phys.\ Lett.\  B {\bf 649} (2007) 275
  [arXiv:hep-ph/0612004].
 
\bibitem{Gudnason:2006mk}
  S.~B.~Gudnason, T.~A.~Ryttov and F.~Sannino,
  Phys.\ Rev.\  D {\bf 76} (2007) 015005
  [arXiv:hep-ph/0612230].

\bibitem{Ma:2005he}
  E.~Ma,
  Phys.\ Lett.\  B {\bf 625} (2005) 76
  [arXiv:hep-ph/0508030].

\bibitem{Schwetz:2007my}
  T.~Schwetz,
  AIP Conf.\ Proc.\  {\bf 981}, 8 (2008)
  [arXiv:0710.5027 [hep-ph]].

\bibitem{Casas:2001sr}
  J.~A.~Casas and A.~Ibarra,
  Nucl.\ Phys.\  B {\bf 618} (2001) 171
  [arXiv:hep-ph/0103065].
 
\bibitem{Ibarra:2003up}
  A.~Ibarra and G.~G.~Ross,
  Phys.\ Lett.\  B {\bf 591}, 285 (2004)
  [arXiv:hep-ph/0312138].

\bibitem{Guo:2006qa}
  W.~L.~Guo, Z.~Z.~Xing and S.~Zhou,
  Int.\ J.\ Mod.\ Phys.\  E {\bf 16} (2007) 1
  [arXiv:hep-ph/0612033].

\bibitem{Abada:2008ea}
  A.~Abada, C.~Biggio, F.~Bonnet, M.~B.~Gavela and T.~Hambye,
  Phys.\ Rev.\  D {\bf 78} (2008) 033007
  [arXiv:0803.0481 [hep-ph]].

\bibitem{He:2009tf}
  X.~G.~He and S.~Oh,
  arXiv:0902.4082 [hep-ph].
 
\bibitem{Arhrib:2009xf}
  A.~Arhrib, R.~Benbrik and C.~H.~Chen,
  arXiv:0903.1553 [hep-ph].

\bibitem{miha09}
J.~Kamenik and M.~Nemev\v sek, to appear.
  
\bibitem{Amsler:2008zz}
  C.~Amsler {\it et al.}  [Particle Data Group],
  Phys.\ Lett.\  B {\bf 667} (2008) 1.

\bibitem{Ibe:2006de}
  M.~Ibe, T.~Moroi and T.~T.~Yanagida,
  Phys.\ Lett.\  B {\bf 644} (2007) 355
  [arXiv:hep-ph/0610277].

\bibitem{Strumia:2006db}
  A.~Strumia and F.~Vissani,
  arXiv:hep-ph/0606054.

\bibitem{cteq6l} CTEQ Collaboration, 
  J.~Pumplin, D.~R.~Stump, J.~Huston, H.~L.~Lai, P.~M.~Nadolsky and W.~K.~Tung,
  JHEP {\bf 0207}, 012 (2002)
  [arXiv:hep-ph/0201195].
  
\bibitem{Han:2006ip}
  T.~Han and B.~Zhang,
  Phys.\ Rev.\ Lett.\  {\bf 97} (2006) 171804
  [arXiv:hep-ph/0604064];
  A.~Atre, T.~Han, S.~Pascoli and B.~Zhang,
  arXiv:0901.3589 [hep-ph].

\bibitem{del Aguila:2007em}
  F.~del Aguila, J.~A.~Aguilar-Saavedra and R.~Pittau,
  JHEP {\bf 0710} (2007) 047
  [arXiv:hep-ph/0703261].

\bibitem{Cheung:2005ba}
  K.~Cheung and C.~W.~Chiang,
  Phys.\ Rev.\  D {\bf 71} (2005) 095003
  [arXiv:hep-ph/0501265].

\bibitem{Abazov:2007ev}
  V.~M.~Abazov {\it et al.}  [D0 Collaboration],
  Phys.\ Rev.\ Lett.\  {\bf 99} (2007) 191802
  [arXiv:hep-ex/0702005].

\bibitem{Perez:2008ha}
  P.~Fileviez Perez, T.~Han, G.~Y.~Huang, T.~Li and K.~Wang,
  Phys.\ Rev.\  D {\bf 78} (2008) 015018
  [arXiv:0805.3536 [hep-ph]].
 
\bibitem{TeVe}
  D.~E.~Acosta {\it et al.}  [CDF Collaboration],
  Phys.\ Rev.\  D {\bf 71}, 052003 (2005)
  [arXiv:hep-ex/0410041]; 
  V.~M.~Abazov {\it et al.}  [D0 Collaboration],
  Nucl.\ Instrum.\ Meth.\  A {\bf 565}, 463 (2006)
  [arXiv:physics/0507191].

\bibitem{LHCe}
  G.~L.~Bayatian {\it et al.}  [CMS Collaboration],
  J.\ Phys.\ G {\bf 34}, 995 (2007);  \\
  G.~Aad {\it et al.}  [ATLAS Collaboration],
  arXiv:0901.0512.
%

\bibitem{Alwall:2007st}
  J.~Alwall, P.~Demin, S.~de Visscher, R.~Frederix, M.~Herquet, F.~Maltoni, 
  T.~Plehn, D.L.~Rainwater and T.~Stelzer, 
  JHEP {\bf 0709} (2007) 028
  [arXiv:0706.2334 [hep-ph]].

\bibitem{Kribs:2007nz}
  G.~D.~Kribs, T.~Plehn, M.~Spannowsky and T.~M.~P.~Tait,
  Phys.\ Rev.\  D {\bf 76} (2007) 075016
  [arXiv:0706.3718 [hep-ph]];
  W.~S.~Hou,
  arXiv:0803.1234 [hep-ph].

\bibitem{ILCLHC}  
G.~Weiglein {\it et al.}  [LHC/LC Study Group],
  Phys.\ Rept.\  {\bf 426} (2006) 47
  [arXiv:hep-ph/0410364].

\bibitem{Derendinger:1990tj}
  J.~P.~Derendinger, 
  ``Lecture Notes On Globally Supersymmetric Theories In Four-Dimensions And
  Two-Dimensions,'' in Proceedings of the Hellenic School of Particle Physics, Corfu,
  Greece, September 1989, edited by G. Zoupanos and N. Tracas; also available at
  http://www.unine.ch/phys/hepth/Derend/SUSY\_nd.pdf.

\bibitem{Dreiner:2008tw}
  H.~K.~Dreiner, H.~E.~Haber and S.~P.~Martin,
  arXiv:0812.1594 [hep-ph].
  
\end{thebibliography}
\end{document}